\documentclass[a4paper,12pt,twoside]{UGthesis}
\usepackage{fancyhdr}
\usepackage{cite}
\usepackage{enumerate}
\usepackage{amsmath}
\usepackage{amssymb}

\newcommand {\ph}[1]{\phantom{#1}}
\newcommand {\ie}{i.e.~}
\newcommand {\eg}{e.g.~}
\newcommand {\cf}{cf.~}

\newcommand {\sss} {\scriptscriptstyle}

\newcommand{\ordo}{\ensuremath{\mathcal{O}}}

\newcommand{\rfive}{\ensuremath{\mathbb{R}^5}}
\newcommand{\rsix}{\ensuremath{\mathbb{R}^6}}

\newcommand{\sone}{\ensuremath{\mathbf{S}^1}}

\newcommand{\tfour}{\ensuremath{\mathbf{T}^4}}
\newcommand{\kthree}{\ensuremath{\mathrm{K3}}}

\newcommand{\al}{\ensuremath{\alpha}}
\newcommand{\be}{\ensuremath{\beta}}
\newcommand{\ga}{\ensuremath{\gamma}}
\newcommand{\Ga}{\ensuremath{\Gamma}}
\newcommand{\de}{\ensuremath{\delta}}
\newcommand{\De}{\ensuremath{\Delta}}
\newcommand{\eps}{\ensuremath{\epsilon}}
\newcommand{\veps}{\ensuremath{\varepsilon}}
\newcommand{\ka}{\ensuremath{\kappa}}
\newcommand{\la}{\ensuremath{\lambda}}

\newcommand{\La}{\ensuremath{\Lambda}}
\newcommand{\si}{\ensuremath{\sigma}}
\newcommand{\Si}{\ensuremath{\Sigma}}
\newcommand{\om}{\ensuremath{\omega}}
\newcommand{\Om}{\ensuremath{\Omega}}

\newcommand{\Ups}{\ensuremath{\Upsilon}}

\newcommand{\alh}{\ensuremath{\hat{\alpha}}}
\newcommand{\beh}{\ensuremath{\hat{\beta}}}
\newcommand{\gah}{\ensuremath{\hat{\gamma}}}

\newcommand{\muh}{\ensuremath{\hat{\mu}}}
\newcommand{\nuh}{\ensuremath{\hat{\nu}}}
\newcommand{\rhoh}{\ensuremath{\hat{\rho}}}
\newcommand{\sih}{\ensuremath{\hat{\si}}}
\newcommand{\tauh}{\ensuremath{\hat{\tau}}}
\newcommand{\vepsh}{\ensuremath{\hat{\veps}}}

\newcommand{\Bf}[2]{b_{#1}^{\ph{#1}#2}}
\newcommand{\kde}[2]{\de_{#1}^{\ph{#1}#2}}
\newcommand{\dia}[3]{{#1}_{#2}^{\ph{#2}#3}}

\newcommand{\mathHb}[1]{{\mathop{\kern0pt#1}\limits^{\,\sss
      \prime\prime}\vphantom{#1}}}

\newcommand{\phivev}{\ensuremath{\langle \phi \rangle}}

\newcommand{\com}[2]{\left[ #1 , #2 \right]}
\newcommand{\acom}[2]{\left\{ #1 , #2 \right\}}

\newcommand {\we} {\wedge}
\newcommand {\pa} {\partial}

\newcommand{\eqnlab}[1]{\label{eqn:#1}}

\newcommand{\eqnref}[1]{(\ref{eqn:#1})}

\newcommand{\Eqnref}[1]{Eq.~(\ref{eqn:#1})}

\newcommand{\Eqsref}[1]{Eqs.~(\ref{eqn:#1})}

\newcommand\ffam{\sffamily}
\newcommand\fser{\bfseries}
\newcommand\fsh{\upshape}

\RequirePackage{hyperref}

\begin{document}

\pagenumbering{roman}
\setcounter{page}{1}

%
%

\thispagestyle{empty}

\begin{center}
  {\sc\small Thesis for the degree of Doctor of Philosophy}
\end{center}



\begin{center}
\vspace{5mm}
{\upshape\sffamily\bfseries\huge Superconformal Theories \\[4mm]
in Six Dimensions} \\[5mm]
\end{center}

\vspace*{2mm}

\vspace*{4mm}
\begin{center}
  {\fsh\ffam\fser\Large P\"ar Arvidsson}\\
\end{center}
\vfill


\vfill
\begin{center}
        {\ffam\fsh Department of Fundamental Physics\\*[2mm]
        CHALMERS UNIVERSITY OF TECHNOLOGY \\*[2mm]
        G\"oteborg, Sweden 2006}
\end{center}

\clearpage

%
%

\pagestyle{empty}
\vspace*{15mm}
\noindent
        {\ffam
        {\fser\Large Superconformal Theories in Six Dimensions} \\[6mm]
        {\normalsize P\"ar Arvidsson  \\
        Department of Fundamental Physics \\
        Chalmers University of Technology \\
        SE-412 96 G\"oteborg, Sweden} \\[5mm]}

\noindent {\ffam\fser Abstract} \newline
\normalsize
\noindent
This thesis consists of an introductory text, which is divided into two parts, and six appended research papers.

The first part contains a general discussion on conformal and superconformal symmetry in six dimensions, and treats how the corresponding transformations act on space-time and superspace fields. We specialize to the case with chiral $(2,0)$ supersymmetry. A formalism is presented for incorporating these symmetries in a manifest way.

The second part of the thesis concerns the so called $(2,0)$ theory in six dimensions. The different origins of this theory in terms of higher-dimensional theories (Type IIB string theory and $M$-theory) are treated, as well as compactifications of the six-dimensional theory to supersymmetric Yang-Mills theories in five and four space-time dimensions. The free $(2,0)$ tensor multiplet field theory is introduced and discussed, and we present a formalism in which its superconformal covariance is made manifest. We also introduce a tensile self-dual string and discuss how to couple this string to the tensor multiplet fields in a way that respects superconformal invariance.

\vspace{15mm}
\noindent {\ffam\fser Keywords} \newline
\normalsize
\noindent Superconformal symmetry, Field theories in higher dimensions, String theory.

\clearpage

\pagestyle{plain}
\noindent This thesis consists of an introductory text and
the following six appended research papers, henceforth referred to
as {\sc Paper I-VI}:

\bigskip

\begin{enumerate}
\def\theenumi{\Roman{enumi}}
\item
P.~Arvidsson, E.~Flink and M.~Henningson, {\it Thomson scattering of
  chiral tensors and scalars against a self-dual string}, J.~High
  Energy Phys. {\bf 12}(2002)010, [hep-th/0210223].
\item
P.~Arvidsson, E.~Flink and M.~Henningson, {\it Free tensor multiplets
  and strings in spontaneously broken six-dimensional (2,0) theory},
J.~High Energy Phys. {\bf 06}(2003)039, [hep-th/0306145].
\item
P.~Arvidsson, E.~Flink and M.~Henningson, {\it Supersymmetric coupling
  of a self-dual string to a (2,0) tensor multiplet background},
J.~High Energy Phys. {\bf 11}(2003)015, [hep-th/0309244].
\item
P.~Arvidsson, E.~Flink and M.~Henningson, {\it The (2,0) supersymmetric theory of tensor multiplets and self-dual strings in six dimensions}, J.~High Energy Phys. {\bf 05}(2004)048, [hep-th/0402187].
\item
P.~Arvidsson, {\it Superconformal symmetry in the interacting theory of (2,0) tensor multiplets and self-dual strings}, J.~Math.~Phys. {\bf 47}(2006)042301, [hep-th/0505197].
\item
P.~Arvidsson, {\it Manifest superconformal covariance in six-dimensional $(2,0)$ theory}, J.~High Energy Phys. {\bf 03}(2006)076, [hep-th/0602193].
\end{enumerate}

%
%
\clearpage
\tableofcontents

\pagestyle{empty}
\clearpage
\pagestyle{fancy}
\renewcommand{\chaptermark}[1]{\markboth{Chapter \thechapter\ \ \ #1}{#1}}
\renewcommand{\sectionmark}[1]{\markright{\thesection\ \ #1}}
\lhead[\fancyplain{}{\sffamily\thepage}]%
  {\fancyplain{}{\sffamily\rightmark}}
\rhead[\fancyplain{}{\sffamily\leftmark}]%
  {\fancyplain{}{\sffamily\thepage}}
\cfoot{}
\setlength\headheight{14pt}


\setcounter{page}{1}
\pagenumbering{arabic}

\chapter*{Outline}
\addcontentsline{toc}{chapter}{\sffamily\bfseries Outline}

This thesis consists of two parts. {\bf Part I} contains a review of space-time and superspace symmetries in physical theories, together with a description of how to make these manifest. {\bf Part II} treats the so called $(2,0)$ theory in six dimensions, using many of the results from Part I.

{\bf Part I} starts with {\bf Chapter~\ref{ch:conformal}}, which contains an introduction to conformal symmetry and the associated coordinate transformations. We also find the action of a conformal transformation on a general space-time field using the method of induced representations. {\bf Chapter~\ref{ch:superspace}} generalizes this to the $(2,0)$ superspace with six bosonic and sixteen fermionic dimensions. The supersymmetry and the superconformal algebras, along with the corresponding coordinate transformations, are introduced. We also find the transformation of a general superfield.

{\bf Chapter~\ref{ch:manifest}} introduces Dirac's formalism, in which conformal symmetry is made manifest. We show how to formulate fields on a projective hypercone, defined such that a linear transformation in an eight-dimensional space yields a conformal transformation of the fields on the hypercone. Similarly, we generalize this to the superconformal case and show how to formulate manifestly superconformally covariant fields.

{\bf Part II} begins with {\bf Chapter~\ref{ch:origin}}, which contains a motivation for the study of $(2,0)$ theory in six dimensions. We also describe its different origins in terms of higher-dimensional theories. The degrees of freedom are derived in a non-technical manner and we make some connections to supersymmetric Yang-Mills theories in five and four dimensions.

In {\bf Chapter~\ref{ch:TM}}, we introduce the $(2,0)$ tensor multiplet as a representation of the $(2,0)$ supersymmetry algebra. We present an on-shell superfield formulation and state the complete superconformal transformation laws for these superfields. The superconformally covariant formalism of Chapter~\ref{ch:manifest} is applied to this theory, and we derive a superfield which transforms linearly under superconformal transformations and contains all tensor multiplet fields in an interesting way. Finally, we show that the differential constraint, which is required for the consistency of the superfield formulation, may be expressed in a compact way in the superconformally covariant notation. {\bf Chapter~\ref{ch:freestring}} introduces the free self-dual string, which is another representation of the supersymmetry algebra. We derive the superconformal transformation laws and find an action for the free string that respects both supersymmetry and $\ka$-symmetry.

Finally, {\bf Chapter~\ref{ch:coupling}} discusses the coupling between the tensor multiplet fields and the self-dual string mentioned above. We do this separately in the bosonic and in the superconformal case; the latter problem requires the introduction of superforms. We focus on the coupling to a background of on-shell tensor multiplet fields, meaning that we take the tensor multiplet fields to obey their free equations of motion. Finally, we discuss how to formulate the complete interacting theory.

The thesis also contains six appended research papers. {\sc Paper I} contains
a non-technical outline of our research program, but also treats
the coupling of the bosonic part of the tensor multiplet to a
self-dual bosonic string in the low-energy limit. The resulting
model is then used to compute the classical amplitude for scattering
tensor multiplet fields against a string. {\sc Paper II} deals
with the supersymmetric model (including fermions), but without
couplings. Actions for the free tensor multiplet and for the free
self-dual string are constructed and we study the process of
spontaneous breaking of $R$-symmetry. {\sc Paper III} discusses the superspace for $(2,0)$ theory, which is
used to derive a supersymmetric and $\kappa$-symmetric action for the
self-dual string coupled to an on-shell tensor multiplet background. {\sc Paper IV} discusses how to formulate the complete interacting theory of $(2,0)$ tensor multiplets and self-dual strings. This is done by making use of a local symmetry that contains $\ka$-symmetry, but is also related to how the choice of Dirac membranes affects the theory.

{\sc Paper V} introduces superconformal symmetry in the theory of tensor multiplets and self-dual strings. The coordinate transformations are found, both from a six-dimensional perspective and by considering a projective supercone in a higher-dimensional space. The transformation laws for the superfields are derived and expressed in a compact way using superspace-dependent parameter functions. It is shown that the interacting theory found in {\sc Paper III} is superconformally invariant. We also consider a superspace analogue to the Poincar\'e dual of the string world-sheet.

{\sc Paper VI} contains a generalization of Dirac's hypercone formalism to the superconformal case. We find a graded symmetric superfield, defined on a supercone, which contains the tensor multiplet fields. A linear transformation of the new superfield implies the known superconformal transformation laws for the tensor multiplet fields, and we show that the differential constraint may be expressed in a very compact way in this notation.

\part{Manifest symmetries}

\chapter{Space-time symmetries}
\label{ch:conformal}

It is hard to overestimate the significance and importance of symmetries in physics. Problems that seem insurmountable at first may be simplified, or even trivialized, by the use of symmetry considerations. A plethora of possible physical theories may be reduced to a single possibility by imposing a large enough symmetry constraint.

The purpose of this chapter is to introduce some of the most common and important symmetries in modern particle physics. We will restrict ourselves to continuous symmetries of space and time. This journey will take us from the simple concepts of rotational invariance and translational symmetry to the conformal group. We will also show how general fields transform under these transformations.

For concreteness, we will work in six dimensions with coordinates denoted by $x^{\mu}$, where the index $\mu$ takes values in the range $(0,\ldots,5)$. In this notation, $x^0$ denotes the time coordinate and we let $x^i$, $i=(1,\ldots,5)$, be the spatial coordinates. Indices repeated twice are summed over, in accordance with Einstein's summation convention. The results of this chapter can easily be generalized to other dimensions.

\section{Translations and rotations}
\label{sec:transrot}

Firstly, we will introduce what perhaps is the simplest symmetry of them all, spatial translational invariance. This involves transformations of the type
\begin{equation}
x^i \rightarrow x^i + a^i,
\end{equation}
where $a^i$ is the parameter of the transformation, a vector which does not depend on $x^\mu$ and encodes in which direction and how far the coordinates are translated.

Translational invariance means that the laws of physics look the same at all points in space. Similarly, we may have translational invariance in time, meaning that the laws look the same at all times. We will denote the parameter corresponding to such translations by $a^0$, so that spatial and temporal translations may be collected into a single vector $a^\mu$ with six components.

In terms of group theory, the translational group is abelian and the corresponding Lie algebra generator is conventionally denoted by $P_{\mu}$. In a coordinate representation\footnote{We use a coordinate representation in which an infinitesimal transformation is written as $\de = a \cdot G$, rather than $\de= i a \cdot G$ which is common in other texts. In these expressions, $G$ is the generator while $a$ is the parameter.}, this becomes $P_{\mu}=\pa_{\mu}$, \ie a derivative with respect to $x^{\mu}$. This representation is used to generate infinitesimal coordinate transformations according to
\begin{equation}
\de x^{\mu} = \com{a^\nu P_\nu}{x^\mu} = \left( a^\nu \pa_\nu \right) x^\mu - x^\mu \left( a^\nu \pa_\nu \right) = a^\nu \left( \pa_\nu x^\mu \right) = a^\mu,
\end{equation}
where $a^\mu$ now is an infinitesimal quantity. The bracket denotes a commutator; its second term cancels the term where the derivative has passed through $x^\mu$. It is clear that the translational group is six-dimensional, since $a^\mu$ has six components.

Next, let us move on to the concept of rotational invariance. This symmetry means that the laws of physics are the same in all directions. Concretely, the coordinates are transformed as
\begin{equation}
\eqnlab{spatial_rot}
x^i \rightarrow \La^i_{\ph{i}j} x^j,
\end{equation}
where the matrix $\La^i_{\ph{i}j}$ contains the parameters of the transformation. The transformation is linear, meaning that the matrix $\La^i_{\ph{i}j}$ cannot depend on the coordinates.

However, the transformation~\eqnref{spatial_rot} is far too general; we must constrain $\La^i_{\ph{i}j}$ in some way to make it a true rotation. The restriction to be imposed is that an infinitesimal line element, defined as
\begin{equation}
\eqnlab{interval}
ds^2 \equiv dx_i dx^i \equiv \eta_{ij} dx^j dx^i,
\end{equation}
should be left invariant by the transformation. In this relation, $\eta_{ij}$ is the metric (a unit matrix in Cartesian coordinates), which may be used to lower spatial indices. Conversely, its inverse $\eta^{ij}$ may be used to raise indices.

The restriction~\eqnref{interval} implies that
\begin{equation}
\eqnlab{La-rel}
\La^i_{\ph{i}k} \La^j_{\ph{j}l} \eta_{ij} = \eta_{kl},
\end{equation}
which means that $\La^i_{\ph{i}j}$ must be an orthogonal matrix. If we require the transformations to preserve the orientation of space, we also have to impose that $\La^i_{\ph{i}j}$ should have positive determinant. This tells us that the rotational group is isomorphic to the special orthogonal group $SO(5)$, which is ten-dimensional.

However, we are mostly interested in infinitesimal transformations. Expand the parameter matrix as
\begin{equation}
\La^i_{\ph{i}j} = \de^i_{\ph{i}j} + \om^i_{\ph{i}j},
\end{equation}
where $\om^i_{\ph{i}j}$ is the infinitesimal rotation parameter and $\de^i_{\ph{i}j}$ is the Kronecker delta symbol (a unit matrix). According to \Eqnref{La-rel}, $\om_{ij}$ must be antisymmetric rather than orthogonal. The corresponding Lie algebra generator is denoted by $M_{ij}=-M_{ji}$ and may be written as
\begin{equation}
  M_{ij}= - x_{[i} \pa_{j]} \equiv - \frac{1}{2} \left( x_i \pa_j - x_j \pa_i \right)
\end{equation}
in a coordinate representation. The second relation in this equation defines the antisymmetrization bracket. The generator obeys the standard commutation relations
\begin{equation}
\com{M_{ij}}{M_{kl}} = \eta_{k[i} M_{j]l} - \eta_{l[i} M_{j]k}
\end{equation}
of the Lie algebra $\mathfrak{so}(5)$ and the induced infinitesimal coordinate transformation is
\begin{equation}
\de x^i = \com{ \om^{jk} M_{jk} }{ x^i } = \om^{ij} x_j.
\end{equation}
The antisymmetry of $\om^{ij}$ makes the invariance of the scalar product of two coordinate vectors trivial.

Having treated the classical symmetries of space and time in this section, the stage is set to generalize these concepts and discuss larger (and more interesting) symmetry groups.

\section{Lorentz and Poincar\'e invariance}
\label{sec:lorentz}

One of the greatest scientific achievements of the 20th century was without doubt the special theory of relativity. As properly understood by A.~Einstein, this discovery changed the way we look upon space and time by uniting them into a single concept, the \emph{space-time}.

The lorentzian space-time transformations, on which the special theory of relativity is founded, were first discovered as an intriguing symmetry of Maxwell's equations of electrodynamics. Einstein, however, realized that the transformations are of a much more fundamental nature; they originate from two basic assumptions concerning space and time. Firstly, there is a principle of relativity stating that the laws of physics are identical in all inertial frames. Secondly, the speed of light is a fundamental constant in nature and has the same value for all observers in all inertial frames.

In its simplest form, special relativity encodes the transformation between two different inertial frames. If the primed coordinate frame moves with a constant velocity $v$ in $x$-direction of the unprimed frame, the relations between the coordinates are
\begin{equation}
  \eqnlab{boost}
  \begin{split}
     t' & = \ga(v) \left(t - vx/c^2 \right) \\
     x' & = \ga(v) (x - vt),
  \end{split}
\end{equation}
while all other components in $x^\mu$ are unaltered. The Lorentz factor is defined by
\begin{equation}
  \ga(v) \equiv \frac{1}{\sqrt{1-v^2/c^2}},
\end{equation}
and $c$ denotes the speed of light. This should be compared with Galilean relativity, which relates two inertial frames in classical mechanics and involves an absolute time. Galilean relativity is restored in the classical limit $v \ll c$. In the following, we will adopt units such that $c=1$, which means that we treat space and time on equal footing, using the same units.

In modern usage, Lorentz transformations are regarded as space-time rotations. The general expression for such a transformation is
\begin{equation}
\eqnlab{Lorentz_rot}
x^\mu \rightarrow \La^\mu_{\ph{\mu}\nu} x^\nu,
\end{equation}
which should be compared with \Eqnref{spatial_rot} for a spatial rotation. The Lorentz transform contains the spatial rotations as a subgroup, but also includes Lorentz boosts such as the transformation in~\Eqnref{boost}.

As in the spatial case, the parameter matrix $\La^\mu_{\ph{\mu}\nu}$ cannot be chosen arbitrarily. We demand the infinitesimal proper time interval
\begin{equation}
\eqnlab{prop_time_int}
d\tau^2 \equiv - dx_\mu dx^\mu \equiv - \eta_{\mu\nu} dx^\mu dx^\nu
\end{equation}
to be invariant under the transformation. Again, $\eta_{\mu\nu}$ is the metric, which in Cartesian coordinates is diagonal with elements $(-1,1,1,1,1,1)$. The metric satisfies
\begin{equation}
\eqnlab{inv_metric}
\eta_{\mu\nu} = \dia{\La}{\mu}{\rho} \dia{\La}{\nu}{\si} \eta_{\rho\si},
\end{equation}
which tells us that the Lorentz group is isomorphic to the special orthogonal group $SO(5,1)$, which is 15-dimensional.

An infinitesimal Lorentz coordinate transformation is generated by
\begin{equation}
  \eqnlab{Mmunu}
  M_{\mu\nu}= - x_{[\mu} \pa_{\nu]},
\end{equation}
which obeys the standard commutation relations
\begin{equation}
\com{M_{\mu\nu}}{M_{\rho\si}} = \eta_{\rho[\mu} M_{\nu]\si} - \eta_{\si[\mu} M_{\nu]\rho}
\end{equation}
of the Lie algebra $\mathfrak{so}(5,1)$. The induced coordinate transformation is
\begin{equation}
\de x^\mu = \com{ \om^{\nu\rho} M_{\nu\rho} }{x^\mu} = \om^{\mu\nu} x_\nu,
\end{equation}
where the antisymmetric matrix $\om^{\mu\nu}$ contains the infinitesimal transformation parameters.

We should also mention the Poincar\'e group (sometimes called the inhomogeneous Lorentz group), which is the semidirect product of the Lorentz group and the group of space-time translations. The corresponding Lie algebras mix in the sense that
\begin{equation}
\com{P_{\mu}}{M_{\nu \rho}}  =  - \eta_{\mu [\nu} P_{\rho]},
\end{equation}
while all operators $P_{\mu}$ commute with each other as required for an abelian group. We note that also translations leave the infinitesimal proper time interval in \Eqnref{prop_time_int} invariant.

\section{Conformal symmetry}
\label{sec:confalg}

In the preceding section, we found that the largest symmetry group that preserved the proper time interval in \Eqnref{prop_time_int} was the Poincar\'e group, consisting of space-time translations and rotations. In the present section, we will relax this requirement slightly, thereby deriving the transformations of the \emph{conformal group}.

The conformal symmetries are not respected by nature in the same way as Lorentz symmetry is. The presence of objects with a certain size or mass breaks conformal symmetry, since it implies (as we will see below) scale invariance. However, theories such as electrodynamics involving only massless point-like particles may be conformally invariant. Indeed, it can be shown~\cite{Cunningham:1910,Bateman:1910} that Maxwell's equations respect conformal invariance.

Let us consider a transformation such that
\begin{equation}
\eqnlab{prop_time_conf}
d \tau^2 \rightarrow \Om^2(x) d \tau^2.
\end{equation}
This means that the proper time interval \eqnref{prop_time_int} is required to be invariant modulo a space-time dependent scale factor. The word \emph{conformal} comes from the fact that such a transformation preserves the angle between two intersecting curves.

The next step is to derive the most general infinitesimal transformation consistent with \Eqnref{prop_time_conf}. Start from
\begin{equation}
x^\mu \rightarrow x^\mu + \xi^\mu(x),
\end{equation}
where $\xi^\mu(x)$ is a space-time dependent vector containing the parameters of the transformation. The corresponding change in $d\tau^2$ is
\begin{equation}
d\tau^2 \rightarrow d\tau^2 + \eta_{\mu\nu} \left( \frac{\pa \xi^\mu}{\pa x^\rho} d x^\rho d x^\nu + dx^\mu \frac{\pa \xi^\nu}{\pa x^\rho} d x^\rho \right),
\end{equation}
where we have neglected terms of order $\xi^2$, since we are working with an infinitesimal transformation. For this to be consistent with \Eqnref{prop_time_conf}, we must have
\begin{equation}
\pa_\mu \xi_\nu(x) + \pa_\nu \xi_\mu(x) = \left( \Om^2(x)-1 \right) \eta_{\mu \nu}.
\end{equation}
Contracting the free indices in this equation by multiplication with the inverse metric $\eta^{\mu\nu}$, we find that
\begin{equation}
2 \pa_\mu \xi^\mu(x) = 6 \left( \Om^2(x)-1 \right).
\end{equation}
This means that
\begin{equation}
\pa_\mu \xi_\nu(x) + \pa_\nu \xi_\mu(x) = \frac{1}{3} \pa \cdot \xi \, \eta_{\mu \nu},
\end{equation}
where the scalar product is defined in the standard way by $\pa \cdot \xi \equiv \pa_{\mu} \xi^\mu$. The general solution~\cite{DiFrancesco:1997} to this equation (the conformal Killing equation) is given by
\begin{equation}
\eqnlab{xi_x}
\xi^\mu(x) = \de x^\mu = a^\mu + \om^\mu_{\ph{\mu}\nu} x^\nu + \la x^\mu - 2 c \cdot x x^\mu + c^\mu x \cdot x,
\end{equation}
where we recognize $a^\mu$ as the parameter for translations, while $\om_{\mu\nu} = - \om_{\nu \mu}$  corresponds to Lorentz rotations. The new ingredients are $\la$, which is the parameter for dilatations or scale transformations, and $c^{\mu}$, which denotes the so called special conformal transformations. The corresponding change of the infinitesimal proper time interval $d\tau^2$ is
\begin{equation}
d \tau^2 \rightarrow \left[ 1 + 2 (\la - 2 c \cdot x) \right] d \tau^2,
\end{equation}
which implies that $\Om^2(x) = 1 + 2 (\la - 2 c \cdot x)$ for an infinitesimal transformation.

The differential operators that generate these new coordinate transformations are
\begin{equation}
D = x \cdot \pa,
\end{equation}
which corresponds to dilatations, and
\begin{equation}
K_\mu = x \cdot x \pa_\mu - 2 x_\mu x \cdot \pa,
\end{equation}
corresponding to special conformal transformations. The generators of the conformal algebra obey the commutation relations
\begin{equation}
\eqnlab{bosonic_comm_coord}
\begin{aligned}
\com{P_{\mu}}{M_{\nu \rho}} &= - \eta_{\mu [\nu} P_{\rho]} & \com{P_{\mu}}{D} &= P_{\mu} \\
\com{K_{\mu}}{M_{\nu \rho}} &= - \eta_{\mu [\nu} K_{\rho]} &  \com{K_{\mu}}{D} &= -K_{\mu} \\
\com{P_{\mu}}{K_{\nu}} &= - 4 M_{\mu \nu} - 2 \eta_{\mu \nu} D &
\com{M_{\mu \nu}}{D} &= 0 \\
\com{P_{\mu}}{P_{\nu}} &= 0 & \com{K_{\mu}}{K_{\nu}} &= 0 \\
\com{M_{\mu \nu}}{M_{\rho \si}} &= \eta_{\rho[\mu} M_{\nu]\si} - \eta_{\si[\mu} M_{\nu]\rho}, \\
\end{aligned}
\end{equation}
which are easily verified using the differential expressions above.
The conformal algebra in six dimensions is, as we will see explicitly in Chapter~\ref{ch:manifest}, isomorphic to $\mathfrak{so}(6,2)$, which is 28-dimensional.

\section{Field transformations}
\label{sec:conformal_field}

So far, we have only considered coordinate transformations of space-time, so called \emph{passive} transformations. However, to formulate physical theories we need to populate space-time with \emph{fields}. These are continuous functions of the space-time coordinates, but may also carry indices of some kind, indicating how (in which representation) they transform under Lorentz transformations. Throughout this section, we will adopt an \emph{active} viewpoint on transformations, meaning that we fix the coordinates but transform the fields instead.

The purpose of this section is to derive how a general field transforms under a conformal transformation; from this all simpler transformation laws may be deduced. We will use the method of induced representations and follow the paper~\cite{Mack:1969} by G.~Mack and A.~Salam quite closely.

Consider a field $\varphi^i(x)$, where $i$ may be any set of indices consistent with a representation of the Lorentz group (not to be confused with the notation used for spatial vector indices in Section~\ref{sec:transrot}). For example, it might be a vector field $\varphi^\mu(x)$ or a scalar field $\varphi(x)$. The field must transform according to a representation of the conformal group, which means that a conformal transformation of $\varphi^i(x)$ may be written as
\begin{equation}
\eqnlab{rep}
\varphi^i(x) \rightarrow S^{i}_{\ph{i}j}(g,x) \varphi^j(g^{-1} x),
\end{equation}
where $g$ denotes an element of the conformal group. We see that the matrix $S^{i}_{\ph{i}j}(g,x)$ depends on $g$, but it must also depend on the space-time coordinate $x^\mu$. It is important to note that, in the active picture, the variation of the field is evaluated in a point $g^{-1} x$. This is necessary since we wanted to transform the fields, not the coordinate system.

The next step, which is the basis for the method of induced representations, is to identify the so called \emph{little group}.
We note from \Eqnref{rep} that $S^{i}_{\ph{i}j}(g,0)$ must be a representation of the stability subgroup of $x^\mu=0$, which is the group of transformations that leave the point $x^\mu=0$ invariant. This is the little group, and the sought representation of the conformal group is induced by a representation of this group.

From \Eqnref{xi_x}, we see that the little group consists of Lorentz transformations, dilatations and special conformal transformations. Denote the corresponding generators by $\Si_{\mu\nu}$, $\De$ and $\ka_\mu$, respectively. They obey the commutation relations
\begin{equation}
\eqnlab{little_comm}
\begin{aligned}
\com{\Si_{\mu\nu}}{\ka_{\rho}} &=  \eta_{\rho [\mu} \ka_{\nu]}
& \com{\Si_{\mu \nu}}{\De} &= 0 \\
\com{\Si_{\mu \nu}}{\Si_{\rho \si}} &= \eta_{\rho[\mu} \Si_{\nu]\si} - \eta_{\si[\mu} \Si_{\nu]\rho} & \com{\ka_{\mu}}{\De} &= - \ka_{\mu}, \\
\end{aligned}
\end{equation}
which are similar to those in \Eqnref{bosonic_comm_coord}. We note that the Lie algebra of the little group is isomorphic to the Poincar\'e algebra together with the dilatations, where the special conformal transformations act as translations.

By assumption, the action\footnote{Note that in this expression and onwards, the generators should be seen as abstract operators rather than the differential expressions stated above.} of the conformal generators $M_{\mu\nu}$, $D$ and $K_{\mu}$ on $\varphi^i(0)$ is known and may be written as
\begin{equation}
  \eqnlab{littleaction}
  \begin{aligned}
  \com{M_{\mu\nu}}{\varphi^i(0)} &= \left( \Si_{\mu \nu} \varphi \right)^i(0) \\
 \com{ D}{\varphi^i(0) } &= \left( \De \varphi \right)^i(0) \\
 \com{ K_{\mu}}{\varphi^i(0)} &= \left( \ka_{\mu} \varphi\right)^i(0).
  \end{aligned}
\end{equation}
The expressions to the right serve as the starting point of the derivation; these are the representations of the little group. $\left( \Si_{\mu \nu} \varphi \right)^i$ tells us how the field transforms under Lorentz transformations (which is intimately connected to the properties of the index $i$), while $\De$ is the so called scaling dimension of the field. Finally, $\left( \ka_{\mu} \varphi\right)^i$ denotes the field's intrinsic properties under special conformal transformations. We will see explicit examples of how these representations may look later in this thesis.

There are some comments to be made here concerning the various possible representations of the little group~\cite{Mack:1969,DiFrancesco:1997}. If the field $\phi^i(x)$ belongs to an irreducible representation of the Lorentz group, then any matrix that commutes with the generator $\Si_{\mu\nu}$ must be a multiple of the identity operator, according to Schur's lemma. This applies to the generator $\De$ and explains why we call it the scaling dimension --- it acts on fields by a multiplicative constant. From \Eqnref{little_comm}, it follows that $\ka_\mu$ must vanish in this case. Such fields are conventionally called \emph{primary} fields.

Let us return to the goal of this section: to find the action of the generators on $\varphi^i(x)$, i.e., on the field at an arbitrary space-time point. We choose a basis in index space so that the translation operator acts only differentially on the field, which means that
\begin{equation}
\eqnlab{Pmu}
\com{ P_{\mu} }{ \varphi^i(x) }= \pa_\mu \varphi^i(x).
\end{equation}
$P_\mu$ generates infinitesimal translations, but a finite translation may be constructed by exponentiation. This means that we may write
\begin{equation}
\eqnlab{fintrans}
\varphi^i(x) = e^{x \cdot P} \varphi^i(0) e^{-x \cdot P},
\end{equation}
which reduces to \Eqnref{Pmu} if we apply a derivative $\pa_{\mu}$ to both sides. This gives a hint to how we can find the action of the other generators on $\varphi^i(x)$. If $\ordo$ is one of the other generators of the conformal group ($M_{\mu\nu}$, $D$ or $K_{\mu}$), we have that
\begin{equation}
\eqnlab{ordoaction}
e^{x \cdot P} \big[ \ordo,\varphi^i(0)\big] e^{-x \cdot P} = \big[ \tilde{\ordo} , \varphi^i(x) \big],
\end{equation}
where the operator $\tilde{\ordo}$ is given by
\begin{equation}
\eqnlab{ordotilde}
\tilde{\ordo} \equiv e^{x \cdot P} \ordo e^{-x \cdot P}.
\end{equation}
This expression may be evaluated using the commutation relations~\eqnref{bosonic_comm_coord} for the conformal group and the identity
\begin{equation}
\eqnlab{Haussdorff}
e^{\sss -A} B e^{\sss A} = B + \com{B}{A} + \frac{1}{2!} \com{\com{B}{A}}{A} + \frac{1}{3!} \com{\com{\com{B}{A}}{A}}{A} + \ldots,
\end{equation}
where $A$ and $B$ are arbitrary operators. It is essential to remember that we are in the active picture, where the generators and the coordinates commute. This affects the signs in the commutation relations; this is motivated nicely \eg in Ref.~\cite{VanProeyen:1999}.

The resulting sum from \Eqsref{ordotilde} and \eqnref{Haussdorff} converges and we find that
\begin{equation}
  \begin{aligned}
    \tilde{M}_{\mu\nu} &= M_{\mu\nu} - x_{[\nu} P_{\mu]} \\
    \tilde{D} &= D - x \cdot P  \\
    \tilde{K}_\mu &= K_\mu - 4 x^\nu M_{\mu\nu} +2 x_\mu D + x \cdot x P_\mu - 2 x_\mu x \cdot P.
  \end{aligned}
\end{equation}
Inserting these expressions into \Eqnref{ordoaction}, using \Eqsref{littleaction} and \eqnref{fintrans}, we get
\begin{equation}
\eqnlab{little}
  \begin{aligned}
    \left(\Si_{\mu\nu} \varphi\right)^i(x) &= \com{M_{\mu\nu}}{\varphi^i(x)} - x_{[\nu} \com{P_{\mu]}}{\varphi^i(x)} \\
    \left(\De \varphi \right)^i(x) &= \com{D}{\varphi^i(x)} - x^\mu \com{P_{\mu}}{\varphi^i(x)} \\
    \left(\ka_\mu \varphi \right)^i(x) &=  \com{K_\mu}{\varphi^i(x)} - 4 x^\nu \com{M_{\mu\nu}}{\varphi^i(x)} + 2 x_\mu \com{D}{\varphi^i(x)} + {} \\
& \quad + x \cdot x \com{P_{\mu}}{\varphi^i(x)} - 2 x_\mu x^\nu \com{P_\nu}{\varphi^i(x)}.
  \end{aligned}
\end{equation}
These equations may be solved for the actions of the generators $M_{\mu\nu}$, $D$ and $K_{\mu}$ on the field $\varphi^i(x)$, given \Eqnref{Pmu} stating the action of $P_{\mu}$ on the field.

In the general case, the conformal variation of the field is given by
\begin{equation}
\de_{\sss C} \varphi^i(x) = \com{ a^\mu P_\mu + \om^{\mu\nu} M_{\mu\nu} + \la D + c^\mu K_\mu }{\varphi^i(x)}.
\end{equation}
Using \Eqnref{little}, this becomes
\begin{multline}
\de_{\sss C} \varphi^i(x) = a^\mu \pa_\mu \varphi^i(x) + \om^{\mu
  \nu} \Big\{ x_{\nu} \pa_{\mu} \varphi^i(x) + \left( \Si_{\mu \nu} \varphi
\right)^i(x) \Big\} + {} \\
\quad + \la \Big\{ x^\mu \pa_\mu \varphi^i(x) + \left( \De
\varphi \right)^i(x) \Big\} + c^{\mu} \Big\{ x \cdot x \pa_{\mu} \varphi^i(x) - 2 x_{\mu} x \cdot \pa \varphi^i(x) + {} \\
{} + 4 x^{\nu} \left( \Si_{\mu \nu} \varphi \right)^i(x) - 2 x_{\mu} \left( \De \varphi \right)^i(x) + (\ka_{\mu} \varphi)^i(x) \Big\},
  \eqnlab{mack_transf}
\end{multline}
which tells us how a space-time field transforms under a conformal transformation, given its properties with respect to the little group. We recognize the differential pieces from the generators of coordinate transformations given previously in this chapter. It is also clear from this expression what the roles of the little group generators are in a general field transformation.

Fortunately, there is a more compact way of expressing this transformation~\cite{Claus:1998,VanProeyen:1999}, which will prove to be useful in the following. We may write
\begin{equation}
  \begin{split}
\de_{\sss C} \varphi^i(x) &= \xi^{\mu}(x) \pa_{\mu} \varphi^i(x) + \Om^{\mu
  \nu}(x) \left( \Si_{\mu \nu} \varphi \right)^i(x) + {} \\
& \quad + \La(x) \left( \De \varphi \right)^i(x) + c^{\mu} \left(\ka_{\mu} \varphi\right)^i(x),
  \end{split}
\eqnlab{general_transf}
\end{equation}
where the space-time dependent parameter functions are defined by
\begin{equation}
  \eqnlab{OmLa_bos}
  \begin{aligned}
\xi^{\mu}(x) &\equiv a^{\mu} + \om^{\mu \nu} x_{\nu} + \la x^{\mu} +
c^{\mu} x^2 - 2 c \cdot x x^{\mu} \\
\Om^{\mu \nu}(x) &\equiv \om^{\mu \nu} + 4 c^{[\mu} x^{\nu]} \\
\La(x) &\equiv \la - 2 c \cdot x.
  \end{aligned}
\end{equation}
Note that the expression for $\xi^\mu(x)$ above coincides with the expression for the variation of $x^\mu$ as given in \Eqnref{xi_x}, indicating that this part corresponds to the change in the field due to its dependence on the space-time coordinates. Observing that
\begin{equation}
\pa^{\nu} \xi^\mu(x) = \La(x) \eta^{\mu\nu} + \Om^{\mu\nu}(x),
\end{equation}
one may solve for the parameter functions in terms of derivatives of $\xi^\mu(x)$.

We also note that these space-time dependent parameter functions may be used when considering coordinate transformations as well. The variation of the coordinate differential $dx^\mu$ induced by \Eqnref{xi_x} may be written as
\begin{equation}
\eqnlab{bos_diff}
\de (dx^\mu) = d \xi^{\mu}(x) = \Om^{\mu \nu}(x) d x_\nu + \La(x) dx^\mu,
\end{equation}
meaning that the differential, in a certain sense, transforms as a primary field with unit scaling dimension. This viewpoint indicates some important properties of differential forms and will be useful in the following.

Summing up this chapter, we have defined the conformal transformations and derived their action on the space-time coordinates as well as on general space-time fields in six dimensions. In the next chapter, we will generalize these concepts further and finally arrive at the superconformal group.

\chapter{Superspace symmetries}
\label{ch:superspace}

In 1967, S.~Coleman and J.~Mandula presented a theorem~\cite{Coleman:1967} stating which symmetries one might have in relativistic quantum field theories. According to this theorem, the largest possible Lie algebra of symmetry operators in a general case is the Poincar\'e algebra. If all particles in the theory happen to be massless, the conformal algebra is allowed. In both these cases, one might also have some internal symmetry algebra.

If we take this theorem seriously, we should end this chapter here. Since the conformal symmetry transformations appear to be the most general ones compatible with physics, there is no point in trying to formulate larger symmetry algebras than those discussed in the previous chapter. Or is there? As the reader might know, there is a possibility not considered in the Coleman-Mandula theorem: It only treats symmetries that take bosons (particles with integer spin) to bosons and fermions (particles with half-integer spin) to fermions.

In 1975, R.~Haag, J.T.~Lopuszanski and M.~Sohnius presented another theorem~\cite{Haag:1975}, which allowed for a new symmetry taking bosons to fermions and vice versa, thereby evading the no-go theorem of Coleman and Mandula. This new symmetry is what we today call supersymmetry. It had actually appeared before 1975 in the context of string theory~\cite{Ramond:1971,Neveu:1971,Gervais:1971}, as a symmetry in a two-dimensional field theory. The supersymmetry algebra in four dimensions was discovered in the Soviet Union in 1971~\cite{Golfand:1971,Volkov:1972}, but it was not until J.~Wess and B.~Zumino~\cite{Wess:1974a,Wess:1974b} extended the idea of supersymmetry from two dimensions to a quantum field theory in four dimensions that it became widely known.

As in the previous chapter, we specialize to a six-dimensional space-time. The different possible supersymmetry algebras in this case were classified by W.~Nahm~\cite{Nahm:1978}. We will focus on the $\mathcal{N}=(2,0)$ case, which is chiral and consistent with superconformal symmetry (which is the topic of Section~\ref{sec:superconformal}). The reason for this choice is mainly that the details of supersymmetry depend very much on the dimensionality of space-time and on the amount of supersymmetry, and our main concern in this thesis is the so called $(2,0)$ theory in six dimensions. However, many of the concepts developed here can easily be generalized to other dimensions and other supersymmetry algebras. For a more general text on supersymmetry, we refer to the textbooks by Wess and Bagger~\cite{WessBagger} or Weinberg~\cite{Weinberg_3}.

\section{Supersymmetry}
\label{sec:susy}

In terms of particles, supersymmetry takes bosons to fermions and vice versa. However, we postpone the action on fields and particles to Section~\ref{sec:superfields} and focus on the coordinate transformations of space and time instead. This is made possible by extending the space-time to a \emph{superspace} with both bosonic and fermionic coordinates. The distinction between these is the following: Bosonic (Grassmann even) coordinates are ordinary commuting quantities, such that $x y = y x$. On the other hand, fermionic (Grassmann odd) coordinates are anti-commuting, such that $\theta \eta = - \eta \theta$. The superspace concept is motivated by and simplifies the treatment of supersymmetric theories.

Before introducing the supersymmetry generators and the superspace coordinates, let us take some time to introduce the relevant representations of the Lorentz algebra. We start from the fundamental representations of $\mathfrak{so}(5,1)\simeq \mathfrak{su}^*(4)$, from which all other representations may be built. These are the four-dimensional chiral spinor representation, denoted by ${\bf 4}$, and the likewise four-dimensional anti-chiral spinor representation, denoted by ${\bf 4'}$. We denote quantities transforming in these representations by a subscript or a superscript spinor index $\al=(1,\ldots,4)$, respectively.

All other representations may be built from tensor products of these. The simplest are
\begin{equation}
  \eqnlab{lorentzreps}
  \begin{aligned}
    {\bf 4} \otimes {\bf 4} & \simeq {\bf 6} \oplus {\bf 10}_+ \\
    {\bf 4'} \otimes {\bf 4'} & \simeq {\bf 6} \oplus {\bf 10}_- \\
    {\bf 4} \otimes {\bf 4'} & \simeq {\bf 1} \oplus {\bf 15}, \\
  \end{aligned}
\end{equation}
where all the representations appearing on the right-hand side have simple interpretations and will be useful in the following. ${\bf 6}$ is antisymmetric in the spinor indices and denotes the vector representation. This means that we may (and we will use this convention) denote the space-time coordinate vector as $x^{\al\be}=-x^{\be\al}$, but also as $x_{\al\be}=-x_{\be\al}$. The relation between these equivalent representations is
\begin{equation}
x^{\al\be} = \frac{1}{2} \veps^{\al\be\ga\de} x_{\ga\de},
\end{equation}
where $\veps^{\al\be\ga\de}$ is the totally antisymmetric invariant tensor, defined such that $\veps^{1234}=1$. Furthermore, the symmetric pieces ${\bf 10}_+$ and ${\bf 10}_-$ are self-dual and anti self-dual three-forms, respectively, while the traceless representation ${\bf 15}$ is a two-form. Finally, the singlet ${\bf 1}$ corresponds to $\kde{\al}{\be}$, which is a Kronecker delta function.

We should also introduce the so called $R$-symmetry algebra, which rotates the different supersymmetry generators into each other. In the case of $(2,0)$ supersymmetry, this algebra is isomorphic to $\mathfrak{so}(5)$ and the fundamental representation is again four-dimensional and a spinor, denoted by ${\bf 4}$. We denote quantities transforming in this representation by an index $a=(1,\ldots,4)$, which may be raised or lowered from the left using the antisymmetric invariant tensors $\Om^{ab}$ and $\Om_{ab}$, respectively. This demands that they satisfy $\Om_{ab} \Om^{bc} = \kde{a}{c}$.

The relevant tensor product is
\begin{equation}
 \eqnlab{rsymreps}
 {\bf 4} \otimes {\bf 4}\simeq {\bf 1} \oplus {\bf 5} \oplus {\bf 10},
\end{equation}
where the singlet denotes the invariant tensor $\Om^{ab}$ introduced above. The ${\bf 5}$ representation is antisymmetric and traceless with respect to $\Om_{ab}$, while ${\bf 10}$ is symmetric. They correspond to a vector and a two-form, respectively.

After these preliminaries, we are ready to extend the Poincar\'e algebra with fermionic generators, aiming at the construction of the $(2,0)$ super-Poincar\'e algebra. The supersymmetry generators must belong to a spinorial representation of the Lorentz algebra~\cite{Haag:1975}; we choose them to be chiral spinors. Furthermore, we let them transform in the spinor representation of the $R$-symmetry algebra. Following the conventions developed above, we denote the supersymmetry generators by the fermionic (Grassmann odd) operators $Q^a_\al$.

As fermionic generators, $Q^a_\al$ obey anticommutation relations rather than commutation relations. The symmetric tensor product of two such representations of $\mathfrak{so}(5,1) \times \mathfrak{so}(5)$ is
\begin{equation}
\left[ ({\bf 4};{\bf 4}) \otimes ({\bf 4};{\bf 4}) \right]_{\mathrm{sym}} \simeq ({\bf 6};{\bf 1}) \oplus ({\bf 6};{\bf 5}) \oplus ({\bf 10}_+;{\bf 10}),
\end{equation}
which means that the most general anticommutation relation between two supersymmetry generators is~\cite{Howe:1998t}
\begin{equation}
 \eqnlab{QQ}
\acom{Q^a_\al}{Q^b_\be} = - 2i \Om^{ab} P_{\al\be} + Z^{ab}_{\al\be} + W^{ab}_{\al\be}.
\end{equation}
In this relation, the antisymmetric $P_{\al\be}$ is the generator of translations from Section~\ref{sec:transrot} while $Z^{ab}_{\al\be}$ and $W^{ab}_{\al\be}$ are central\footnote{Note that these charges are not ''true'' central charges of the algebra, since they transform non-trivially under both Lorentz and $R$-symmetry transformations.} charges of the algebra. They obey the relations $Z^{ab}_{\al\be} = -Z^{ba}_{\al\be} = -Z^{ab}_{\be\al}$, $\Om_{ab} Z^{ab}_{\al\be}=0$ and $W^{ab}_{\al\be} = W^{ba}_{\al\be} = W^{ab}_{\be\al}$.

Generally, $p$-form central charges in the supersymmetry algebra correspond to $p$-dimensional extended objects in the theory~\cite{deAzcarraga:1989}. We therefore expect the central charges $Z^{ab}_{\al\be}$ and $W^{ab}_{\al\be}$ to correspond to a one-dimensional object (a string) and a three-dimensional object (a brane), respectively. We will have more to say about this in the second part of this thesis, where we discuss the degrees of freedom of $(2,0)$ theory in six dimensions.

We should also pay some attention to the reality properties of the supersymmetry generators. They obey a symplectic Majorana reality condition~\cite{Kugo:1983}, which states that
\begin{equation}
\eqnlab{real_Q}
\left( Q^a_\al \right)^* = C_\al^{\ph{\al}\be} \Om_{ab} Q^b_\be,
\end{equation}
where $C_\al^{\ph{\al}\be}$ is the charge conjugation matrix. Note that complex conjugation raises or lowers $R$-symmetry spinor indices, but does not affect Lorentz spinor indices. The latter statement tells us that the complex conjugate of a chiral spinor is another chiral spinor.

For consistency, we require that
\begin{equation}
  \begin{aligned}
    {C^*}_\al^{\ph{\al}\be} C_\be^{\ph{\be}\ga} & = - \kde{\al}{\ga} \\
    \left( \Om_{ab} \right)^* & = - \Om^{ab},
  \end{aligned}
\end{equation}
which implies that $\left(\left( Q^a_\al \right)^*\right)^* = Q^a_\al$.

The next step is to introduce the fermionic superspace coordinates. Supersymmetry is supposed to generate translations in superspace, which means that the fermionic coordinates should transform in the conjugate representation compared to $Q^a_\al$. Thus, we take these to be anti-chiral spinors with respect to Lorentz transformations and spinors under $R$-symmetry rotations and denote them by the Grassmann odd $\theta_a^\al$. They obey the reality condition
\begin{equation}
\eqnlab{real_th}
\left(\theta_a^\al \right)^* = - {C^*}_\be^{\ph{\be}\al} \Om^{ab} \theta_b^\be,
\end{equation}
\cf \Eqnref{real_Q}. In this notation, the bosonic coordinates obey
\begin{equation}
 \eqnlab{real_x}
 \left(x^{\al\be}\right)^* = {C^*}_\ga^{\ph{\ga}\al} {C^*}_\de^{\ph{\de}\be} x^{\ga\de}.
\end{equation}
This means that they are not real, as one might have anticipated. The translation from the vector notation with real coordinates $x^\mu$ to the bispinor notation with coordinates $x^{\al\be}$, obeying \Eqnref{real_x}, may be expressed explicitly using gamma matrices.

Having introduced the superspace coordinates, we need to find out how supersymmetry acts on them. We take the central charges $Z^{ab}_{\al\be}$ and $W^{ab}_{\al\be}$ to be zero, since we want to describe an empty superspace. The coordinate representation of the supersymmetry generators is
\begin{equation}
Q^a_\al = \pa^a_\al - i \Om^{ab} \theta^\be_b \pa_{\al\be},
\end{equation}
where $\pa^a_\al$ is the derivative with respect to $\theta^\al_a$ and $\pa_{\al\be}$ is with respect to $x^{\al\be}$. The conventions are such that
\begin{equation}
 \pa^a_\al \theta^\be_b = \kde{\al}{\be} \de^a_{\ph{a}b}, \qquad \pa_{\al\be} x^{\ga \de} = \kde{[\al}{\ga} \kde{\be]}{\de}.
\end{equation}
Taking the generator of translations to be $P_{\al\be}=\pa_{\al\be}$, in accordance with the conventions in Section~\ref{sec:transrot}, we see that this representation of $Q^a_\al$ satisfies the anticommutation relations \eqnref{QQ}.

Let the infinitesimal supersymmetry parameters be denoted by $\eta^\al_a$, which obviously is a fermionic quantity and obeys the same reality condition~\eqnref{real_th} as $\theta^\al_a$. Under supersymmetry, the superspace coordinates transform according to
\begin{equation}
  \begin{aligned}
    \de x^{\al\be} & = \com{\eta^\ga_c Q_\ga^c}{x^{\al\be}} = - i \Om^{ab} \eta_{a\ph{b}}^{[\al} \theta_b^{\be]} \\
    \de \theta^\al_a & = \com{\eta^\ga_c Q_\ga^c}{\theta^\al} = \eta^\al_a,
  \end{aligned}
\end{equation}
where we note that $\eta^\al_a Q^a_\al$ is real. We adopt the standard convention that
\begin{equation}
 \left( \eta^\al_a Q^a_\al \right)^* = \left( Q^a_\al \right) ^* \left( \eta^\al_a \right)^*,
\end{equation}
which is valid for the complex conjugate of any product of fermionic quantities.

We want to reformulate the generator of Lorentz transformations in terms of spinor indices, as we did for the coordinate vector $x^{\al\be}$ and the generator $P_{\al\be}$. $M_{\mu\nu}$ in \Eqnref{Mmunu} is antisymmetric in its vector indices, meaning that it transforms in the ${\bf 15}$ representation of $\mathfrak{so}(5,1)$. This representation can also be built, according to \Eqnref{lorentzreps}, using one chiral and one anti-chiral spinor index.

However, we also have to take into account that Lorentz transformations act non-trivially on the fermionic coordinate $\theta_a^\al$, being an anti-chiral Lorentz spinor. This is accomplished by adding a piece to the generator involving fermionic coordinates and derivatives. The end result is that we may take
\begin{equation}
\dia{M}{\al}{\be} = 2 x^{\be\ga} \pa_{\al \ga} - \frac{1}{2} \kde{\al}{\be} x \cdot \pa + \theta^{\be}_c \pa^c_\al - \frac{1}{4} \dia{\de}{\al}{\be} \theta \cdot \pa,
\end{equation}
where the scalar products are defined as $x \cdot \pa \equiv x^{\al\be} \pa_{\al\be}$ and $\theta \cdot \pa \equiv \theta^\al_a \pa_\al^a$. Note that the trace $\dia{M}{\al}{\al}=0$, as required for a generator transforming in the ${\bf 15}$ representation of the Lorentz algebra. The corresponding parameter is denoted $\dia{\om}{\al}{\be}$ and is also traceless.

Summing up this section, we have found that the $(2,0)$ super-Poincar\'e algebra in six dimensions is generated by $\dia{M}{\al}{\be}$, $Q^a_\al$ and $P_{\al\be}$. It is easy to verify that the generators obey the relations
\begin{equation}
  \begin{aligned}
\com{P_{\al\be}}{M_{\ga}^{\ph{\ga}\de}} &= -2 \kde{[\al}{\de}
   P_{\be]\ga} - \frac{1}{2} \kde{\ga}{\de} P_{\al\be} \,\, &  \com{P_{\al \be}}{Q_{\ga}^c} &= 0 \\
\com{M_{\al}^{\ph{\al}\be}}{M_{\ga}^{\ph{\ga}\de}} &=
   \kde{\al}{\de} \dia{M}{\ga}{\be} - \kde{\ga}{\be} \dia{M}{\al}{\de} & \com{P_{\al\be}}{P_{\ga\de}} &= 0 \\
\com{M_{\al}^{\ph{\al}\be}}{Q^a_{\ga}} &= - \kde{\ga}{\be}
   Q^a_{\al} + \frac{1}{4} \kde{\al}{\be} Q^a_{\ga} &
\acom{Q^a_\al}{Q^b_\be} &= - 2i \Om^{ab} P_{\al\be},
  \end{aligned}
\end{equation}
where we have put all central charges to zero. It should be noted that the $R$-symmetry algebra $\mathfrak{so}(5)$ is an internal symmetry that affects the supersymmetry generators but has no influence on (it commutes with) the bosonic generators of the Poincar\'e algebra. We will have more to say about the $R$-symmetry in the following section.

The generators induce the infinitesimal transformations
\begin{equation}
\eqnlab{superPtransf}
 \begin{aligned}
   \de x^{\al\be} &= a^{\al\be} + \dia{\om}{\ga}{\al} x^{\ga\be} + \dia{\om}{\ga}{\be} x^{\al \ga} - i \Om^{ab} \eta_a^{[\al} \theta_b^{\be]} \\
   \de \theta^\al_a &= \dia{\om}{\ga}{\al} \theta^\ga_a + \eta^a_\al
 \end{aligned}
\end{equation}
when acting on the superspace coordinates $x^{\al\be}$ and $\theta^\al_a$.

\section{Superconformal symmetry}
\label{sec:superconformal}

The preceding section discussed how to extend the Poincar\'e algebra of Section~\ref{sec:lorentz} to a super-Poincar\'e algebra including supersymmetry generators. Our next topic concerns the extension of the conformal algebra of Section~\ref{sec:confalg} to a \emph{superconformal} algebra. We will start with the commutation relations defining the algebra, thereafter we will consider explicit coordinate transformations.

The superconformal algebra is very restrictive and there are, in fact, quite few examples of physical theories that respect this symmetry. These were classified by Nahm~\cite{Nahm:1978} in 1978. He found that the largest space-time dimension consistent with superconformal symmetry is six. In this case, there are two possibilities with relevance to physics: the minimal $\mathcal{N}=(1,0)$ and the extended $\mathcal{N}=(2,0)$, which both are chiral. As stated above, we will focus on the latter, but the truncation to the $\mathcal{N}=(1,0)$ case should be straight-forward. In the minimal case, the $R$-symmetry algebra is $\mathfrak{su}(2)$ rather than $\mathfrak{so}(5)$.

This means that we are considering a superconformal theory with the maximal amount of supersymmetry in the highest possible dimension --- this alone is a motivation for the study of these concepts. One might also consider unitarity restrictions on these theories~\cite{Minwalla:1998,Dobrev:2002}; this constrains the possible values for the scaling dimensions of the fields involved.

Obviously, the superconformal algebra in six dimensions includes the generators of translations ($P_{\al\be}$), Lorentz rotations ($\dia{M}{\al}{\be}$) and supersymmetry transformations ($Q^a_\al$) inherited from the super-Poincar\'e algebra. We also want it to contain the generators of dilatations ($D$) and special conformal transformations (denoted $K^{\al\be}$ in terms of spinor indices) from the bosonic conformal algebra, since we expect that algebra to appear as a factored subalgebra of the full superconformal algebra.

Generally, for these generators to be part of a superconformal algebra, their commutation relations must satisfy the super-Jacobi identity~\cite{Kac:1977}. This is a generalization of the usual Jacobi identity for Lie algebras, and is conventionally written as
\begin{equation}
\eqnlab{superjacobi}
(-1)^{\sss AC} \left[ \left[ A,B \right\} ,C \right\} + (-1)^{\sss AB} \left[ \left[ B, C \right\} ,A \right\} + (-1)^{\sss BC} \left[ \left[ C,A \right\} ,B \right\} = 0.
\end{equation}
In this expression, the $[\cdot,\cdot \}$ bracket denotes an anti-commutator if its entries both are fermionic, otherwise it is a commutator. The signs are such that $(-1)^{\sss A}$ is positive if $A$ is a bosonic operator, and negative if $A$ is fermionic. In other words, one may take $A=0$ in the exponent if the corresponding operator is bosonic, and $A=1$ if it is fermionic.

It is interesting to note that the super-Jacobi identity is impossible to satisfy in $d>6$, which is the origin of the statement above concerning the different possible superconformal theories. This is so because the identities require certain gamma matrix properties, that are true only in low dimensions. The very existence of a superconformal algebra in six dimensions is intimately connected to the triality property of the algebra $\mathfrak{so}(8)$; this algebra has three essentially equivalent eight-dimensional representations: the vector, the chiral spinor and the anti-chiral spinor.

By considering the super-Jacobi identity with $(Q^a_\al,Q^b_\be,K^{\ga\de})$, it turns out that the commutator of $K^{\ga\de}$ and $Q^a_\al$ must involve a new fermionic generator, which we denote by the anti-chiral spinor $S_a^\al$. This generates so called special supersymmetry transformations; the corresponding infinitesimal parameter is denoted by the fermionic chiral spinor $\rho^a_\al$.

Similarly, the super-Jacobi identity involving $(Q^a_\al,Q^b_\be,S_c^\ga)$ forces us to introduce the $R$-symmetry rotation generator $U^{ab}=U^{ba}$, which is a bosonic $\mathfrak{so}(5)$ two-form (transforming in the ${\bf 10}$ representation). The corresponding infinitesimal parameter is denoted by $v_{ab}$.

Finally, we are forced to remove the central charges $Z_{\al\be}^{ab}$ and $W_{\al\be}^{ab}$ appearing in \Eqnref{QQ}, since they do not seem compatible with the superconformal algebra, \cf Refs.~\cite{Bedding:1984,Bedding:1985,Meessen:2003,Peeters:2003}. This may be understood in a rather simple way from the manifestly covariant formalism that will be developed in Chapter~\ref{ch:manifest}.

From the super-Jacobi identity in \Eqnref{superjacobi} and the known commutation relations for the conformal algebra from Section~\ref{sec:confalg}, the explicit commutation relations for the superconformal algebra may be derived. The non-zero relations are found to be (similar expressions appear \eg in Refs.~\cite{Claus:1998,Park:1998})
 \begin{equation}
  \eqnlab{super_comm}
  \begin{split}
  \begin{aligned}
   \com{P_{\al\be}}{M_{\ga}^{\ph{\ga}\de}} &= -2 \kde{[\al}{\de}
   P_{\be]\ga}^{\ph{\de}} - \frac{1}{2} \kde{\ga}{\de} P_{\al\be} &
   \acom{Q_{\al}^a}{Q_{\be}^b} &= -2i \Om^{ab} P_{\al \be} \\
   \com{K^{\al\be}}{M_{\ga}^{\ph{\ga}\de}} &= 2 \kde{\ga}{[\al}
   K^{\be]\de} + \frac{1}{2} \kde{\ga}{\de} K^{\al\be} &
   \big\{ S^{\al}_a, S^{\be}_b \big\} &= -2i \Om_{ab} K^{\al \be} \\
   \com{M_{\al}^{\ph{\al}\be}}{M_{\ga}^{\ph{\ga}\de}} &=
   \kde{\al}{\de} \dia{M}{\ga}{\be} - \kde{\ga}{\be} \dia{M}{\al}{\de} &
   \com{P_{\al \be}}{D} &= P_{\al \be} \\
   \com{U^{ab}}{U^{cd}} &= - \Om^{a(c} U^{d)b} - \Om^{b(c}
   U^{d)a} &
   \com{K^{\al \be}}{D} &= - K^{\al \be} \\
   \com{M_{\al}^{\ph{\al}\be}}{Q^a_{\ga}} &= - \kde{\ga}{\be}
   Q^a_{\al} + \frac{1}{4} \kde{\al}{\be} Q^a_{\ga} &
   \com{Q_{\al}^a}{D} &= \frac{1}{2} Q_{\al}^a \\
   \com{\dia{M}{\al}{\be}}{S_a^{\ga}} &= \kde{\al}{\ga} S_a^{\be} -
   \frac{1}{4} \kde{\al}{\be} S_a^{\ga} &
   \com{S^{\al}_a}{D} &= -\frac{1}{2} S^{\al}_a \\
   \com{K^{\al \be}}{Q_{\ga}^a} &= - 2 \Om^{ac}
   \de_{\ga}^{\ph{\ga}[\al} S_c^{\be]} &
   \com{U^{ab}}{Q_{\ga}^c} &= \Om^{c(a} Q^{b)}_{\ga} \\
   \com{P_{\al \be}}{S^{\ga}_a} &= 2 \Om_{ac}
   \de_{[\al}^{\ph{[\al}\ga} Q^{c\ph{\ga}}_{\be]} &
   \com{U^{ab}}{S^{\ga}_c} &= \kde{c}{(a}\Om^{b)d} S_d^{\ga} \\
   \end{aligned} \\
   \begin{aligned}
   \big\{Q_{\al}^a,S^{\be}_b\big\} &= i \kde{\al}{\be} \Big( \de^a_{\ph{a}b} D - 4 \Om_{bc} U^{ac}
   \Big) + 2i \de^a_{\ph{a}b} \dia{M}{\al}{\be}  \qquad \quad \\
   \com{P_{\al \be}}{K^{\ga\de}} &= - 4
   \kde{[\al}{[\ga} \dia{M}{\be]}{\de]} - 2 \kde{[\al}{\ga}
   \kde{\be]}{\de} D.
  \end{aligned}
  \end{split}
 \end{equation}
As we will see in Chapter~\ref{ch:manifest}, this algebra is isomorphic to the superalgebra $\mathfrak{osp}(8^*|4)$. The bosonic subalgebra of this superalgebra is isomorphic to $\mathfrak{so}^*(8) \times \mathfrak{so}(5)$, where we recognize the real form $\mathfrak{so}^*(8) \simeq \mathfrak{so}(6,2)$ as the bosonic conformal algebra, as required.

The next step is to find out how these generators act on the superspace coordinates. It turns out that the coordinate representations
\begin{equation}
  \eqnlab{super_P}
  \begin{aligned}
  P_{\al \be} &= \pa_{\al \be} \\
\dia{M}{\al}{\be} &= 2 x^{\be\ga} \pa_{\al \ga} - \frac{1}{2} \kde{\al}{\be} x \cdot \pa + \theta^{\be}_c \pa^c_\al - \frac{1}{4} \dia{\de}{\al}{\be} \theta \cdot \pa \\
  D &= x \cdot \pa + \frac{1}{2} \theta \cdot \pa \\
  K^{\al \be} &= -4 x^{\al\ga} x^{\be\de} \pa_{\ga\de}
  - \theta^{\ga} \cdot \theta^{[\al} \theta^{\be]} \cdot \theta^{\de} \pa_{\ga\de} +2 \theta_c^{[\al} \Big( 2 x^{\be] \ga} - i \theta^{\be]} \cdot \theta^{\ga} \Big) \pa^c_{\ga}
\end{aligned}
\end{equation}
for the bosonic conformal group generators,
\begin{equation}
  \eqnlab{Q_super}
  \begin{aligned}
  Q^a_{\al} &= \pa^a_{\al} - i \Om^{ac} \theta_c^{\ga} \pa_{\al\ga}
  \\
  S_a^{\al} &= \Om_{ac} \Big( 2 x^{\al \ga} - i \theta^{\al} \cdot \theta^{\ga} \Big) \pa^c_{\ga} + 2i \theta_a^{\ga} \theta_c^{\al} \pa^c_{\ga} - i \theta_a^{\ga} \Big( 2 x^{\de \al} - i \theta^{\de} \cdot \theta^{\al} \Big) \pa_{\ga \de}
  \end{aligned}
\end{equation}
for the fermionic generators, and
\begin{equation}
  \eqnlab{super_U}
  U^{ab} = \frac{1}{2} \Big( \Om^{ac} \theta_c^{\ga}
  \pa_{\ga}^b + \Om^{bc} \theta_c^{\ga} \pa_{\ga}^a \Big)
\end{equation}
for the $R$-symmetry generator satisfy the relations in \Eqnref{super_comm}. In these equations, we have introduced a dot product between fermionic coordinates as
\begin{equation}
  \theta^\al \cdot \theta^\be \equiv \Om^{ab} \theta_a^\al \theta_b^\be,
\eqnlab{thetadot}
\end{equation}
which obviously is symmetric in the indices $\al$ and $\be$.

If we let these generators act on the coordinates $x^{\al\be}$ and $\theta^\al_a$, we find that the corresponding coordinate transformations are
\begin{align}
\eqnlab{transf_superx}
\begin{split}
\de x^{\al \be} &= a^{\al \be} - \dia{\om}{\ga}{[\al} x^{\be] \ga} +
+ \la x^{\al \be} + 4 c_{\ga \de} x^{\ga\al} x^{\be\de} - i \Om^{ab} \eta^{[\al}_{a\ph{b}} \theta^{\be]}_b - {} \\
& \quad - c_{\ga \de} \theta^\ga \cdot \theta^{[\al}
\theta^{\be]} \cdot \theta^{\de} -i \rho^c_{\ga} \theta^{[\al}_c
    \left( 2 x^{\be]\ga} - i \theta^{\be]} \cdot \theta^{\ga} \right)
\end{split} \\
\eqnlab{transf_superth}
\begin{split}
\de \theta^{\al}_a &= (\dia{\om}{\ga}{\al} - 4 c_{\ga \de} x^{\al \de}
- 2i c_{\ga \de} \theta^{\al} \cdot \theta^{\de} + 2i
\rho^c_{\ga} \theta^{\al}_c) \theta^{\ga}_a + \frac{1}{2} \la \theta^{\al}_a + {} \\
& \quad + \eta^{\al}_a - \Om_{ac} \rho^c_{\ga} \left( 2 x^{\ga \al} - i \theta^{\ga} \cdot \theta^{\al} \right) + v_{ac} \Om^{cd} \theta^{\al}_d,
\end{split}
\end{align}
which tell us how a general superconformal transformation acts on superspace. These are presented in {\sc Paper V}, and similar transformations, but in a different notation, appear in Ref.~\cite{Park:1998}. We recognize some pieces of these transformations from our previous results --- the bosonic conformal transformations (although the notation has been changed) from \Eqnref{xi_x} and the super-Poincar\'e transformations from \Eqnref{superPtransf}.

\section{Superfield transformations}
\label{sec:superfields}

In the preceding sections of this chapter, we have developed the notion of a superspace, having both bosonic and fermionic coordinates. The logical next step is to formulate \emph{superfields} living in this superspace. They are functions of the superspace coordinates (both bosonic and fermionic) and may carry indices indicating how they transform under Lorentz transformations, like the space-time fields, but also indices showing how $R$-symmetry acts on them.

Consider an arbitrary superfield $\Phi^i(x,\theta)$, where $i$ is a collective notation for Lorentz indices and/or $R$-symmetry indices. The field may be either bosonic or fermionic (commuting or anti-commuting). As in the bosonic case in \Eqnref{Pmu}, we require the generator of bosonic translations to act only differentially, i.e., according to
\begin{equation}
\eqnlab{Paction}
\com{P_{\al\be}}{ \Phi^i(x,\theta)} = \pa_{\al\be} \Phi^i(x,\theta).
\end{equation}
The important point here is that translations do not affect the indices.

Likewise, we demand that supersymmetry acts only differentially, according to
\begin{equation}
\eqnlab{Qaction}
\big[Q^a_\al,\Phi^i(x,\theta) \big\}= \left( \pa^a_\al - i \Om^{ab} \theta_b^\be \pa_{\al\be} \right) \Phi^i(x,\theta),
\end{equation}
where the bracket is a commutator if $\Phi^i(x,\theta)$ is a bosonic field and an anti-commutator if $\Phi^i(x,\theta)$ is a fermionic field. This relation actually defines what we mean by a superfield, and contains a lot of information about the superfield and its properties. If we expand $\Phi^i(x,\theta)$ in powers of $\theta$ (this expansion will always be finite, since the fermionic coordinates are anticommuting, and contains 16 terms at the most) as
\begin{equation}
\Phi^i(x,\theta) = \phi^i_0(x) + \theta^\al_a (\phi^i_1)^a_\al (x) + \ldots,
\end{equation}
then a supersymmetry transformation, given by
\begin{equation}
\eqnlab{susy_transf}
\de \Phi^i(x,\theta) \equiv \com{\eta \cdot Q }{\Phi^i(x,\theta)},
\end{equation}
tells us that the variation of one component field depends on other component fields. For example, it follows that the lowest component of $\Phi^i(x,\theta)$ transforms according to
\begin{equation}
\de \phi^i_0(x) = \eta^\ga_c (\phi^i_1)^c_\ga(x),
\end{equation}
which relates the variation of a bosonic field to a fermionic or vice versa, depending on whether the superfield $\Phi^i(x,\theta)$ is bosonic or fermionic.


The next step, and the main goal of this section, is to derive how a general superconformal transformation acts on a superfield. We will do this by generalizing the method of induced representations from Section~\ref{sec:conformal_field} to superspace.

The little group in this case is the group that leaves both $x=0$ and $\theta=0$ invariant. From \Eqsref{transf_superx}--\eqnref{transf_superth} we see that this group consists of all superconformal generators except supersymmetry transformations (generated by $Q^a_\al$) and translations (generated by $P_{\al\be}$). The commutation relations for the little group follow from \Eqnref{super_comm}.

The action of the little group generators on $\Phi(x=0,\theta=0)$ is given by
\begin{equation}
  \begin{aligned}
  \com{\dia{M}{\al}{\be}}{\Phi^i(0,0)} &= \left( \dia{\Si}{\al}{\be} \Phi \right)^i(0,0) & \com{S_a^\al}{\Phi^i(0,0)} &= \left( \si_a^\al \Phi \right)^i(0,0) \\
 \com{D}{\Phi^i(0,0) } &= \left( \De \Phi \right)^i(0,0) & \,\, \com{U^{ab}}{\Phi^i(0,0)} &= \left( u^{ab} \Phi \right)^i(0,0) \\
 \com{K^{\al\be}}{\Phi^i(0,0)} &= \left( \ka^{\al\be} \Phi\right)^i(0,0),
  \end{aligned}
\end{equation}
in analogy with \Eqnref{littleaction} from the bosonic case. The little group generators $\dia{\Si}{\al}{\be}$, $\De$ and $\ka^{\al\be}$ are analogous to the corresponding quantities in the bosonic case, while $\si_a^\al$ and $u^{ab}$ obviously correspond to special supersymmetry and $R$-symmetry.

Next, we choose a basis in index space so that \Eqsref{Paction} and \eqnref{Qaction} are valid. This means that we may translate the superfield according to
\begin{equation}
\Phi^i(x,\theta) = e^{x \cdot P + \theta \cdot Q} \Phi^i(0,0) e^{- x \cdot P - \theta \cdot Q},
\end{equation}
where it should be noted that
\begin{equation}
\com{\theta \cdot Q}{\Phi^i(x,\theta)} = \theta^\al_a \pa^a_\al \Phi^i(x,\theta).
\end{equation}
In this expression, the term involving the bosonic derivative $\pa_{\al\be}$ has vanished algebraically.

Following the recipe from the bosonic case, we need to evaluate
\begin{equation}
\eqnlab{superordotilde}
\tilde{\ordo} \equiv e^{x \cdot P + \theta \cdot Q} \ordo e^{-x \cdot P - \theta \cdot Q},
\end{equation}
where $\ordo$ denotes any of the little group generators. The result is
\begin{equation}
  \begin{aligned}
    \dia{\tilde{M}}{\al}{\be} &= \dia{M}{\al}{\be} - 2 x^{\be\ga} P_{\al\ga} + \frac{1}{2} \kde{\al}{\be} x \cdot P - \theta^\be_c Q^c_\al + \frac{1}{4} \theta \cdot Q - i \theta^\be \cdot \theta^\ga P_{\al\ga} \\
    \tilde{D} &= D - x \cdot P - \frac{1}{2} \theta \cdot Q \\[2mm]
    \tilde{K}^{\al\be} &= K^{\al\be} - \left( 4 x^{\ga[\al} - 2i \theta^\ga \cdot \theta^{[\al} \right) \dia{M}{\ga}{\be]} + 2 x^{\al\be} D - 2 \Om^{cd} \theta_{c\ph{d}}^{[\al} S_d^{\be]} - {} \\
& \quad - 4i \theta^\al_c \theta^\be_d U^{cd} - 4 x^{\al\ga} x^{\be\de} P_{\ga\de} + 4i x^{\ga[\al} \theta^{\be]} \cdot \theta^\de P_{\ga\de} + {} \\
& \quad + \theta^{\al} \cdot \theta^\ga \theta^\be \cdot \theta^\de P_{\ga\de} + 4 \theta^{[\al}_c x^{\be]\ga}_{\ph{c}} Q^c_\ga - 2i \theta^\ga \cdot \theta^{[\al}_{\ph{c}} \theta^{\be]}_c Q^c_\ga
  \end{aligned}
\end{equation}
for the generators of the bosonic conformal group, and
\begin{equation}
  \begin{aligned}
    \tilde{S}_a^\al &= S_a^\al - 2i \theta_a^\ga \dia{M}{\ga}{\al} - i \theta^\al_a D + 4i \Om_{ac} \theta_d^\al U^{cd} -2i \theta_c^\al \theta_a^\ga Q^c_\ga- {} \\ & \quad - \left(4i x^{\al\ga} - 2 \theta^\al \cdot \theta^\ga \right)\theta_a^\de P_{\ga\de} - \Om_{ac} \left( 2 x^{\al\ga} + i \theta^\al \cdot \theta^\ga \right) Q^c_\ga \\
    \tilde{U}^{ab} &= U^{ab} - \frac{1}{2} \Om^{ac} \theta_c^\ga Q^{b}_\ga - \frac{1}{2} \Om^{bc} \theta_c^\ga Q^{a}_\ga - i \Om^{ac} \Om^{bd} \theta^\ga_c \theta^\de_d P_{\ga\de}
  \end{aligned}
\end{equation}
for special supersymmetry and $R$-symmetry. This yields that the superconformal variation of a superfield $\Phi^i(x,\theta)$ becomes
\begin{multline}
\eqnlab{transf_superfield}
  \de_{\sss C} \Phi^i(x,\theta) = a^{\al\be} \pa_{\al\be} \Phi^i +  \la \left\{ (\De \Phi )^i + \left[ x \cdot \pa + \frac{1}{2} \theta \cdot \pa \right] \Phi^i  \right\} + {} \\
\shoveleft{ \quad + \eta^\al_a \left\{ \pa^a_\al - i \Om^{ab} \theta^\be_b \pa_{\al\be} \right\} \Phi^i + v_{ab} \left\{ (u^{ab} \Phi)^i + \Om^{ac} \theta_c^\ga \pa_\ga^b \Phi^i  \right\} + {} } \\
\shoveleft{ \quad + \dia{\om}{\be}{\al} \left\{ (\dia{\Si}{\al}{\be} \Phi )^i + \left[2 x^{\be\ga} \pa_{\al\ga} + \theta^\be_c \pa^c_\al \right] \Phi^i  \right\} + {} } \\
\shoveleft{ \quad + c_{\al\be} \Big\{ (\ka^{\al\be} \Phi )^i +
\big[ 4 x^{\ga\al} x^{\be\de} \pa_{\ga\de} - \theta^\al \cdot \theta^\ga \theta^\be \cdot \theta^\de \pa_{\ga\de} + 4 \theta^\al_c x^{\be\ga} \pa^c_\ga - {} } \\
\shoveleft{ \quad - 2i \theta_c^\al \theta^\be \cdot \theta^\ga \pa^c_\ga \big] \Phi^i -
2 x^{\al\be} ( \De \Phi )^i  - \left( 4 x^{\al\ga} - 2i \theta^\al \cdot \theta^\ga \right) ( \dia{\Si}{\ga}{\be} \Phi )^i  + {} } \\
\shoveleft{ \quad + 2 \Om^{cd} \theta_c^\al ( \si_d^\be \Phi )^i -
 4i \theta_c^\al \theta_d^\be (u^{cd} \Phi )^i  \Big\} + \rho^a_\al \Big\{ (\si_a^\al \Phi)^i + {} } \\
\shoveleft{ \quad + \big[ \Om_{ac} (2 x^{\al\ga} - i \theta^\al \cdot \theta^\ga ) \pa^c_\ga - i \theta^\ga_a ( 2 x^{\de\al} - i \theta^\de \cdot \theta^\al ) \pa_{\ga \de} + 2i \theta^\ga_a \theta^\al_c \pa^c_\ga \big] \Phi^i + {} } \\
{} + i \theta_a^\al (\De \Phi)^i + 2i \theta_a^\ga (\dia{\Si}{\ga}{\al} \Phi) - 4i \Om_{ac} \theta_d^\al ( u^{cd} \Phi) \Big\},
\end{multline}
where all fields on the right-hand side are evaluated in the general superspace point $(x,\theta)$. This expression looks very complicated, but fortunately there is a nice way of compactifying the notation, developed in {\sc Paper V}. In analogy with \Eqnref{general_transf}, we write
\begin{multline}
\eqnlab{transf_func}
  \de_{\sss C} \Phi^i(x,\theta) = \xi^{\al\be}(x,\theta) \pa_{\al\be} \Phi^i + \xi^\al_a(x,\theta) \pa^a_\al \Phi^i + \dia{\Om}{\al}{\be}(x,\theta) (\dia{\Si}{\be}{\al} \Phi)^i + {}\\
{} + \La(x,\theta) (\De \Phi)^i + V_{ab}(\theta) (u^{ab} \Phi)^i + R^a_\al(\theta) (\si_a^\al \Phi)^i + c_{\al\be} (\ka^{\al\be} \Phi)^i,
\end{multline}
where the superspace-dependent parameter functions\footnote{Note that the notation is slightly different compared to the relations given in \textsc{Paper V} and \textsc{Paper VI}. Hopefully, the notation in this text is more logical and transparent.} are given by
\begin{align}
\begin{split}
\eqnlab{xi_bos}
\xi^{\al\be}(x,\theta) &= a^{\al\be} - 2 \dia{\om}{\ga}{[\al} x^{\be]\ga}_{\ph{\ga}} + \la x^{\al\be} + 4 c_{\ga\de} x^{\ga\al} x^{\be\de} - i \eta^{[\al} \cdot \theta^{\be]} -{} \\ & \quad
- c_{\ga\de} \theta^\ga \cdot \theta^\al \theta^\be \cdot \theta^\de - 2i \rho^c_\ga \theta^{[\al}_c x^{\be]\ga}_{\ph{c}} - \rho^c_\ga \theta^{[\al}_c \theta^{\be]}_{\ph{c}} \cdot \theta^\ga_{\ph{c}}
\end{split} \\
\begin{split}
\xi^\al_a(x,\theta) &= \dia{\om}{\ga}{\al} \theta^\ga_a + 2 c_{\ga\de} \theta^\ga_a (2 x^{\de \al} - i \theta^\de \cdot \theta^\al) + \frac{1}{2} \la \theta^\al_a + \eta^\al_a + {} \\
&\quad + 2i \rho^c_\ga \theta^\al_c \theta^\ga_a - \Om_{ac} \rho^c_\ga ( 2 x^{\ga\al} - i \theta^\ga \cdot \theta^\al) + v_{ac} \Om^{cd} \theta_d^\al
\end{split} \\
\begin{split}
\dia{\Om}{\al}{\be}(x,\theta) &= \dia{\om}{\al}{\be} - 4 c_{\al\ga} x^{\be\ga} + c \cdot x \kde{\al}{\be} - 2i c_{\al\ga} \theta^\be \cdot \theta^\ga + {} \\
& \quad + 2i \rho^c_\al \theta^\be_c - \frac{i}{2} \rho \cdot \theta \kde{\al}{\be}
\end{split}\\
\La(x,\theta) &= \la - 2 c \cdot x + i \rho \cdot \theta \\
V_{ab}(\theta) &= v_{ab} - 4i c_{\ga\de} \theta_a^\ga \theta^\de_b + 2i \Om_{ac} \rho^c_\ga \theta^\ga_b + 2i \Om_{bc} \rho^c_\ga \theta^\ga_a \\
\eqnlab{Raal}
R^a_\al(\theta) &= \rho^a_\al + 2 \Om^{ac} c_{\al\ga} \theta^\ga_c.
\end{align}
We recognize the function $\xi^{\al\be}(x,\theta)$ as the variation of the coordinate $x^{\al\be}$ from \Eqnref{transf_superx} and $\xi^\al_a(x,\theta)$ as the variation of $\theta^\al_a$ from \Eqnref{transf_superth}. The superspace-dependent parameter functions $\dia{\Om}{\al}{\be}(x,\theta)$, $\La(x,\theta)$, $R^a_\al(\theta)$ and $V_{ab}(\theta)$ all have intuitive meanings and this notation will make things considerably simpler (at least notationally) in the following. Again, we call a field with $\ka^{\al\be}=0$ \emph{primary}, while a field with both $\ka^{\al\be}=0$ and $\si^\al_a=0$ is called \emph{superprimary}.

As a final comment in this chapter, we would like to present the coordinate variations of the superspace differentials, in the same manner as we did in the bosonic case in \Eqnref{bos_diff}. In superspace, the relevant differentials are
\begin{equation}
\eqnlab{super_diffs}
\begin{aligned}
e^{\al \be} &= d x^{\al \be} + i \Om^{ab} \theta_a^{[\al} d \theta_b^{\be]} \\
e^\al_a &= d\theta^\al_a,
\end{aligned}
\end{equation}
which both are invariant under supersymmetry. These are introduced in {\sc Paper III} and we will have more to say about them in Chapter~\ref{ch:coupling}.

It is rather easily shown that the induced transformation of the superspace differentials, when the coordinates are transformed according to \Eqsref{transf_superx} and \eqnref{transf_superth}, is given by
\begin{equation}
\eqnlab{transf_e}
\begin{split}
\de e^{\al \be} &= \dia{\Om}{\ga}{\al}(x,\theta) e^{\ga \be} +
\dia{\Om}{\ga}{\be}(x,\theta) e^{\al \ga} + \La(x,\theta) e^{\al \be} \\
\de e_a^\al &= \dia{\Om}{\ga}{\al}(x,\theta) e_a^\ga +
\frac{1}{2} \La(x,\theta) e_a^\al + V_{ac}(\theta) \Om^{cd}
e_d^\al + 2 \Om_{ac} R^c_{\ga}(\theta) e^{\al \ga},
\end{split}
\end{equation}
where the superspace-dependent parameter functions are the same as in \Eqsref{xi_bos}--\eqnref{Raal}.

The transformations in \Eqnref{transf_e} are very similar to those for a general superfield and contain the expected Lorentz, dilatation and $R$-symmetry parts (with
generalized superspace-dependent parameters), but also a term
connecting the variation of $e_a^\al$ to $e^{\al \be}$ with the parameter function $R^a_{\al}(\theta)$, containing the (constant) parameters $c_{\al\be}$ and $\rho^a_\al$. This separates special conformal and special supersymmetry transformations
from the other transformations, which yield no such possibilities. This new ingredient corresponds to the action of the little group generator $\si^\al_a$ on a field, and will be useful in the second part of this thesis.

It is also interesting to note that the infinitesimal supersymmetric interval length is preserved up to a superspace-dependent scale factor under superconformal transformations, i.e.,
\begin{equation}
\de \left( \frac{1}{2}\eps_{\al\be\ga\de} e^{\al\be} e^{\ga\de} \right) = \La(x,\theta) \eps_{\al\be\ga\de} e^{\al\be} e^{\ga\de}.
\end{equation}
This relation may in fact be seen as a definition of the superconformal transformations, in analogy with \Eqnref{prop_time_conf} in the bosonic case.

\chapter{Manifest invariance}
\label{ch:manifest}

It is sometimes stated that the only good symmetries in physics are manifest symmetries. Such symmetries are built into the physical models notationally, in the sense that everything you may write down, which is consistent with some general rules, will respect the symmetry in question. The most common and well-known example of this concept is Lorentz invariance, which is made manifest by the use of Lorentz tensors. These are quantities that transform in a specific and linear way under Lorentz transformations.

The purpose of this chapter is to describe a formalism in which not only Lorentz symmetry, but also conformal and superconformal symmetry is made manifest.

%
%

\section{Conformal symmetry}
\label{sec:manifest_conformal}

For simplicity, we will start with the bosonic conformal symmetry group. As in Chapter~\ref{ch:conformal}, we will present the formalism in a space-time with six dimensions, but everything should be easily generalizable to a space-time with arbitrary dimensionality.


\subsection{Coordinates}

Consider a six-dimensional space-time with Minkowski signature and coordinates $x^\mu$, $\mu=(0,\ldots,5)$. The conformal group in this space-time is the special orthogonal group $SO(6,2)$, which also happens to be the isometry group of a space with six space-like and two time-like dimensions, the \emph{conformal space}. This observation is the basis of the manifestly conformally covariant formalism, introduced by P.~Dirac~\cite{Dirac:1936} in 1936.
Since then, his work has been refined and extended~\cite{Kastrup:1966,Marnelius:1979,Marnelius:1979b,Marnelius:1980}, most notably by G.~Mack and A.~Salam~\cite{Mack:1969}.

Let us denote the coordinates in the conformal space by the eight-component vector $y^{\muh}=(y^\mu,y^+,y^-)$, where obviously $y^+$ and $y^-$ are the additional dimensions compared to the underlying six-dimensional Minkowski space-time. These are in "light-cone form", which means that the conformal space metric is
\begin{equation}
\eta_{\muh\nuh} = \left(
\begin{array}{ccc}
 \eta_{\mu \nu} & 0 & 0 \\
 0 & 0 & -1 \\
 0 & -1 & 0
\end{array}
\right),
\end{equation}
where $\eta_{\mu\nu}$ is the space-like Minkowski metric in six dimensions. The metrics can, in the usual way, be used to raise and lower vector indices on coordinates and tensors.

The next step is to relate the generators of the conformal algebra to rotation generators in the conformal space, denoted in the standard way by $J_{\muh\nuh}$. It turns out that, if we let
\begin{equation}
\begin{aligned}
 M_{\mu\nu} &= J_{\mu\nu} \\
 P_{\nu} &= 2 J_{+ \nu} \\
 K_{\nu} &= 4 J_{- \nu} \\
 D &= 2 J_{-+},
\end{aligned}
\eqnlab{J-pieces}
\end{equation}
where the numerical factors are purely conventional, the commutation relations for the conformal algebra in \Eqnref{bosonic_comm_coord} may be summarized in a single relation, namely
\begin{equation}
\left[ J_{\muh\nuh},J_{\rhoh\sih} \right] = \eta_{\rhoh[\muh} J_{\nuh]\sih} - \eta_{\sih[\muh} J_{\nuh]\rhoh}.
\eqnlab{SO(D,2)}
\end{equation}
This is the standard lorentzian relation, corresponding to the algebra $\mathfrak{so}(6,2)$. In this way, we have shown that the conformal group in six dimensions is isomorphic to $SO(6,2)$ as claimed. Consequently, conformal transformations (from a six-dimensional point of view) act as Lorentz rotations in the eight-dimensional conformal space. This is the key observation behind the manifestly covariant formalism, which enables us to make conformal symmetry manifest, in exactly the same way as we make Lorentz symmetry manifest, using ordinary tensor notation. We should also mention that the conformal algebra is isomorphic to the anti-deSitter algebra in seven dimensions; this is an essential ingredient in the so called AdS/CFT correspondence~\cite{Maldacena:1998}.

However, we still want to describe physics in six dimensions. Therefore, the next step is to connect this eight-dimensional conformal space to the ordinary six-dimensional Minkowski space-time. Following the argument in Refs.~\cite{Dirac:1936,Mack:1969}, we regard the six-dimensional space-time as the surface of a \emph{projective hypercone} in eight dimensions. This hypercone is defined through the relation
\begin{equation}
y^2 \equiv \eta_{\muh\nuh} y^{\muh} y^{\nuh} = 0,
\eqnlab{hypercone}
\end{equation}
which by definition is invariant under the (linear) $SO(6,2)$ transformations. The connection to the coordinates in six-dimensional Minkowski space-time is made through the parametrization
\begin{equation}
\begin{aligned}
y^{\mu} &= \ga x^\mu \\
y^+ &= \frac{1}{2} \ga x \cdot x \\
y^- &= \ga,
\end{aligned}
\eqnlab{y-x}
\end{equation}
where $x^{\mu}$ denotes the usual six-dimensional coordinate vector and $\ga$ is a projective parameter, which requires $\ga \neq 0$. It is apparent that this parametrization satisfies the hypercone condition~\eqnref{hypercone}. In fact, the conformal space coordinates $y^{\muh}$ should really be regarded as homogeneous coordinates in a seven-dimensional projective space.

A priori, it is not evident (although likely) that the quantities $x^\mu$ in \Eqnref{y-x} really are the usual coordinates in the six-dimensional space-time. To show this, we need to consider how they transform under conformal transformations.

By definition, the eight-dimensional coordinate vector $y^{\muh}$ transforms linearly under an $SO(6,2)$ transformation. Comparing with the Lorentz generator in~\Eqnref{Mmunu}, it is clear that the usual choice
\begin{equation}
L_{\muh\nuh} = y_{[\nuh} \pa_{\muh]},
\eqnlab{diff_J}
\end{equation}
where $\pa_{\muh}$ is the partial derivative with respect to $y^{\muh}$, satisfies the commutation relations~\eqnref{SO(D,2)}. We denote the differential generator by $L_{\muh\nuh}$, for reasons which will be apparent when we consider the transformation of fields in the next section. This means that $y^{\muh}$ transforms according to
\begin{equation}
\de y^{\muh} = \com{\pi^{\rhoh\sih} L_{\rhoh\sih}}{ y^{\muh} }= \pi^{\muh\sih} y_{\sih},
\end{equation}
where $\pi^{\rhoh\sih}$ is a matrix containing the infinitesimal parameters of the transformation. We want to identify the components of this matrix with the usual parameters of the conformal group, in the same way as we did for $J_{\muh\nuh}$ in \Eqnref{J-pieces}. This is done by imposing that
\begin{equation}
\pi^{\rhoh\sih} J_{\rhoh \sih} = \om^{\rho \si} M_{\rho \si} + a^\rho
  P_\rho + c^\rho K_\rho + \la D,
\end{equation}
which tells us that we need to take
\begin{equation}
\begin{aligned}
 \pi^{\mu\nu} &= \om^{\mu\nu} \\
 \pi^{+ \nu} &= a^{\mu} \\
 \pi^{- \nu} &= 2 c^{\mu} \\
 \pi^{-+} &= \la,
\end{aligned}
\eqnlab{pi-pieces}
\end{equation}
where the quantities $\om^{\mu\nu}$, $a^{\mu}$, $c^{\mu}$ and $\la$ are the (infinitesimal) parameters of Lorentz transformations, translations, special conformal transformations and dilatations, respectively, as defined in Chapter~\ref{ch:conformal}.

Putting all this together, we see that we are required to take
\begin{equation}
\eqnlab{delta_x}
\begin{aligned}
\de x^{\mu} &= a^{\mu} + \om^{\mu \nu} x_{\nu} +
\la x^{\mu} + c^{\mu} x^2 - 2 c\cdot x x^{\mu} \\
\de \ga &= \left( 2 c \cdot x - \la \right) \ga,
\end{aligned}
\end{equation}
in order to make \Eqnref{y-x} consistent.
The first of these two equations is the well-known conformal coordinate transformation from \Eqnref{xi_x}, as we expected. The second transformation tells us what happens to the projective parameter $\ga$, but this information is more or less irrelevant since we are on a projective hypercone. Thus, the quantity $x^{\mu}$ from \Eqnref{y-x} can consistently be interpreted as the six-dimensional coordinate vector.

\subsection{Fields}
\label{sec:manifest_fields}

Having defined and parametrized the projective hypercone in the preceding subsection, the next step is to construct \emph{fields} in the conformal space and consider what they look like in terms of the coordinates $x^{\mu}$ on the hypercone. The motivation for this is simply that such a construction enables a manifestly conformally covariant formulation, since the conformal group acts linearly on fields in the eight-dimensional space, in the same way as the Lorentz group acts linearly on tensors in ordinary Minkowski space.

Consider a field $\Ups^i(y)$ restricted to the hypercone~\eqnref{hypercone}, where the superscript $i$ denotes some index or indices, depending on the type of field. Note that $i$ should correspond to a representation of the conformal group $SO(6,2)$. We also require the field to be a \emph{homogeneous} function of the coordinates $y^{\muh}$ in the sense that
\begin{equation}
y^{\muh} \pa_{\muh} \Ups^i(y) =  n \Ups^i(y),
\eqnlab{homogeneity_general}
\end{equation}
where $n$ denotes the \emph{degree of homogeneity}. Noting that
\begin{equation}
y^{\muh} \frac{\pa}{\pa y^{\muh}} = \ga \frac{\pa}{\pa \ga},
\end{equation}
which follows from \Eqnref{y-x}, this allows us to remove the dependence on the projective parameter $\ga$ from the field. Thus, we define the new field
\begin{equation}
\tilde{\Phi}^i(x) = \ga^{-n} \Ups^i(y),
\end{equation}
where it is clear that $\tilde{\Phi}^i(x)$ depends only on the coordinates $x^{\mu}$ (remember that we defined the field $\Ups^i(y)$ on the hypercone, not in the entire eight-dimensional space).

However, this is not enough to reproduce consistent fields in six dimensions. This failure can be understood by considering the generators of the conformal group. Since we are acting on fields, the conformal generator is divided into two parts according to
\begin{equation}
\eqnlab{split-operator}
\tilde{J}_{\muh\nuh} = L_{\muh\nuh} + s_{\muh\nuh},
\end{equation}
where the tilde on $\tilde{J}_{\muh\nuh}$ denotes that the operator is supposed to act on the field $\tilde{\Phi}^i(x)$, but it also acts on $\Ups^i(y)$. The generator has the same form in these two cases (it commutes with $\ga^{-n}$), given that we use~\Eqnref{homogeneity_general} to introduce the degree of homogeneity $n$ in the expressions whenever possible.

Moreover, $L_{\muh\nuh}$ is the differential (orbital) piece, explicitly given by \Eqnref{diff_J}, while $s_{\muh\nuh}$ is the intrinsic (spin) piece, acting only on the indices of the field. The form of the latter depends on which field it acts upon, or rather, in which representation the field transforms. It satisfies the commutation relations~\eqnref{SO(D,2)} and is completely analogous to the quantity $\Si_{\mu\nu}$ used in Section~\ref{sec:conformal_field} when discussing the Lorentz group. This means that the field $\Ups^i(y)$ transforms according to
\begin{equation}
\de \Ups^i(y) = \pi^{\muh\nuh} \tilde{J}_{\muh \nuh} \Ups^i(y) = \pi^{\muh\nuh} y_{\nuh} \pa_{\muh} \Ups^i(y) + \pi^{\muh\nuh} \left( s_{\muh\nuh} \Ups \right)^i(y),
\end{equation}
where $\pi^{\muh\nuh}$ again denotes the parameter matrix.

If we decompose $\tilde{J}_{\muh\nuh}$ according to \Eqnref{J-pieces} and translate to the variables $x^\mu$, we find that its components are
\begin{equation}
\begin{aligned}
 \tilde{M}_{\mu\nu} &= x_{[\nu} \pa_{\mu]} + s_{\mu\nu} \\
 \tilde{P}_{\mu} &= \pa_{\mu} + 2 s_{+\mu}\\
 \tilde{K}_{\mu} &= x \cdot x \pa_{\mu} - 2 x_{\mu} \left( x \cdot \pa - n \right) + 4 s_{-\mu}\\
 \tilde{D} &= x \cdot \pa - n + 2 s_{-+},
\end{aligned}
\eqnlab{MPKD_tilde}
\end{equation}
where it should be noted that the vectorial derivatives are with respect to $x^{\mu}$. We see that the differential pieces of the generators agree with the standard result, but the expressions also include additional pieces in $\tilde{K}_{\mu}$ and in $\tilde{D}$ containing the degree of homogeneity $n$. We also note that the components of the intrinsic generator $s_{\muh\nuh}$ appear on the right-hand side. This is the origin of the problem mentioned above --- we expect the generator $P_{\mu}$ of space-time translations to act only differentially, i.e., without any intrinsic piece, on any field, \cf \Eqnref{Pmu}!

We will try to remedy this problem by defining a new field according to
\begin{equation}
\Phi^i(x) = V(x) \tilde{\Phi}^i(x) = \ga^{-n} V(x) \Ups^i(y),
\eqnlab{Phi-Phitilde}
\end{equation}
where the operator $V(x)$ is defined by
\begin{equation}
V(x) \equiv \exp(2 x^\mu s_{+\mu}).
\eqnlab{V(x)}
\end{equation}
Since we are dealing with transformations of \emph{fields} (active transformations), we demand that
\begin{equation}
\de \Phi^i(x) = \ga^{-n} V(x) \de \Ups^i(y)
\eqnlab{dede}
\end{equation}
under a conformal transformation. Specifically, for this relation to be valid for translations, we need
\begin{equation}
P_{\mu} \Phi^i(x) = \ga^{-n} V(x) \tilde{P}_{\mu} \Ups^i(y),
\end{equation}
where $P_{\mu}$ is the generator of translations acting on the field $\Phi^i(x)$. This implies that
\begin{equation}
P_{\mu} = V(x) \tilde{P}_{\mu} V^{-1}(x),
\eqnlab{P-Ptilde}
\end{equation}
where
\begin{equation}
V^{-1}(x) = \exp(-2 x^\mu s_{+\mu}).
\eqnlab{Vbosinv}
\end{equation}
This inverse is well-defined, since all operators $s_{+\mu}$ commute with each other according to \Eqnref{SO(D,2)}. Thus, we find that
\begin{equation}
P_{\mu} \Phi^i(x) =  \pa_{\mu} \Phi^i(x),
\end{equation}
which is the desired action of the translation operator on a field.

We also need to transform the other operators in \Eqnref{MPKD_tilde} according to \Eqnref{P-Ptilde}, to find the generators acting on $\Phi^i(x)$. Using \Eqnref{SO(D,2)}, the result of this calculation is
\begin{equation}
\begin{aligned}
 M_{\mu\nu} &= x_{[\nu} \pa_{\mu]} + s_{\mu\nu} \\
 P_{\mu} &= \pa_{\mu} \\
 K_{\mu} &= x \cdot x \pa_{\mu} - 2 x_{\mu} \left( x \cdot \pa - n + 2 s_{-+} \right) + 4 x^\nu s_{\mu\nu} + 4 s_{-\mu} \\
 D &= x \cdot \pa -n + 2 s_{-+}.
\end{aligned}
\eqnlab{MPKD}
\end{equation}
Comparing this with the action of a conformal transformation on a general field in \Eqnref{mack_transf}, we see that we should identify
\begin{equation}
\begin{aligned}
  \Si_{\mu\nu} &= s_{\mu\nu} \\
  \ka_{\mu} &= 4 s_{-\mu} \\
  \De &= 2 s_{-+} - n,
\end{aligned}
\eqnlab{Siwk}
\end{equation}
where $\Si_{\mu\nu}$, $\De$ and $\ka_{\mu}$ are the generators of the little group that leaves the point $x=0$ invariant. These quantities determine the transformation properties of the field unambiguously, as described in Chapter~\ref{ch:conformal}.

The final step is to project out the unphysical components~\cite{Mack:1969} that may be present since $i$ is an $SO(6,2)$ index. For example, if $i$ is a vector index $\muh$, we get two additional components ($\Phi^{+}$ and $\Phi^{-}$), compared to a vector field in six dimensions. This projection is done by demanding that
\begin{equation}
(s_{-\mu} \Phi)^{i}=0
\eqnlab{physical}
\end{equation}
for all physical values of $i$ and all values of $\mu$. The notation in this equation encodes the action of the intrinsic $SO(6,2)$ generator $s_{-\mu}$ (corresponding to special conformal transformations) on a field. In other words, we demand that all fields in six dimensions should be primary, i.e., have $\ka_{\mu}=0$.

Following the recipe of Mack and Salam~\cite{Mack:1969}, we have in this way recovered the fields in six dimensions and their transformation laws from a manifestly conformally covariant perspective in eight dimensions.

At this point, it is worthwhile to consider a specific example, illustrating these concepts. This will also lead us to an important extension of this formalism. Consider a self-dual three-form field strength $h_{\mu\nu\rho}$, transforming in the ${\bf 10}_+$ representation of the Lorentz group $SO(5,1)$. There is also an associated chiral two-form gauge field $b_{\mu\nu}$, but we will mainly consider the field strength in this discussion. The self-duality condition is
\begin{equation}
\eqnlab{self-dual}
h_{\mu\nu\rho} = \frac{1}{6} \veps_{\mu\nu\rho\si\tau\la} h^{\si\tau\la},
\end{equation}
where $\veps^{\mu\nu\rho\ka\si\tau}$ is the totally antisymmetric invariant tensor, defined such that $\veps^{012345}=1$.

We note that $h_{\mu\nu\rho}$ has ten independent components (not to be confused with the number of physical polarizations, which is three), therefore we expect the corresponding field in the eight-dimensional conformal space to have ten algebraic degrees of freedom. We will see below that the natural field in conformal space corresponding to $h_{\mu\nu\rho}$ has more independent components than ten, meaning that we will have to impose some algebraic conditions on the conformally covariant field.

There are two basic ways of formulating this manifestly covariant field corresponding to $h_{\mu\nu\rho}$. The first option, which is similar to the one Dirac used for the Maxwell field~\cite{Dirac:1936}, is to consider a three-form $H_{\muh\nuh\rhoh}$ in the conformal space. It has 56 components, which means that we have to impose some condition on $H_{\muh\nuh\rhoh}$ to reduce the number of algebraic degrees of freedom. The condition has to be conformally covariant, i.e., it has to be expressed in terms of tensors in the conformal space.

A natural choice is to impose that
\begin{equation}
y^{\muh} H_{\muh\nuh\rhoh} = 0,
\end{equation}
which gives 21 independent conditions on the components in $H_{\muh\nuh\rhoh}$. This is most easily seen in the point where $x^{\mu}=0$, appealing to translational invariance. We may also remove 15 components by considering the arbitrariness caused by the possibility to add a term multiplied by $y^2$ to the associated potential $B_{\muh\nuh}$. Such an addition does not change the value of $B_{\muh\nuh}$ on the hypercone (where $y^2=0$), but still yields non-zero terms in $H_{\muh\nuh\rhoh}$. These correspond to unphysical components that should be projected out, \cf \Eqnref{physical}.

These considerations leave us with 20 components, which are twice as many as we required. This was expected, since we have not taken self-duality into account, which effectively halves the number of degrees of freedom in the field strength. However, this formulation is obviously not very well suited for handling self-dual fields --- the dual of a three-form in eight dimensions is a five-form.

Instead, there is a second formulation~\cite{Siegel:1988,Siegel:1989,Siegel:1994} where we define a homogeneous \emph{four-form} field strength $\Ups_{\muh\nuh\rhoh\sih}$ on the hypercone. This has the obvious advantage that we may impose a self-duality condition similar to \Eqnref{self-dual}.

A priori, a four-form in eight dimensions has 70 components. We may remove 35 of these by the condition
\begin{equation}
y^{\muh} \Ups_{\muh\nuh\rhoh\sih}  =  0.
\eqnlab{ydotH}
\end{equation}
Furthermore, 15 additional conditions are given by
\begin{equation}
y_{[\muh} \Ups_{\nuh\rhoh\sih\tauh]}  =  0,
\eqnlab{yasymH}
\end{equation}
which leave us with 20 independent components in $\Ups_{\muh\nuh\rhoh\sih}$. This number may by halved by imposing the self-duality condition; this will be further discussed below. It should be noted that $\Ups_{\muh\nuh\rhoh\sih}$ is \emph{not} a field strength in the usual sense (it is not the exterior derivative of some potential), which means that we need not take the arbitrariness consideration (due to the hypercone) into account. Therefore, we will not have to deal with unphysical components.

Next, let us see how we may relate the field $\Ups_{\muh\nuh\rhoh\sih}$ in conformal space to the field strength $h_{\mu\nu\rho}$ in Minkowski space. As discussed in the previous paragraph, the number of independent components is the same for both fields (let us postpone the question of self-duality for both fields for a while). This means that we may parametrize the solution to the algebraic equations~\eqnref{ydotH} and \eqnref{yasymH} by a three-form field $h_{\mu\nu\rho}$, which a priori need not be the three-form we are looking for in Minkowski space. We find that the unique solution (apart from normalization factors) is given by
\begin{equation}
\begin{aligned}
\Ups_{\mu\nu\rho\si} &= \ga^n x_{[\mu} h_{\nu\rho\si]} \\
\Ups_{+\nu\rho\si} &= - \frac{1}{4}\ga^n h_{\nu\rho\si} \\
\Ups_{-\nu\rho\si} &= - \frac{1}{8}\ga^n x\cdot x h_{\nu\rho\si} + \frac{3}{4}\ga^n x^\tau x_{[\nu} h_{\rho\si]\tau} \\
\Ups_{+-\rho\si} &= \frac{1}{4}\ga^n x^\tau h_{\rho\si\tau},
\end{aligned}
\eqnlab{H_selfdual_bos}
\end{equation}
where we have included factors of the projective parameter $\ga$ in order to make $h_{\mu\nu\rho}$ a function of only $x^\mu$.

It turns out that \Eqnref{H_selfdual_bos} may be summarized in
\begin{equation}
\Ups_{\muh\nuh\rhoh\sih} = \ga^n V^{-1}(x) \mathcal{H}_{\muh\nuh\rhoh\sih},
\eqnlab{Ups-H_bos}
\end{equation}
where the operator $V^{-1}(x)$ is given by \Eqnref{Vbosinv} and the only non-zero components in $\mathcal{H}_{\muh\nuh\rhoh\sih}$ are $\mathcal{H}_{\mu\nu\rho+}=\frac{1}{4}h_{\mu\nu\rho}$. It should be stressed that, despite its appearance, $\mathcal{H}_{\muh\nuh\rhoh\sih}$ is \emph{not} a covariant tensor, i.e., it does \emph{not} transform linearly under conformal transformations; the conformal generators acting on it are those in \Eqnref{MPKD}. The action of the intrinsic generator $s_{\muh\nuh}$ built into $V(x)$ is given by the standard relation
\begin{equation}
s_{\muh\nuh} \Ups_{\rhoh \sih \tauh \vepsh} = - \eta_{\rhoh[\muh} \Ups_{\nuh]\sih\tauh\vepsh} + \eta_{\sih[\muh} \Ups_{\nuh]\tauh\vepsh\rhoh} - \eta_{\tauh[\muh} \Ups_{\nuh]\vepsh\rhoh\sih} +
\eta_{\vepsh[\muh} \Ups_{\nuh]\rhoh\sih\tauh},
\eqnlab{SonUps}
\end{equation}
when acting on any field with four $SO(6,2)$ vector indices. This is consistent with the commutation relations~\eqnref{SO(D,2)}.

The relation \eqnref{Ups-H_bos} should be compared with \Eqnref{Phi-Phitilde}; it turns out that we have, by imposing two algebraic constraints, recovered the general expression relating a manifestly conformally covariant field to the associated field in the ordinary Minkowski space-time. This tells us that, if we require $\Ups_{\muh\nuh\rhoh\sih}$ to transform linearly (as a regular four-form) in eight dimensions, we recover the standard transformation laws for the three-form $h$ in six dimensions by using \Eqnref{H_selfdual_bos}.

Another important aspect of this formulation is that if we require $\Ups_{\muh\nuh\rhoh\sih}$ to be self-dual in eight dimensions, \Eqnref{H_selfdual_bos} implies that $h_{\mu\nu\rho}$ is self-dual in six dimensions.


\section{Superconformal symmetry}
\label{sec:manifest_superconf}

The purpose of the present section is to generalize the methods of the preceding to the superconformal symmetry algebra. Again, we will restrict ourselves to a six-dimensional space-time with $(2,0)$ supersymmetry.

\subsection{Coordinates}

In the bosonic model, we managed to collect all conformal generators into a single rotation generator $J_{\muh\nuh}$. A similar thing is possible in the superconformal case, but requires the introduction of \emph{superindices}.

Consider a superspace with eight bosonic and four fermionic dimensions, having the natural isometry supergroup $OSp(8^*|4)$. We will call this the \emph{superconformal space}, in analogy with Dirac's notion of a conformal space in the bosonic case. The bosonic subgroup of the supergroup is $SO^*(8) \times USp(4)$, where we recognize $SO^*(8)\simeq SO(6,2)$ as the bosonic conformal group in six dimensions; $SO^*(8)$ denotes the real form of the special orthogonal Lie group $SO(8)$. Furthermore, $USp(4)$ is the $R$-symmetry group, whose Lie algebra is isomorphic to $\mathfrak{so}(5)$.

The fundamental (anti)commutation relations defining the superalgebra corresponding to $OSp(8^*|4)$ are~\cite{Kac:1977,Claus:1998}
\begin{equation}
\eqnlab{superalgebra}
\begin{aligned}
\Big[J_{\sss AB}, J_{\sss CD} \Big\} &= -\frac{1}{2} \Big( I_{\sss BC} J_{\sss AD} - (-1)^{\sss AB} I_{\sss AC} J_{\sss BD} - {} \\
& \quad - (-1)^{\sss CD} I_{\sss BD} J_{\sss AC} +
(-1)^{\sss AB+CD} I_{\sss AD} J_{\sss BC} \Big),
\end{aligned}
\end{equation}
which should be compared with~\Eqnref{SO(D,2)} for the bosonic case. $J_{\sss AB}$ is the generator and the bracket in the left hand side is an anticommutator if both entries in it are fermionic, otherwise it is a commutator. The indices $A$ and $B$ are superindices involving both bosonic and fermionic pieces. They will be further explained and decomposed below. $J_{\sss AB}$ is graded antisymmetric, which is denoted in the standard way by the relation
\begin{equation}
J_{\sss AB} = - (-1)^{\sss AB} J_{\sss BA}.
\end{equation}
This means that $J_{\sss AB}$ is antisymmetric unless both $A$ and $B$ are fermionic, in which case it is symmetric. The superspace metric $I_{\sss AB}$ is graded symmetric and the induced scalar product between vectors is invariant under an $OSp(8^*|4)$ transformation, which can be seen as a definition of the supergroup.

By using the concept of triality in eight dimensions~\cite{Claus:1998}, we regard the superindex $A$ to be composed of $\alh=(1,\ldots,8)$, which is a chiral $SO(6,2)$ spinor index, and $a=(1,\ldots,4)$, which is a fundamental $USp(4)$ index (or equivalently, an $SO(5)$ spinor index) as before. In this notation, $\alh$ is a bosonic index while $a$ is a fermionic index. Furthermore, $\alh$ can naturally be decomposed into one chiral $SO(5,1)$ spinor index $\al=(1,\ldots,4)$ (subscript) and one anti-chiral $SO(5,1)$ spinor index $\al=(1,\ldots,4)$ (superscript) in agreement with the notation developed in Section~\ref{sec:susy}. Using this decomposition, we may make contact with the generators used in Section~\ref{sec:superconformal}. We find that the definitions
\begin{equation}
  J_{\sss AB} =
  \left( \begin{array}{ccc}
  \frac{1}{2} P_{\al\be} &  \frac{1}{2} \dia{M}{\al}{\be} + \frac{1}{4} \kde{\al}{\be} D & \frac{i}{2\sqrt{2}} Q^b_\al \\
  - \frac{1}{2} \dia{M}{\be}{\al} - \frac{1}{4} \kde{\be}{\al} D & - \frac{1}{2} K^{\al\be} & \frac{i}{2\sqrt{2}} \Om^{bc} S^\al_c \\
  - \frac{i}{2\sqrt{2}} Q^a_\be & - \frac{i}{2\sqrt{2}} \Om^{ac} S^\be_c & i U^{ab}
  \end{array} \right)
  \eqnlab{J_AB}
\end{equation}
and
\begin{equation}
  I_{\sss AB} =
  \left( \begin{array}{ccc}
  0 &  \kde{\al}{\be} & 0 \\
  \de^\al_{\ph{\al}\be} & 0 & 0 \\
  0 & 0 & i \Om^{ab}
  \end{array} \right),
  \eqnlab{superspacemetric}
\end{equation}
together with \Eqnref{superalgebra}, reproduce all commutation relations of the superconformal algebra in six dimensions, as given in \Eqnref{super_comm}. This indicates that the superconformal group found in Section~\ref{sec:superconformal} indeed is isomorphic to $OSp(8^*|4)$, as claimed above.

It may also be seen from this formulation that we cannot have central charges like the ones in \Eqnref{QQ} in a superconformal theory. They are simply not compatible with the superalgebra and may not be introduced in a covariant way, cf.~Refs.~\cite{Bedding:1984,Bedding:1985,Meessen:2003,Peeters:2003}. This is most easily seen from the super-Jacobi identity: Consider the anti-commutator of the two fermionic generators $J_{\alh}^{\ph{\alh}b}$ and $J_{\gah}^{\ph{\gah}d}$, which contain both supersymmetry and special supersymmetry as components. This anti-commutator involves the bosonic conformal generator $J_{\alh \gah}$ and the $R$-symmetry generator $U^{bd}$, but should also include the central charges if such were present. However, since the central charges cannot commute with $J_{\alh \gah}$ or $U^{bd}$, the super-Jacobi identity~\eqnref{superjacobi} cannot be fulfilled.

Apart from the metric given in \Eqnref{superspacemetric}, it is convenient to define an inverse superspace metric, i.e., a metric with superscript indices. This becomes
\begin{equation}
  I^{\sss AB} =
  \left( \begin{array}{ccc}
  0 &  \de^\al_{\ph{\al}\be} & 0 \\
  \kde{\al}{\be} & 0 & 0 \\
  0 & 0 & -i \Om_{ab}
  \end{array} \right),
  \eqnlab{superspacemetric_upper}
\end{equation}
which makes the relation
\begin{equation}
I_{\sss AB} I^{\sss BC} = \de^{\sss C}_{\sss A}
\end{equation}
valid (which is essential for consistency reasons if we want to raise and lower indices from the left).

The next step is to introduce coordinates in the superconformal space and relate them to the ones used in the ordinary $(2,0)$ superspace. The most common choice would be to choose the vector representation of $SO(6,2)$ for the bosonic coordinates. However, we will instead use another eight-dimensional representation, the chiral spinor, for these coordinates. This is motivated by our convention to denote the $SO(6,2)$ generator by a bispinor $J_{\alh\beh}$. The fermionic coordinates are in the fundamental four-dimensional representation of $USp(4)$, as expected.

Thus, the coordinates in the superconformal space are denoted by $y_{\sss A}=(y_{\alh},y^a)=(y_\al,y^\al,y^a)$. It should be emphasized that $y_{\alh}$ is a commuting quantity (Grassmann even), while $y^a$ is anti-commuting (Grassmann odd). We also introduce a derivative $\pa_{\sss A}$ such that
\begin{equation}
\pa_{\sss A} y_{\sss B} = I_{\sss AB}.
\end{equation}
The use of a chiral spinor instead of the usual coordinate vector introduces a subtlety concerning reality. There is no Majorana-Weyl spinor in eight dimensions with signature $(6,2)$, which means that the components of $y_{\alh}$ cannot all be real~\cite{Kugo:1983}. The standard way of introducing a reality condition is instead to attach a fundamental $SU(2)$ index $i=(1,2)$ to $y_{\alh}$ and impose a symplectic $SU(2)$ Majorana condition. This additional $R$-symmetry can be motivated from the quaternionic structure of the conformal group $SO(6,2) \simeq SO(4;\mathbb{H})$. However, we will treat $y_{\alh}$ as an ordinary complex chiral spinor and not impose any reality condition.

The next step is to introduce a projective \emph{supercone}, inspired by \Eqnref{hypercone}. The corresponding expression in the superconformal space should be superconformally invariant and the most natural such definition of a projective supercone is given by
\begin{equation}
\eqnlab{supercone}
y^2 \equiv I^{\sss AB} y_{\sss A} y_{\sss B} = 0.
\end{equation}
In the bosonic case, we introduced coordinates on the hypercone in a straight-forward manner in \Eqnref{y-x}. In the present superconformal case this is hard to do. Instead, we will introduce the ordinary superspace coordinates in a more implicit way, as fields relating the different components of the coordinate vector $y_{\sss A}$ on the supercone.

Consider a point on the supercone, with coordinates $y_{\sss A}$. One solution to the condition \eqnref{supercone} may be found by introducing a fermionic field $\theta_a^\al(y)$ such that for any point on the supercone,
\begin{equation}
\eqnlab{y^a}
y^a = \sqrt{2} \Om^{ab} \theta_b^\be y_\be,
\end{equation}
where the factor $\sqrt{2}$ is purely conventional. By requiring $\theta_a^\al$ to transform as an anti-Weyl spinor under $SO(5,1)$,
this field is well-defined in all points on the supercone. In the same manner, we may introduce the bosonic field $x^{\al\be}(y)=-x^{\be\al}(y)$ such that
\begin{equation}
\eqnlab{y^al}
y^\al = \left( 2 x^{\al\be} - i \Om^{ab} \theta^\al_a \theta^\be_b \right) y_\be
\end{equation}
for any point on the supercone.

It is easily verified that all points $y_{\sss A}$ of this form lie on the supercone defined by \Eqnref{supercone}. Due to the projectiveness, we may always multiply the coordinate $y_{\sss A}$ by a constant and still remain on the supercone; this feature is obvious in \Eqsref{y^a} and \eqnref{y^al} as well. Note that the projective parameter $\ga$ introduced in \Eqnref{y-x} is absent in the equations, but still has a natural place in the formalism.

We should also mention that the relations~\eqnref{y^a} and \eqnref{y^al} look just like the super-twistor relations introduced by Ferber~\cite{Ferber:1978}. However, twistors~\cite{Penrose:1967,Penrose:1984,Penrose:1986} are conventionally used to describe the phase-space of on-shell particles while our aim is to consider space-time itself, without conjugate momenta. Twistors in six dimensions have been used in complex notation in Ref.~\cite{Hughston:1987} and in quaternionic notation in Ref.~\cite{Bengtsson:1988}. Our usage of twistors is rather similar to Witten's in the context of MHV amplitudes~\cite{Witten:2003,Sinkovics:2004}. We regard the coordinate $y_{\sss A}$ as a homogeneous coordinate in a projective super-twistor space. The latter space is a copy of the supermanifold $\mathbb{CP}^{7|4}$. We cannot take the coordinates to be real, but we can do the next best thing: we will only consider functions that depend on $y_{\alh}$ and $y^a$, not on their complex conjugates.

Returning to the twistor relations~\eqnref{y^a} and \eqnref{y^al}, it remains to be shown that the quantities $x^{\al\be}$ and $\theta_a^\al$ may be identified with the coordinates of the $(2,0)$ superspace, with six bosonic and sixteen fermionic dimensions. As in the bosonic case, we show that $x^{\al\be}$ and $\theta_a^\al$ transform as we expect the corresponding coordinates to do.

A superconformal transformation of the coordinates $y_{\sss A}$ is naturally generated by the differential operator
\begin{equation}
L_{\sss AB} = - y_{\sss [A} \pa_{\sss B]} = (-1)^{\sss AB} y_{\sss [B} \pa_{\sss A]},
\eqnlab{L_AB}
\end{equation}
which, as is easily verified, satisfies the commutation relations
\eqnref{superalgebra}. This generator acts on the coordinates according to
\begin{equation}
\de y_{\sss A} = \pi^{\sss CD} L_{\sss CD} y_{\sss A} = - \pi^{\sss CD} y_{\sss C} I_{\sss DA} = (-1)^{\sss C(A+D)} I_{\sss AC} \pi^{\sss CD} y_{\sss D},
\end{equation}
where the matrix $\pi^{\sss AB}$ contains the parameters of the transformation. Again, to make contact with the underlying six-dimensional theory, we demand that
\begin{equation}
\pi^{\sss AB} J_{\sss AB} = \dia{\om}{\al}{\be} \dia{M}{\be}{\al} + a^{\al\be} P_{\al\be}
+ c_{\al\be} K^{\al\be} + \la D + \eta^\al_a Q^a_\al + \rho^a_\al S_a^\al + v_{ab} U^{ab},
\eqnlab{pidotJ}
\end{equation}
where $J_{\sss AB}$ is decomposed in \Eqnref{J_AB}. This implies that the parameter matrix $\pi^{\sss AB}$ must be given by
\begin{equation}
  \pi^{\sss AB} =
  \left( \begin{array}{ccc}
  2 a^{\al\be} & \dia{\om}{\be}{\al} + \frac{1}{2} \la \kde{\be}{\al} &
    -i \sqrt{2} \eta^\al_b \\
  - \dia{\om}{\al}{\be} - \frac{1}{2} \la \kde{\al}{\be} &
    -2 c_{\al\be} & -i \sqrt{2} \rho^c_\al \Om_{cb} \\
  i \sqrt{2} \eta^\be_a & i \sqrt{2} \rho^c_\be \Om_{ca} & -i v_{ab}
  \end{array} \right),
\end{equation}
in terms of the parameters for the transformations of the superconformal group, as defined in Section~\ref{sec:superconformal}.

Since \Eqnref{supercone} is invariant under a transformation of this type, we may require the left-hand and the right-hand sides of \Eqsref{y^a} and \eqnref{y^al} to transform equally.
The implicated transformation properties of the fields $x^{\al\be}(y)$ and $\theta_a^\al(y)$ when the $y$-coordinates are transformed in this way are
\begin{align}
\begin{split}
\de x^{\al \be} &= a^{\al \be} - \dia{\om}{\ga}{[\al} x^{\be] \ga}
+ \la x^{\al \be} + 4 c_{\ga \de} x^{\ga\al} x^{\be\de} - i \Om^{ab} \eta^{[\al}_{a\ph{b}} \theta^{\be]}_b - {} \\
& \quad - c_{\ga \de} \theta^\ga \cdot \theta^{[\al}
\theta^{\be]} \cdot \theta^{\de} -i \rho^c_{\ga} \theta^{[\al}_c
    \left( 2 x^{\be]\ga} - i \theta^{\be]} \cdot \theta^{\ga} \right)
\end{split} \\
\begin{split}
\de \theta^{\al}_a &= (\dia{\om}{\ga}{\al} - 4 c_{\ga \de} x^{\al \de}
- 2i c_{\ga \de} \theta^{\al} \cdot \theta^{\de} + 2i
\rho^c_{\ga} \theta^{\al}_c) \theta^{\ga}_a + \frac{1}{2} \la \theta^{\al}_a + {} \\
& \quad + \eta^{\al}_a - \Om_{ac} \rho^c_{\ga} \left( 2 x^{\ga \al} - i \theta^{\ga} \cdot \theta^{\al} \right) + v_{ac} \Om^{cd} \theta^{\al}_d,
\end{split}
\end{align}
and agree exactly with the superconformal coordinate transformations as given in \Eqsref{transf_superx} and \eqnref{transf_superth}. This explains the choice of notation, and implies that the rather complicated transformation laws for $x^{\al\be}$ and $\theta_a^\al$ are a mere consequence of a simple rotation, in a superspace with eight bosonic and four fermionic dimensions!

\subsection{Fields}

In Section~\ref{sec:manifest_conformal}, we found a manifestly conformally covariant formulation of space-time fields. The purpose of this section is to generalize this formalism to superspace, in order to formulate theories with manifest superconformal invariance.

Consider a field $\Ups^i(y)$, defined on the supercone~\eqnref{supercone}. In this case, the superscript $i$ denotes a set of indices corresponding to a representation of the superconformal group $OSp(8^*|4)$. Inspired by \Eqnref{homogeneity_general}, we require the field to be a homogeneous function of the coordinates in the superconformal space, such that
\begin{equation}
\frac{1}{2} I^{\sss CD} y_{\sss C} \pa_{\sss D} \Ups^i(y) = n \Ups^i(y).
\eqnlab{homo_Ups}
\end{equation}
The factor of $\frac{1}{2}$ on the left-hand side of this equation is purely conventional and introduced to make the degree of homogeneity agree with the bosonic model.

Again, this allows us to remove the dependence on the projective parameter $\ga$. We define a new superfield according to
\begin{equation}
\tilde{\Phi}^i(x,\theta) = \ga^{-n} \Ups^i(y),
\end{equation}
which should depend only on the ordinary superspace coordinates $x^{\al\be}$ and $\theta^\al_a$.

The next step is to find the action of a general superconformal transformation on these fields. The generator is divided into two parts according to
\begin{equation}
\tilde{J}_{\sss AB} = L_{\sss AB} + s_{\sss AB},
\end{equation}
where the tilde on $\tilde{J}_{\sss AB}$ indicates that the generator is supposed to act on the fields $\tilde{\Phi}^i(x,\theta)$ or $\Ups^i(y)$, \cf \Eqnref{split-operator}. In the same way as in the bosonic case, $s_{\sss AB}$ is the intrinsic piece that acts only on the indices of the field in question. Its explicit form depends on the actual representation of the superconformal group, and it satisfies \Eqnref{superalgebra}.

We may now decompose $\tilde{J}_{\sss AB}$ according to \Eqnref{J_AB} and translate to the variables $x^{\al\be}$ and $\theta^\al_a$ and the corresponding derivatives. This process is slightly more complicated than in the bosonic case, due to our implicit description of the supercone. However, guided by previous experience and the results from the bosonic case we find that
\begin{equation}
\begin{aligned}
 \dia{\tilde{M}}{\al}{\be} &= 2 x^{\be\ga} \pa_{\al\ga} + \theta^\be_c \pa^c_\al - \frac{1}{2} \kde{\al}{\be} \left(x \cdot \pa + \frac{1}{2} \theta \cdot \pa \right)  + {}  \\
& \quad + 2 \dia{s}{\al}{\be} - \frac{1}{2} \kde{\al}{\be} \dia{s}{\ga}{\ga} \\
 \tilde{P}_{\al\be} &= \pa_{\al\be} + 2 s_{\al\be}\\[2mm]
 \tilde{K}^{\al\be} &= -4 x^{\al\ga} x^{\be\de} \pa_{\ga\de} - \theta^\al \cdot \theta^\ga \theta^\be \cdot \theta^\de \pa_{\ga\de} + {} \\
& \quad + 2 \theta^{[\al}_c \left( 2 x^{\be]\ga} - i \theta^{\be]} \cdot \theta^\ga \right) \pa^c_\ga + 2n x^{\al\be} - 2 s^{\al\be} \\
 \tilde{D} &= x \cdot \pa + \frac{1}{2} \theta \cdot \pa - n + \dia{s}{\ga}{\ga} \\
\tilde{Q}^a_\al &= \pa^a_\al - i\Om^{ac}\theta^\ga_c \pa_{\al\ga} - i 2 \sqrt{2} \dia{s}{\al}{a} \\
\tilde{S}_a^\al &= \Om_{ac} \left( 2 x^{\al\ga} - i \theta^\al \cdot \theta^\ga \right) \pa^c_\ga - i \theta^\ga_a \left( 2 x^{\de\al} - i \theta^\de \cdot \theta^\al \right) \pa_{\ga\de} + {} \\
& \quad + 2i \theta^\ga_a \theta^\al_c \pa^c_\ga - i n \theta^\al_a - i 2 \sqrt{2} \Om_{ab} s^{\al b} \\
\tilde{U}^{ab} &= \frac{1}{2} \left( \Om^{ac} \theta^\ga_c \pa^b_\ga + \Om^{bc} \theta^\ga_c \pa^a_\ga \right) -i s^{ab},
\end{aligned}
\eqnlab{tildegenerators}
\end{equation}
where we have used~\Eqnref{homo_Ups} to introduce the degree of homogeneity $n$ in the expressions. The differential pieces agree with \Eqnref{transf_superfield}, and we note an unwanted intrinsic $s$-piece in $P_{\al\be}$ as in the bosonic case. However, in a supersymmetric theory, we expect also the generator $Q^a_\al$ of supersymmetry transformations (which really are superspace translations) to act only differentially on fields, \cf \Eqnref{Qaction}.

In the bosonic case, we remedied this problem by multiplying the functions by an operator $V(x)$ defined in \Eqnref{V(x)}. In the superconformal case, we generalize this as
\begin{equation}
\Phi^i(x,\theta) = V(x,\theta) \tilde{\Phi}^i(x,\theta) = \ga^{-n} V(x,\theta) \Ups^i(y),
\eqnlab{manifest_Phi}
\end{equation}
where the operator $V(x,\theta)$ is given by
\begin{equation}
V(x,\theta) \equiv \exp\left(2 x^{\ga\de} s_{\ga\de} - 2 \sqrt{2} i \theta^\ga_d \dia{s}{\ga}{d} \right).
\end{equation}
This forces us to transform the generators according to
\begin{equation}
J_{\sss AB} = V(x,\theta) \tilde{J}_{\sss AB} V^{-1}(x,\theta),
\end{equation}
\cf \Eqnref{P-Ptilde}. The inverse $V^{-1}(x,\theta)$ is well-defined since the commutators $[x\cdot s,x\cdot s]$, $[x\cdot s, \theta \cdot s]$ and $[\theta \cdot s,\theta \cdot s]$ all are zero.

Using the identity~\eqnref{Haussdorff} and the commutation relations~\eqnref{superalgebra}, we find the expressions
\begin{equation}
\begin{aligned}
 \dia{M}{\al}{\be} &= 2 x^{\be\ga} \pa_{\al\ga} + \theta^\be_c \pa^c_\al - \frac{1}{2} \kde{\al}{\be} \left(x \cdot \pa + \frac{1}{2} \theta \cdot \pa \right)  + {}  \\
& \quad + 2 \dia{s}{\al}{\be} - \frac{1}{2} \kde{\al}{\be} \dia{s}{\ga}{\ga} \\
 P_{\al\be} &= \pa_{\al\be} \\
 K^{\al\be} &= -4 x^{\al\ga} x^{\be\de} \pa_{\ga\de} - \theta^\al \cdot \theta^\ga \theta^\be \cdot \theta^\de \pa_{\ga\de} + {} \\
& \quad + 2 \theta^{[\al}_c \left( 2 x^{\be]\ga} - i \theta^{\be]} \cdot \theta^\ga \right) \pa^c_\ga - 2 s^{\al\be} + 2n x^{\al\be} -2 x^{\al\be} \dia{s}{\ga}{\ga} - {} \\
& \quad - \left( 2x^{\al\ga} -i \theta^\al \cdot \theta^\ga \right) \left( 2 \dia{s}{\ga}{\be} - \frac{1}{2} \kde{\ga}{\be} \dia{s}{\de}{\de} \right) - 4 \sqrt{2} i \theta_c^{[\al} s^{\be]c} + {} \\
& \quad + \left( 2x^{\be\ga} -i \theta^\be \cdot \theta^\ga \right) \left(2 \dia{s}{\ga}{\al} - \frac{1}{2} \kde{\ga}{\al} \dia{s}{\de}{\de} \right) - 4 \theta^\al_c \theta^\be_d s^{cd} \\[2mm]
 D &= x \cdot \pa + \frac{1}{2} \theta \cdot \pa - n + \dia{s}{\ga}{\ga}
\end{aligned}
\end{equation}
for the bosonic conformal generators,
\begin{equation}
\begin{aligned}
Q^a_\al &= \pa^a_\al - i\Om^{ac}\theta^\ga_c \pa_{\al\ga} \\
S_a^\al &= \Om_{ac} \left( 2 x^{\al\ga} - i \theta^\al \cdot \theta^\ga \right) \pa^c_\ga - i \theta^\ga_a \left( 2 x^{\de\al} - i \theta^\de \cdot \theta^\al \right) \pa_{\ga\de} + {} \\
& \quad + 2i \theta^\ga_a \theta^\al_c \pa^c_\ga - i n \theta^\al_a - i 2 \sqrt{2} \Om_{ab} s^{\al b} + i \theta^\al_a \dia{s}{\ga}{\ga} - {} \\
& \quad - 4 \Om_{ab} \theta^\al_c s^{bc} + 2i \theta^\ga_a \left( 2 \dia{s}{\ga}{\al} - \frac{1}{2} \kde{\ga}{\al} \dia{s}{\de}{\de} \right)
\end{aligned}
\end{equation}
for the fermionic generators, and
\begin{equation}
\begin{aligned}
U^{ab} &= \frac{1}{2} \left( \Om^{ac} \theta^\ga_c \pa^b_\ga + \Om^{bc} \theta^\ga_c \pa^a_\ga \right) -i s^{ab}
\end{aligned}
\end{equation}
for the $R$-symmetry generator. It should be emphasized that these are the generators that act on the ordinary superfield $\Phi^i(x,\theta)$, not on the covariant superfield $\Ups^i(y)$.

The generators look very complicated in this notation, but if we insert the expressions in \Eqnref{pidotJ}, we find that
\begin{equation}
\eqnlab{manifest_piJ}
\begin{split}
\pi \cdot J &= \xi^{\al\be}(x,\theta) \pa_{\al\be} + \xi^\al_a(x,\theta) \pa^a_{\al} + \dia{\Om}{\al}{\be}(x,\theta) \left(2\dia{s}{\be}{\al} - \frac{1}{2} \kde{\be}{\al} \dia{s}{\ga}{\ga} \right) + {} \\
& \quad + \La(x,\theta) \left( \dia{s}{\ga}{\ga} - n\right) + V_{ab}(\theta) (-i s^{ab}) + {} \\
& \quad + R^a_{\al}(\theta) ( - i 2\sqrt{2} \Om_{ab} s^{\al b}) - 2 c_{\al\be} s^{\al\be},
\end{split}
\end{equation}
where the coordinate-dependent parameter functions have been defined in \Eqsref{xi_bos}--\eqnref{Raal}.

This result should be compared with \Eqnref{transf_func}, which tells us that if we identify
\begin{equation}
\eqnlab{Si_s}
\begin{aligned}
\dia{\Si}{\al}{\be} &= 2\dia{s}{\al}{\be} - \frac{1}{2} \kde{\al}{\be} \dia{s}{\ga}{\ga} &\quad  u^{ab} &= -i s^{ab} \\
\De &= \dia{s}{\ga}{\ga} - n & \si_a^\al &= - i 2\sqrt{2} \Om_{ab} s^{\al b} \\
\ka^{\al\be} &= -2s^{\al\be},
\end{aligned}
\end{equation}
then the two different methods of deriving a general superconformal field transformation give the same result. \Eqnref{Si_s} is also consistent with the little group commutation relations.

This derivation ends the chapter on manifest symmetries, but also this part of the thesis. In the following, we will focus on a specific theory --- the six-dimensional $(2,0)$ theory and its properties. We will use the methods derived in this part to obtain a better understanding of the symmetries of that particular theory.

\part{$(2,0)$ theory}

\chapter{The origin of $(2,0)$ theory}
\label{ch:origin}

In his contribution to the proceedings of the conference \emph{Strings '95}~\cite{Witten:1995}, E.~Witten investigated what happens to the Type IIB string theory under the circumstances when the Type IIA theory develops an enhanced non-abelian gauge symmetry. This lead to the discovery of a new superconformal six-dimensional string theory without dynamical gravity. The theory has been named $(2,0)$ theory after the supersymmetry algebra under which it is invariant~\cite{Nahm:1978}.

The present chapter is devoted to the study of the different origins
of the $(2,0)$ theory in terms of higher-dimensional theories
(Sections~\ref{sec:TypeIIB} and~\ref{sec:M-theory}), but
also includes motivations coming from a lower-dimensional
perspective in Section~\ref{sec:lowerdim}. We explore the different degrees of
freedom of the theory in Section~\ref{sec:dof} and describe our approach to the theory in Section~\ref{sec:approach}.

\section{The Type IIB perspective}
\label{sec:TypeIIB}

The web of dualities connecting the different string theories to each
other has been a natural part of the research in string theory since
its appearance in the mid 90's. For example, it is well known that
the heterotic string on the space $\rsix \times \tfour$ is equivalent
to a Type IIA string on $\rsix \times \kthree$, see, e.g.,
Ref.~\cite{Witten:1995sd}. The space \kthree{} appearing in the latter
compactification is a four-dimensional hyper-K\"ahler
surface\footnote{See Ref.~\cite{Aspinwall:1996} for a review of \kthree{}
  surfaces in string theory.}, and it is the second simplest compact
Ricci-flat manifold after the torus. The duality implies that the
gauge symmetry of the Type IIA theory is extended (corresponding to
the appearance of extra massless particles) at certain points in the
moduli space of the \kthree. These points are those where a set of two-spheres embedded in the \kthree{} collapse to a singular point~\cite{Witten:1995sd}, and
will therefore be denoted as singular points or singularities of the
moduli space. These singularities obey an $ADE$
classification, i.e., there are two infinite series, $A_r$,
$r=1,2,\ldots$, and $D_r$, $r=3,4,\ldots$, and three exceptional cases
$E_6$, $E_7$ and $E_8$. The subscript denotes the \emph{rank} of the
singular point; the notation is in line with the associated simply
laced Lie algebras.

It is also well known that the Type IIA string theory is dual to the Type
IIB theory, in the sense that they are equivalent when compactified on
circles. These considerations seem to indicate that something should
happen when the Type IIB theory is compactified on a \kthree-manifold;
something that can account for the extra massless particles that
should appear upon compactification on a circle.

Before investigating what actually happens, let us first note that the Type IIB
theory \emph{cannot} develop an enhanced gauge symmetry as the Type IIA
theory did. This is due to the chiral supersymmetry of Type IIB theory
on $\rsix \times \kthree$ which does not admit gauge vector
multiplets~\cite{Witten:1995}.

Let us now focus on the $A_1$ case, where the \kthree{} contains a
single two-sphere which collapses to a singular point as we approach a
certain point in the moduli space of the \kthree. Close to this point,
one can see that the Type IIA theory on $\rsix \times \kthree$
involves $W$ bosons with a certain mass, corresponding to the enhanced
gauge symmetry discussed above. Following Ref.~\cite{Witten:1995},
we note that this mass is unchanged
upon compactification on a circle, after which we end up with $W$
bosons on $\rfive \times \sone \times \kthree$ with a
mass proportional to the distance to the singular point in
the moduli space and inversely proportional to the ten-dimensional
Type IIA string coupling constant. On the Type IIB side (still on
$\rfive \times \sone \times \kthree$), this mass is proportional to
the circumference of the compactification circle and the distance to
the singular point in moduli space, while it is inversely proportional
to the Type IIB string coupling constant. This corresponds naturally
to a \emph{string} wound around the circle \sone{} and indicates that
the Type IIB theory on $\rsix \times \kthree$ involves a string-like
object. The tension of this
string vanishes as we approach the singular point in the moduli space
of the \kthree, where the gauge symmetry gets extended for the Type
IIA theory. The string tension is also inversely proportional to the
Type IIB string coupling constant.

This string, effectively living in six dimensions, is definitely not
the fundamental Type IIB string, whose tension in fact is
\emph{independent} of the Type IIB string coupling constant. Instead, it
corresponds to a self-dual (in the sense that it couples to a
four-form potential with a self-dual five-form field strength, \cf Ref.~\cite{Duff:1991}) Type IIB D3-brane wrapping the two-sphere in the
\kthree. The string tension is proportional to the area of the sphere
times the tension of the D3-brane, which in turn is proportional to
the inverse of the string coupling constant. As we approach the
singular point in the moduli space, the area of the two-sphere vanishes,
and hence the string in question becomes tensionless as was argued in
the previous paragraph. The string we obtain in this way is similar to
the self-dual string first described in Ref.~\cite{Duff:1994}.

The argument given above for the $A_1$ version can be extended to all
kinds of \kthree{} surfaces. This means that there is a string
theory of this kind for each
possible type of isolated singularity ($A$, $D$ or $E$) of the
\kthree. Apart from the choice of singularity, there are no discrete
or continuous parameters in the theory. However, the approach to the
singular point can be described by $4r$ real moduli, where $r$ is the
rank of the singularity. The Ramond-Ramond two-form of Type IIB
supergravity also contributes with $r$ moduli, meaning that we in
total have $5r$ real moduli parameters. The origin of \emph{this}
moduli space (which is the moduli space of the new six-dimensional
string theory) corresponds to the singular point in the moduli space
of the \kthree.

Close to the singular point in the moduli space, the string becomes very
light (meaning that its tension is much below the string and Planck
scales) and cannot influence gravity. This means that the string is
effectively propagating in a \emph{flat} six-dimensional
space-time, decoupled from the bulk degrees of freedom. We call such a string \emph{non-critical}. The existence
of such a theory is somewhat surprising; it was commonly believed that
any quantum theory in more than four dimensions must contain
gravity.

Furthermore, the string is self-dual, due to the
self-duality of the two-spheres (the cohomology class of their Poincar\'e duals is self-dual) and of the D3-brane, in the sense that it couples to a two-form potential with a self-dual three-form field strength; this means that its electric and magnetic charges are
equal. Due to Dirac quantization effects~\cite{Dirac:1931,Nepomechie:1985,Teitelboim:1986,Deser:1998}, these charges cannot be taken to be small; the electromagnetic coupling is therefore of
order unity. These dyonic properties depend heavily on the space-time dimensionality~\cite{Deser:1998}.

\section{The $M$-theory perspective}
\label{sec:M-theory}

The $A$-series of $(2,0)$ theory also has an interpretation in terms
of $M$-theory, the mysterious eleven-dimensional theory that is
supposed to unite the five string theories.
Not much is known about $M$-theory, but it is fairly clear that it has
eleven-dimensional supergravity as its low energy limit and that it
contains a two-dimensional membrane and a five-dimensional brane,
called the $M2$-brane and the $M5$-brane, respectively. These are both
BPS states preserving half of the 32 supersymmetries.
In this interpretation, the $A_r$ series of $(2,0)$
theory is regarded as the world-volume theory on $r+1$ parallel
$M5$-branes; the equivalence of this picture and the one described in
the previous section was shown in Refs.~\cite{Witten:1996o,Dasgupta:1996}; see also Ref.~\cite{Sen:1997}. This gives a natural interpretation of the $5r$ moduli parameters as the relative transverse positions of the $M5$-branes in the eleven-dimensional bulk space. Hence, they can be thought of as vacuum expectation values of Goldstone bosons, originating from the spontaneous symmetry breaking of translational invariance induced by the branes.

The $M$-theory membranes ($M2$-branes) may end on the $M5$-branes, in
the same way as strings end on D-branes in ten-dimensional string
theory. In our model, we may have $M2$-branes that stretch between the
parallel
$M5$-branes~\cite{Strominger:1996,Townsend:1996,Becker:1996}. From the
world-volume perspective on the $M5$-brane, these are perceived as
\emph{strings}. There are two basic possibilities for this: the
membrane may be infinitely extended in the direction parallel to the
$M5$-brane, in which case the string in the six-dimensional theory
will be infinitely long and approximately straight, but the membrane
may also have the shape of a cylinder connecting the $M5$-branes,
which then is seen as a closed string from the six-dimensional point
of view. The string tension is given by the tension of the
$M2$-brane multiplied by the distance between the two $M5$-branes in
question. This indicates a connection between the moduli parameters
and the string tension: at the origin of moduli space, the $M5$-branes
coincide and the strings become tensionless, in agreement with the Type IIB
picture described in Section~\ref{sec:TypeIIB}.

In the same way as the strings were found to be non-critical from the
Type IIB perspective, it can be argued that the strings in the
$M$-theory perspective decouple from gravity when the distance between
the $M5$-branes is small compared to the eleven-dimensional Planck
length. Furthermore, since the gravitons in the eleven-dimensional
theory live in the bulk space, we have no gravitons propagating in the
world-volume of the $M5$-brane~\cite{Witten:1996o}. This again leaves
us with a theory of self-dual strings propagating in a flat
six-dimensional space-time.

It is interesting to note that also intersections between two $M5$-branes are possible; they
are perceived as three-dimensional extended objects (three-branes) from the six-dimensional
perspective~\cite{Papadopoulos:1996}. Their existence was suggested in Section~\ref{sec:susy}, motivated by the presence of an associated central charge in the supersymmetry algebra in \Eqnref{QQ}. There is even an attempt to construct a ''theory of everything'' living on these three-branes~\cite{Smilga:2005}. However, we will not consider these rather exotic objects any further in this thesis.

Similarly, the $D$-series of $(2,0)$ theory may possibly be described
from $M$-theory by including a parallel orientifold
plane. The $E_6$, $E_7$ and $E_8$ versions, however, have no known
$M$-theory realization.

As a final remark in this section, we note that there is
another motivation for the study of $(2,0)$ theory in
terms of the so called $\mathrm{AdS}/\mathrm{CFT}$ correspondence. In
his famous paper on this conjecture~\cite{Maldacena:1998},
J.~Maldacena constructs a duality between $M$-theory on the space
$\mathrm{AdS}_7 \times S^4$ and a superconformal theory with
$(2,0)$ supersymmetry in six dimensions.

\section{Degrees of freedom}
\label{sec:dof}

The $(2,0)$ theory involves a certain number of degrees of freedom. We
will start by analyzing these from the $M$-theory perspective
described in the previous section, thereby focusing on the
$A$-series.

It was suggested in Ref.~\cite{Gibbons:1993} and shown
in Refs.~\cite{Kaplan:1996,Adawi:1998} that the dynamics of a single
five-brane in eleven-dimensional supergravity may be described by a so
called $(2,0)$ tensor multiplet living in the world-volume of the
brane. This multiplet involves five scalar
fields, corresponding to the Goldstone bosons associated with the
spontaneous symmetry breaking of translational invariance induced by
the presence of the brane. These (Lorentz) scalars
transform naturally in the vector representation of $SO(5)$, which is
the $R$-symmetry group of $(2,0)$ supersymmetry \cite{Nahm:1978}. The
vacuum expectation values of these scalars are identified with the
moduli of the theory.

Furthermore, the presence of the brane breaks 16 of the 32
supersymmetries of
$M$-theory, giving rise to 16 Goldstone fermions (corresponding to 8
fermionic degrees of freedom). These are chiral Lorentz spinors and
transform in the spinor representation of $SO(5)$, or equivalently,
in the fundamental representation of $USp(4)$. They
obey a symplectic Majorana
condition~\cite{Kugo:1983} which originates in the Majorana reality
condition imposed on spinors in eleven dimensions.
The fact that the spinors in six dimensions are chiral is a
consequence of the BPS property of the configuration; the right hand
side of the $M$-theory supersymmetry algebra, including a central
charge corresponding to the $M5$-brane, is proportional to a
projection operator onto chiral spinors in six dimensions if the
BPS limit is saturated. This means that the Goldstone spinors, coming
from the broken supersymmetries, must be chiral.

Summing up the bosonic and fermionic degrees of freedom obtained this
far, we see that we lack three bosonic degrees of freedom for the
model to have any chance of being supersymmetric. These are contained
in a chiral two-form potential~\cite{Kaplan:1996,Adawi:1998}, which is related to the breaking of the gauge symmetry of the three-form potential in eleven-dimensional
supergravity. Here chiral means that it has a self-dual three-form
field strength. Both the two-form potential and its field strength are
$R$-symmetry scalars.

In this way, we have found a tensor multiplet consisting of eight bosonic and
eight fermionic degrees of freedom. This agrees with the matter
representation of the $(2,0)$ supersymmetry algebra. The result that
the dynamics of the $M5$-brane is described by a $(2,0)$ tensor
multiplet was also found in Ref.~\cite{Howe:1997sb} using an embedding
approach in superspace. Furthermore, it is known that $M$-theory
compactified on a circle becomes the Type IIA string theory and
indeed, it has been shown~\cite{Callan:1991} that the dynamics of an
NS5-brane in the Type IIA theory is also described by the $(2,0)$
tensor multiplet, see also Ref.~\cite{Aharony:1999}. This observation provides yet another
higher-dimensional interpretation of the tensor multiplet field theory.

Returning to $M$-theory, we may add more $M5$-branes parallel to the
first one, thereby obtaining the
$A_r$-series of $(2,0)$ theory. We get one tensor
multiplet for each $M5$-brane, but a certain linear combination
(namely the sum) decouples from the theory. This means that for the
$A_r$ version, we have $r+1$ parallel $M5$-branes but only $r$ tensor
multiplets. This is analogous to the world-volume theory of $r+1$
parallel D3-branes in Type IIB string theory, which is a
four-dimensional $\mathcal{N}=4$ supersymmetric Yang-Mills theory with
gauge group $\mathrm{SU}(r+1)$ rather than
$\mathrm{U}(r+1)$~\cite{Witten:1996}.

As mentioned in Section~\ref{sec:M-theory}, $M2$-branes stretching between the $M5$-branes are
perceived as strings from the six-dimensional point of view. The
existence of these strings is a consequence of a central charge of the
$(2,0)$ supersymmetry algebra in \Eqnref{QQ}, and they may be viewed
as solitonic from the $M5$-brane point of view~\cite{Howe:1998s}, see also Ref.~\cite{Gustavsson:2002}. This approach involves the equations of
motion~\cite{Howe:1997sb,Howe:1997d,Howe:1997c,Howe:1997s} for the
$M5$-brane coupled to eleven-dimensional supergravity. The string tension is related to the distance between the $M5$-branes in question, and therefore depends on the tensor multiplet fields. When the branes coincide, the string becomes tensionless.

The minimal BPS-saturated
representation corresponds to a multiplet of infinitely
long straight strings~\cite{Gustavsson:2001sr}, while for non-BPS
representations, waves may propagate along the strings. These degrees
of freedom are represented by a set of bosonic embedding fields
describing the world-sheet of the string, along with a set of
fermionic fields coming from the breaking of supersymmetry induced by
the string. Note that the string world-sheet need not be connected;
every connected component corresponds to one string.

Having $r+1$ $M5$-branes, there are in total $r(r+1)/2$ possibilities of
connecting these with $M2$-branes. This means that there are
$r(r+1)/2$ different species of strings in the theory. These
are oriented and the orientation may be changed by interchanging the
roles of the $M5$-branes in question.

It should be mentioned that there is a
viewpoint~\cite{Becker:1996,Dijkgraaf:1997a,Dijkgraaf:1997b}, where
the tensor multiplet fields are considered to originate from massless
states of closed self-dual strings living on the $M5$-brane. This may
very well be the case, but it will not affect our model and will
therefore not be further discussed in this thesis.

The degrees of freedom found above from the $M$-theory perspective may
also be motivated (and generalized to the full $ADE$ classification)
from the Type IIB approach described in Section~\ref{sec:TypeIIB}. When
compactifying the Type IIB theory on \kthree, the zero modes of the
Type IIB supergravity fields will give rise to $r$ tensor multiplets,
where $r$ is the rank of the
$ADE$-singularity~\cite{Townsend:1984}. This means that we may, taking
the $ADE$ classification seriously, associate a $(2,0)$ tensor
multiplet to each of the $r$ Cartan generators of the corresponding
$ADE$-type Lie algebra. Let us see how the different component fields
of the tensor multiplet appear in this perspective: The wedge product
of the two-form potential in six dimensions with a self-dual harmonic
two-form on \kthree{} (with support near the two-sphere) is the
four-form potential of Type IIB theory (which has a self-dual field
strength). This motivates the self-duality of the three-form field
strength in the six-dimensional theory. The fermionic fields
of the $(2,0)$ tensor multiplet have their origin in the fermionic
fields of ten-dimensional Type IIB supergravity. Finally, the scalar
fields correspond to the local values of the moduli parameters. At
spatial infinity, these fields approach their vacuum expectation
values --- the moduli parameters of the theory.

Furthermore, since there is one two-sphere in the \kthree{} for each
positive simple root of the corresponding Lie algebra, the strings mentioned
above (arising as a D3-brane wrapped around such a two-sphere) are
grouped in multiplets associated with the positive root
generators. Considering the $A$-series, we have $r(r+1)/2$
positive roots and consequently equally many species of strings. This
coincides with the number found above from the $M$-theory
perspective. The negative roots of the $ADE$-type Lie algebra
correspond to reversing the orientation of the D3-brane, and thereby
reversing the string in six dimensions.

This discussion provides a deep connection between the $ADE$-type Lie
algebra classifying the singular points in the \kthree{} and the
corresponding realization of $(2,0)$ theory. The connection will be
further elucidated in the next section. We should also mention that the $ADE$-classification may be motivated from a purely six-dimensional perspective by an argument involving anomaly cancellation~\cite{Henningson:2004}.

\section{Connections to lower-dimensional theories}
\label{sec:lowerdim}

It is interesting to consider compactifications of $(2,0)$
theory to five- and four-dimensional space-times. Let us first
compactify on a circle to obtain a five-dimensional
theory~\cite{Seiberg:1998,Flink:2006}. The tensor multiplets (which were
associated with the Cartan generators of the underlying $ADE$-type Lie
algebra) then yield massless vector multiplets, while the strings
(corresponding to root generators) wound around the circle give rise
to \emph{massive} vector multiplets, with mass proportional to the
string tension times the circumference of the compactification
circle. These states correspond to the perturbative degrees of
freedom of a maximally supersymmetric
($\mathcal{N}=4$) Yang-Mills theory in five dimensions, where the
$ADE$-type gauge symmetry has been spontaneously broken. The symmetry
breaking is due to non-zero values of the moduli parameters and is
responsible for the mass of the second group of vector multiplets by
means of the Higgs mechanism.

Strings in other directions, i.e., not wound around the circle,
appear as magnetically charged strings in five dimensions and are
interpreted as non-perturbative solitons of the Yang-Mills theory.
Furthermore, there is a Kaluza-Klein tower of massive multiplets
originating from the massless tensor multiplets in six dimensions.

Approaching the origin of moduli space, the massive vector multiplets
are un-Higgsed and become massless. This clarifies the non-abelian
gauge symmetry of the theory and indicates an interpretation of
$(2,0)$ theory as the ultraviolet completion of maximally
supersymmetric Yang-Mills theory in five dimensions, unifying the
fundamental and the solitonic degrees of freedom.

We may also compactify the $(2,0)$ theory on a
torus~\cite{Witten:1995,Verlinde:1995,Green:1996,Henningson:2000cs,Witten:2002}. This yields
the maximally supersymmetric ($\mathcal{N}=4$)
Yang-Mills theory in four dimensions with $ADE$-type gauge group and a
(complexified) coupling constant~$\tau$, containing the usual coupling constant and the theta angle. The constant $\tau$ is then exactly the geometric modulus of the compactification torus. Strings wound around the $a$-cycle of the torus will be perceived as electrically charged particles while strings wound around the $b$-cycle will have magnetic charge. This interpretation of $\mathcal{N}=4$ super Yang-Mills theory as the compactification of
$(2,0)$ theory on a torus makes $S$-duality~\cite{Montonen:1977,Osborn:1979} manifest --- the modular $SL(2,\mathbb{Z})$ invariance is just a consequence of diffeomorphisms of the torus.

\section{Our approach to $(2,0)$ theory}
\label{sec:approach}

The $(2,0)$ theory does not contain any parameter that can be regarded
as small. This is of course a nuisance, making perturbation theory
impossible. Or so it seems. What we would like to advocate is to
consider the theory at a \emph{generic} point in the moduli space, safely
away from the origin. This breaks the conformal invariance and means that the strings will have a finite
string tension, and amounts to choosing non-zero vacuum expectation
values for the moduli fields. This new theory can hopefully give us
some clues to what the theory at the origin will look like, and also
constitutes an interesting problem in its own right. Anyway, our
limited understanding of the behavior at the origin prevents us from
working there; not much is known about tensionless
strings. Note, however, that we choose the moduli parameters to be
small enough for gravity to decouple from the theory.

This means that we are from now on dealing with a multiplet of tensile
strings in six dimensions. A general state in the theory can be
specified by the number of
infinitely extended approximately straight strings. These have certain
spatial directions and momenta, and also allow for waves propagating
along them. Finally, the state may also contain tensor multiplet
quanta.

If the energy of a typical string excitation is taken
to be vanishingly small compared to the string tension, we are left
with an exactly solvable model
of straight strings with free waves propagating on them, along with free
tensor multiplets. This provides a firm basis for doing perturbation
theory; the dimensionless parameter will then be the squared energy
divided by the string tension. As this parameter is increased to a
finite value, we obtain a model of tensor multiplets interacting
weakly with excitation modes on the strings.

It should also be said that in this limit, the energy difference
between states containing different numbers of strings is
infinite, since the strings have a finite string tension and are
infinitely long. This has important implications for the Hilbert space
of the theory, which is decomposed into superselection sectors
characterized by the number of strings contained therein, meaning that
we can consider configurations with a \emph{fixed} number of strings;
no string creation or annihilation is possible.

There may also be closed strings in the theory, not only the infinitely extended straight strings mentioned above. However, we argue on dimensional grounds that their energy must be of the order of the square root of the string tension and they can therefore be neglected in the limit of large string tension. We expect possible massless states in the spectrum of the closed strings, apart from those contained in the tensor multiplet, to decouple from the rest of the theory. This statement is based on the observation that such modes have no analogue in the supersymmetric Yang-Mills theories
obtained by compactifications to five or four dimensions (as described in Section~\ref{sec:lowerdim}).

For simplicity, we will hereafter focus on the $A_1$ version of
$(2,0)$ theory, containing a single tensor multiplet and a single type
of string. Note that the theory evidently can contain many strings,
but in this version, they are all of the same \emph{type}. The strings
will be treated as \emph{fundamental}, no matter what their actual
origin is.

A configuration with one single string involves only the coupling
between the string and the tensor multiplet fields, while a
configuration with two or more strings allows for direct string-string
interactions. In terms of the lower-dimensional supersymmetric
Yang-Mills theories described in Section~\ref{sec:lowerdim}, these couplings
correspond to trilinear and quadrilinear couplings,
respectively. Pursuing this connection to Yang-Mills theory, we expect
no other couplings to appear as the number of strings is increased.

\chapter{The $(2,0)$ tensor multiplet}
\label{ch:TM}

The purpose of this chapter is to introduce and discuss the free $(2,0)$ tensor multiplet and its properties with respect to supersymmetry and superconformal symmetry. We also introduce a formalism for making the superconformal properties of the tensor multiplet fields manifest.

\section{Field content}

The free massless tensor multiplet is the simplest representation of the $(2,0)$ super-Poincar\'e group~\cite{Strathdee:1987}. Let us see how it may be found: Considering the anti-commutation relations defining the supersymmetry algebra in \Eqnref{QQ}, the tensor multiplet representation has vanishing values for the central charges $Z^{ab}_{\al\be}$ and $W^{ab}_{\al\be}$, while the momentum $P_{\al\be}$ is light-like. This configuration breaks the Lorentz group $SO(5,1)$ to the little group $SO(4) \simeq SU(2) \times SU(2)$, but leaves the $R$-symmetry group $SO(5)$ unbroken.

The sixteen supersymmetry generators $Q^a_{\al}$ transform according to
\begin{equation}
({\bf 1},{\bf 2};{\bf 4}) \oplus ({\bf 2},{\bf 1};{\bf 4}),
\end{equation}
under the bosonic symmetry group $SU(2) \times SU(2) \times SO(5)$. From the anticommutation relations given in \Eqnref{QQ}, it may be seen that the first set of generators is unbroken and annihilate a quantum state of this configuration, while the second set is broken and forms a Clifford algebra. The latter may be represented on a set of states transforming as
\begin{equation}
({\bf 3},{\bf 1};{\bf 1}) \oplus ({\bf 2},{\bf 1};{\bf 4}) \oplus ({\bf 1},{\bf 1};{\bf 5}),
\end{equation}
which form the sixteen states, eight bosonic and eight fermionic, of the $(2,0)$ tensor multiplet in six dimensions. All fields involved are real or obey reality conditions.

In terms of fields, the representation $({\bf 3},{\bf 1};{\bf 1})$ corresponds to a chiral two-form potential with a self-dual three-form field strength, both $R$-symmetry scalars. Utilizing the notation developed in~\Eqnref{lorentzreps}, we denote the two-form, transforming in the ${\bf 15}$ representation under the Lorentz group, by
$\Bf{\al}{\be}$, where the trace $\Bf{\al}{\al}=0$. The
corresponding self-dual field strength, transforming in the ${\bf 10}_{+}$ representation, is given by
\begin{equation}
\eqnlab{hofb}
h_{\al \be} = \pa_{\al \ga} \Bf{\be}{\ga} + \pa_{\be \ga}
\Bf{\al}{\ga}.
\end{equation}
One may also form an anti self-dual field strength (in the ${\bf
  10}_{-}$ representation) as
\begin{equation}
\eqnlab{antihofb}
h^{\al \be} = \pa^{\al \ga} \Bf{\ga}{\be} +
\pa^{\be \ga} \Bf{\ga}{\al}.
\end{equation}
This field is \emph{not} part of the tensor multiplet, but it will
prove to be useful in the following. Both field strengths are invariant under the local gauge transformation
\begin{equation}
\de b_\al^{\ph{\al}\be} = \pa_{\al\ga} \La^{\be\ga} - \frac{1}{4} \kde{\al}{\be} \pa_{\ga\de} \La^{\ga\de},
\end{equation}
where $\La^{\al\be}(x)=-\La^{\be\al}(x)$ is an infinitesimal one-form gauge parameter. The self-dual and anti self-dual field strengths are related by the Bianchi identity
\begin{equation}
\eqnlab{Bianchi}
\pa^{\al \ga} h_{\al \be} - \pa_{\al \be} h^{\al \ga} = 0,
\end{equation}
which is easily verified by using the definitions \eqnref{hofb} and
\eqnref{antihofb} along with the identity $\pa_{\al \ga} \pa^{\be \ga}
\equiv \frac{1}{4} \de_{\al}^{\ph{\al}\be} \pa_{\ga \de} \pa^{\ga \de}$.


Moving on, the representation $({\bf 1},{\bf 1};{\bf 5})$ is a set of
five bosonic Lorentz scalars transforming as an $SO(5)$ vector. Using the results in \Eqnref{rsymreps}, these will be
denoted by the antisymmetric matrix $\phi^{ab}$ fulfilling the tracelessness requirement $\Om_{ab} \phi^{ab} = 0$.
As has been noted in the preceding chapter, the vacuum expectation values of these fields are the moduli of the theory.

Finally, the $({\bf 2},{\bf 1};{\bf 4})$ representation corresponds to
a set of fermionic fields transforming in the four-dimensional chiral spinor representation of the Lorentz group and in the likewise four-dimensional spinor representation of the $R$-symmetry group. They are written as Grassmann odd quantities $\psi^a_{\al}$.


We should also mention that the tensor multiplet representation may be found as a doubleton representation of the superconformal group $OSp(8^*|4)$, using an oscillator construction~\cite{Gunaydin:1984,Gunaydin:1999}.

\section{Dynamics}

The free fields of the tensor multiplet satisfy the standard equations of motion for scalar, spinor and tensor fields, respectively. Explicitly, these are
\begin{equation}
\eqnlab{TM_eom}
\begin{aligned}
\pa^{\al \be} \pa_{\al \be} \phi^{ab} &= 0  \\
\pa^{\al \be} \psi^b_{\be} &= 0 \\
\pa^{\al \ga} h_{\al \be} &= 0,
\end{aligned}
\end{equation}
and can be found by varying the action
\begin{equation}
\eqnlab{TMaction}
S_{\sss TM} = \int d^6 x \left\{ - \Om_{ac} \Om_{bd} \pa_{\al \be}
\phi^{ab} \pa^{\al \be} \phi^{cd} + 2 h_{\al \be} h^{\al \be} - 4i
\Om_{ab} \psi^a_{\al} \pa^{\al \be} \psi^b_{\be} \right\},
\end{equation}
using the Bianchi identity in \Eqnref{Bianchi}. This action was stated in {\sc Paper II} and {\sc Paper III}, and encodes the dynamics of the free tensor multiplet. It is clearly both Lorentz and $R$-symmetry invariant, as required. The dimensionalities of the fields involved are easily deduced from the fact that the action is dimensionless.

Note that the action \eqnref{TMaction} contains the anti
self-dual part $h^{\al \be}$ of the field strength, even though it is
not part of the tensor multiplet. Strictly speaking, it is not
possible to give a lagrangian description of only a self-dual
three-form in six dimensions~\cite{Witten:1997}. We therefore relax
the self-duality requirement at the level of the action and carry the
anti self-dual part of the field strength along as a spectator field. However, we have to be careful when adding interactions; we need to make sure that this part does not couple
to anything, by choosing appropriate coefficients in the interaction terms.

The action \eqnref{TMaction} is unique, with no free parameters, in the sense that the numerical coefficients are uniquely determined modulo rescalings of the fields $\phi^{ab}(x)$ and $\psi^a_\al(x)$. The normalization of the gauge field has a physical meaning, since it really is a connection on a one-gerbe. This means that its normalization cannot be changed. The value of the constant preceding the corresponding kinetic term in the action can be fixed by considering when the decoupling of the anti self-dual part of the field strength is consistent, but we have not made any attempt to do this in this text.

It should be mentioned that there is another
approach~\cite{Pasti:1997,Bandos:1997}, called PST
(named after Pasti, Sorokin and Tonin), where an auxiliary scalar field is introduced,
making it possible to write down an action for the self-dual
three-form. This action then yields the correct equations of
motion, including the self-duality requirement. We will not consider this particular model any further in this text.

\section{Supersymmetry transformations}

The coefficients in the action~\eqnref{TMaction} are chosen such that it is invariant under the global supersymmetry transformations
\begin{equation}
\eqnlab{susy_TM}
\begin{aligned}
\de \phi^{ab} &= - i \eta^{\al}_c \left( \Om^{ca} \psi^b_{\al} +
\Om^{bc} \psi^a_{\al} + \frac{1}{2} \Om^{ab} \psi^c_{\al} \right) \\
\de \psi^a_{\al} &= \Om^{ab} \eta_b^{\be} h_{\al \be} + 2 \pa_{\al
  \be} \phi^{ab} \eta_b^{\be} \\
\de \Bf{\al}{\be} &= -i \eta_a^{\be} \psi^a_{\al} + \frac{i}{4}
\de_{\al}^{\ph{\al}\be} \eta_a^{\ga} \psi^a_{\ga},
\end{aligned}
\end{equation}
where $\eta^{\al}_a$ is the constant fermionic supersymmetry parameter introduced in Section~\ref{sec:susy}. The transformation
of $b_\al^{\ph{\al}\be}$ induces the following transformations of the field strengths:
\begin{equation}
\eqnlab{susy_h}
\begin{aligned}
\de h_{\al \be} &= - i \eta_a^{\ga} \left( \pa_{\al \ga}
\psi^a_{\be} + \pa_{\be \ga} \psi^a_{\al} \right) \\
\de h^{\al \be} &= -i \eta_a^{\al} \pa^{\be \ga} \psi_{\ga}^a - i
\eta_a^{\be} \pa^{\al \ga} \psi_{\ga}^a.
\end{aligned}
\end{equation}
It is important to note that the anti self-dual part, which does not
really belong to the tensor multiplet, seems to transform non-trivially under
a supersymmetry transformation. However, when using the equations of motion \eqnref{TM_eom} for $\psi^a_\al$, the right-hand side of that particular transformation vanishes and thus, $h^{\al \be}$ is a supersymmetry invariant on-shell. It is also essential that $h^{\al \be}$ does not appear on the right-hand side of \Eqnref{susy_TM}. Since the supersymmetry algebra is only expected to close on-shell, we conclude that the transformations are consistent with the fact that $h^{\al \be}$ is not really a part of the tensor multiplet.

One may further verify that the supersymmetry algebra indeed closes
on-shell, i.e., that the commutator of two supersymmetry transformations
acting on a tensor multiplet field yields a translation when the equations of motion \eqnref{TM_eom} are used.

\section{Superfield formulation}

The next step is to describe the free $(2,0)$ tensor multiplet in terms of superfields, in the superspace introduced in Chapter~\ref{ch:superspace}. This will prove to be very useful in Chapter~\ref{ch:coupling}, where we couple the tensor multiplet to a self-dual string.

The on-shell superfield formulation for the $(2,0)$ tensor multiplet~\cite{Howe:1983} involves a scalar superfield $\Phi^{ab}(x,\theta)$. As indicated by the index structure, it transforms in the ${\bf 5}$ representation of the $R$-symmetry group, and therefore obeys the algebraic constraint
\begin{equation}
\eqnlab{algcondS}
\Om_{ab} \Phi^{ab} = 0.
\end{equation}
Furthermore, the superfield has to satisfy the differential constraint
\begin{equation}
\eqnlab{diffcon}
D^a_{\al} \Phi^{bc} + \frac{1}{5} \Om_{de} D^d_{\al} \left( 2 \Om^{ab}
\Phi^{ec} - 2 \Om^{ac} \Phi^{eb} + \Om^{bc} \Phi^{ea} \right) = 0,
\end{equation}
where the superspace covariant derivative is defined as
\begin{equation}
\eqnlab{superd}
D^a_{\al} = \pa^a_{\al} + i \Om^{ab} \theta^{\be}_b \pa_{\al \be}.
\end{equation}
The covariant derivative is fermionic, and has the important property that it anti-commutes with the supersymmetry generator $Q^a_\al$. This means that the covariant derivative of a superfield is another superfield, according to the general requirement~\eqnref{Qaction}.

We should also mention that the differential constraint in \Eqnref{diffcon} has a geometric meaning~\cite{Howe:1997sb}, as a master constraint originating from the embedding of a superfivebrane in eleven-dimensional superspace.

For convenience, we use the covariant derivative to define supplementary superfields according to
\begin{equation}
\eqnlab{Psi_H}
\begin{aligned}
\Psi^c_{\al}(x,\theta) &= - \frac{2i}{5} \Om_{ab} D_{\al}^a \Phi^{bc} \\
H_{\al \be}(x,\theta) &= \frac{1}{4} \Om_{ab} D^a_{\al} \Psi^b_{\be},
\end{aligned}
\end{equation}
but it should be emphasized that these contain no new degrees of freedom compared to $\Phi^{ab}(x,\theta)$. The supplementary superfields satisfy the differential constraints
\begin{equation}
\eqnlab{diffcon2}
\begin{aligned}
D^a_\al \Psi^b_{\be} &= 2 \pa_{\al\be} \Phi^{ab} - \Om^{ab} H_{\al\be} \\
D^a_\al H_{\be\ga} &= 2i \pa^{\ph{a}}_{\al(\be} \Psi^a_{\ga)},
\end{aligned}
\end{equation}
which are derived from the original constraint~\eqnref{diffcon}.

A superfield transforms, by definition, according to \Eqnref{susy_transf} under supersymmetry. After some calculations, this implies that the lowest components in a $\theta$-expansion of the superfields $\Phi^{ab}(x,\theta)$, $\Psi^a_{\al}(x,\theta)$ and $H_{\al \be}(x,\theta)$ obey exactly the same transformation laws as the space-time fields $\phi^{ab}(x)$, $\psi^a_{\al}(x)$ and $h_{\al \be}(x)$, as given in \Eqsref{susy_TM}--\eqnref{susy_h}. This motivates the choice of notation for the superfields --- the superfield corresponding to a certain field in the tensor multiplet (denoted by small letters) is denoted by the corresponding capital letter.

An explicit expression for the lowest components of the superfield $\Phi^{ab}(x,\theta)$ is presented in {\sc Paper III}.

We also note that the differential constraint~\eqnref{diffcon} implies that the component fields $\phi^{ab}(x)$, $\psi^a_{\al}(x)$ and $h_{\al \be}(x)$ must obey the free equations of motion given in \Eqnref{TM_eom}, indicating the on-shell property of the superfield formulation. To the best of our knowledge, no off-shell superfield formulation for the $(2,0)$ tensor multiplet is known.

\section{Superconformal transformations}

So far, we have only found the supersymmetry transformation laws for the tensor multiplet fields. This is not enough for our purposes; we need to know the action of the complete superconformal group on these fields. From the recipe in Section~\ref{sec:superfields}, we learn that this is given by the action of the little group generators on the fields.

Consider the superfield $\Phi^{ab}(x,\theta)$. It is defined to be a Lorentz scalar, meaning that $\left( \dia{\Si}{\al}{\be} \Phi \right)^{ab}=0$. We also require it to be a \emph{superprimary} field, meaning that both $\left( \ka^{\al\be} \Phi \right)^{ab}=0$ and $\left( \si_c^{\al} \Phi \right)^{ab}=0$. Moreover, it transforms as a vector under the $R$-symmetry group, which implies that
\begin{equation}
\left( u^{cd} \Phi \right)^{ab} = - \Om^{c[a} \Phi^{b]d} - \Om^{d[a} \Phi^{b]c}.
\end{equation}
Finally, we demand the scaling dimension to coincide with the physical mass dimension of the field, so that $\left(\De \Phi\right)^{ab}=2 \Phi^{ab}$, which is consistent with the unitarity requirements and the BPS property of the representation~\cite{Minwalla:1998}. This yields the result
\begin{equation}
\eqnlab{supertransf_Phi}
\de \Phi^{ab} = \left[ \xi^{\ga\de}(x,\theta) \pa_{\ga\de} + \xi^\ga_c(x,\theta) \pa^c_\ga + 2 \La(x,\theta) \right] \Phi^{ab} - 2 V_{cd}(\theta) \Om^{c[a} \Phi^{b]d}.
\end{equation}
This was also found in {\sc Paper V} by deriving the most general superconformal field transformation that is consistent with the differential constraint~\eqnref{diffcon}. The agreement of these two viewpoints is quite striking and points to some interesting properties of the differential constraint. This will be further discussed in Section~\ref{sec:diffcon}.

The transformation given in \Eqnref{supertransf_Phi} implies, through the definitions~\eqnref{Psi_H}, that the supplementary superfields transform according to
\begin{equation}
\eqnlab{supertransf_Psi_H}
\begin{aligned}
\de \Psi^a_\al &= \left[ \xi^{\ga\de}(x,\theta) \pa_{\ga\de} + \xi^\ga_c(x,\theta) \pa^c_\ga + \frac{5}{2} \La(x,\theta) \right] \Psi^a_\al + {} \\
& \quad + \dia{\Om}{\al}{\ga}(x,\theta) \Psi^a_\ga - \Om^{ac} V_{cd}(\theta) \Psi^d_\al - 4 \Om_{bc} R^b_\al(\theta) \Phi^{ca} \\
\de H_{\al\be} &= \left[ \xi^{\ga\de}(x,\theta) \pa_{\ga\de} + \xi^\ga_c(x,\theta) \pa^c_\ga + 3 \La(x,\theta) \right] H_{\al\be} + {} \\
& \quad + 2 \dia{\Om}{(\al}{\ga}(x,\theta) H_{\be)\ga} - 6i \Om_{cd} R^c_{(\al}(\theta) \Psi^d_{\be)}.
\end{aligned}
\end{equation}
From these relations, the action of the little group generators on the supplementary superfields may be found. The properties with respect to Lorentz and $R$-symmetry rotations agree with our expectations, and we note that the scaling dimensions again agree with the physical dimensions of the fields. However, it is apparent that the superfields $\Psi^a_\al$ and $H_{\al\be}$ are primary, but not superprimary. This will have some important consequences in the discussion on possible coupling terms in Chapter~\ref{ch:coupling}, and is closely connected to the appearance of the superspace-dependent parameter function $R^a_{\al}(\theta)$ in the transformation law for the fermionic superspace differential $e^\al_a$ in \Eqnref{transf_e}.

Having found the general transformation laws for the superfields of the $(2,0)$ tensor multiplet, it is a simple task to find how the space-time fields $\phi^{ab}(x)$, $\psi^a_\al(x)$ and $h_{\al\be}(x)$ transform. However, these transformation laws are not as compact as the ones given above for the superfields, and less useful in the following. We will therefore not state these in this text and instead refer to {\sc Paper V}.

\section{Manifest superconformal covariance}

We would now like to use the methods developed in Chapter~\ref{ch:manifest} to formulate a new superfield with manifest superconformal covariance. This is the main result of \textsc{Paper VI}, and will also serve as an alternative derivation of the transformation laws of the preceding section.

The aim is to find a superfield that transforms linearly under superconformal transformations, but also contains all the fields of the $(2,0)$ tensor multiplet. Inspired by the discussion in Section~\ref{sec:manifest_conformal}, we want to incorporate self-duality in the superconformal space as well. In the bosonic case, with vector indices, this was accomplished by adding a fourth vector index on the tensor field. How is this carried over to the notation where a self-dual three-form in six dimensions is written as $h_{\al\be}=h_{\be\al}$?

A self-dual four-form in eight dimensions transforms in the ${\bf 35}_+$ representation of $SO(6,2)$. This representation may also be built from two symmetric chiral spinor indices, if we require tracelessness with respect to the eight-dimensional metric. Concretely, this means that we may write the self-dual four-form as $\Ups_{\alh\beh}=\Ups_{\beh\alh}$. The tracelessness is accomplished by requiring that
\begin{equation}
I^{\alh\beh} \Ups_{\alh\beh} = 0,
\end{equation}
where $I^{\alh\beh}$ is the purely bosonic piece of the metric $I^{\sss AB}$ in \Eqnref{superspacemetric_upper}. This notation allows the results from Section~\ref{sec:manifest_conformal} to be written using spinor indices.

Generalizing this to the superconformal space, let $\Ups_{\sss AB}(y)$ be a graded symmetric tensor field defined on the supercone. It should be supertraceless, meaning that
\begin{equation}
I^{\sss AB} \Ups_{\sss AB} = 0.
\eqnlab{supertraceless}
\end{equation}
This is a natural extension of the bosonic field $\Ups_{\alh\beh}$ discussed above.

We also demand the field to be a homogeneous function of $y$ in the same sense as in \Eqnref{homo_Ups}, i.e.,
\begin{equation}
\frac{1}{2} I^{\sss CD} y_{\sss C} \pa_{\sss D} \Ups_{\sss AB} = n \Ups_{\sss AB}.
\eqnlab{homo_super}
\end{equation}
To reduce the number of components in $\Ups_{\sss AB}(y)$, we impose the subsidiary condition
\begin{equation}
I^{\sss AB} y_{\sss A} \Ups_{\sss BC} =0,
\eqnlab{subcon}
\end{equation}
which clearly is superconformally covariant and should be valid on the entire supercone. This constraint should be compared with \Eqsref{ydotH} and \eqnref{yasymH} from the discussion on bosonic tensor fields.

If we expand the constraint \eqnref{subcon} by means of \Eqsref{y^a} and \eqnref{y^al}, we find that
\begin{equation}
 \begin{aligned}
  \dia{\Ups}{\al}{\be} &= \left( 2 x^{\be\ga} + i \theta^\be \cdot \theta^\ga \right) \Ups_{\al\ga} - \sqrt{2}i\theta^\be_b \dia{\Ups}{\al}{b} \\
  \Ups^{a \be} &= \left( 2 x^{\be\ga} + i \theta^\be \cdot \theta^\ga \right) \dia{\Ups}{\ga}{a} + \sqrt{2} i \theta^\be_b \Ups^{ab} \\
  \Ups^{\al\be} &= \left( 2 x^{\be\ga} + i \theta^\be \cdot \theta^\ga \right) \dia{\Ups}{\ga}{\al} - \sqrt{2} i \theta_b^\be \Ups^{\al b},
 \end{aligned}
 \eqnlab{Ups-Ups}
\end{equation}
where the dot product between two $\theta$-coordinates was defined above, in \Eqnref{thetadot}.

From \Eqnref{Ups-Ups}, we see that the most general solution to the algebraic equation \eqnref{subcon} is parametrized by the fields $\Ups_{\al\be}$, $\dia{\Ups}{\al}{b}$ and $\Ups^{ab}$. The supertracelessness condition \eqnref{supertraceless} becomes
\begin{equation}
\Om_{ab} \Ups^{ab} = 2 \theta^\al \cdot \theta^\be \Ups_{\al\be} - 2 \sqrt{2} \theta^\al_a \dia{\Ups}{\al}{a},
\end{equation}
and effectively removes one of the components in the parametrization fields. This suggests that we may use the superfields $\Phi^{ab}$, $\Psi^a_\al$ and $H_{\al\be}$ (or rather, the corresponding tensor multiplet fields, but it is more convenient to work with superfields) defined above to parametrize $\Ups_{\sss AB}$. However, from the supertracelessness condition it follows that we cannot simply identify these with the parametrization components in $\Ups_{\sss AB}$, since $\Phi^{ab}$ is supposed to be traceless according to \Eqnref{algcondS}.

The solution to this problem is to let
\begin{equation}
 \begin{aligned}
\Ups_{\al\be} &= \frac{1}{\ga^2} H_{\al\be} \\
\dia{\Ups}{\al}{b} &= \frac{1}{\ga^2} \left[ - \frac{3}{\sqrt{2}} \Psi^b_\al + \sqrt{2} \Om^{bc} \theta_c^\ga H_{\al\ga} \right] \\
\Ups^{ab} &= \frac{1}{\ga^2} \left[ -6i \Phi^{ab} - 3 \Om^{ac} \theta_c^\ga \Psi_\ga^b + 3 \Om^{bc} \theta_c^\ga \Psi_\ga^a + 2 \Om^{ac} \Om^{bd} \theta_c^\ga \theta_d^\de H_{\ga\de} \right],
 \end{aligned}
 \eqnlab{Ups_H1}
\end{equation}
where we have introduced factors of the projective parameter $\ga$ to take the degree of homogeneity from \Eqnref{homo_super} into account. We have put $n=-2$, a choice that will be motivated in Section~\ref{sec:diffcon}.

The remaining components in $\Ups_{\sss AB}$ follow from \Eqnref{Ups-Ups} and are
\begin{equation}
 \begin{aligned}
\dia{\Ups}{\al}{\be} &= \frac{1}{\ga^2} \left[ (2x^{\be\ga} -i \theta^\be \cdot \theta^\ga ) H_{\al\ga} + 3i \theta^\be_c \Psi^c_\al \right] \\
\Ups^{\al\be} &= \frac{1}{\ga^2} \Big[  (2x^{\al\ga} -i \theta^\al \cdot \theta^\ga ) (2x^{\be\de} -i \theta^\be \cdot \theta^\de ) H_{\ga\de} - 12i \theta^\al_a \theta^\be_b \Phi^{ab} +  \\
& \quad + 6i \theta^{(\al}_a  (2x^{\be)\ga} -i \theta^{\be)} \cdot \theta^\ga ) \Psi^a_\ga \Big] \\
\Ups^{\al b} &= \frac{1}{\ga^2} \bigg[ (2x^{\al\ga} -i  \theta^\al \cdot \theta^\ga ) \left( \sqrt{2} \Om^{bd}\theta_d^\de H_{\ga\de} - \frac{3}{\sqrt{2}} \Psi^b_\ga \right) - {}  \\
& \quad - 6 \sqrt{2} \theta^\al_a \Phi^{ab} + 3 \sqrt{2} i \Om^{bc} \theta^\ga_c \theta^\al_a \Psi^a_\ga \bigg].
 \end{aligned}
 \eqnlab{Ups_H2}
\end{equation}
In this way, we have found a unique expression for the covariant field $\Ups_{\sss AB}$ in terms of some fields $H_{\al\be}$, $\Psi^a_\al$ and $\Phi^{ab}$, which a priori need not be the superfields of the $(2,0)$ tensor multiplet. This result is interesting in its own right, but it would be more useful if we could relate it to the formalism developed in Section~\ref{sec:manifest_superconf}.

Define the graded symmetric matrix
\begin{equation}
\eqnlab{calH}
\mathcal{H}_{\sss AB} =
  \left( \begin{array}{ccc}
    H_{\al\be} & 0 & - \frac{3}{\sqrt{2}} \Psi^b_\al \\
    0 & 0 & 0 \\
    - \frac{3}{\sqrt{2}} \Psi^a_\be & 0 & -6i \Phi^{ab}
  \end{array} \right),
\end{equation}
which only contains the superfields in the $(2,0)$ tensor multiplet (no explicit coordinate dependence). It should be noted that $\mathcal{H}_{\sss AB}$ is \emph{not} a covariant field in the superconformal space; it does not transform linearly under superconformal transformations. It is an analogue of the field $\mathcal{H}_{\muh\nuh\rhoh\sih}$ appearing in \Eqnref{Ups-H_bos} to the superconformal space. With this expression for $\mathcal{H}_{\sss AB}$, it turns out that \Eqsref{Ups_H1} and \eqnref{Ups_H2} may be summarized in
\begin{equation}
\Ups_{\sss AB} = \frac{1}{\ga^2} \exp \left( -2 x^{\ga\de} s_{\ga\de} + 2 \sqrt{2} i \theta^\ga_d \dia{s}{\ga}{d} \right) \mathcal{H}_{\sss AB},
\eqnlab{ups_H}
\end{equation}
given that the action of the intrinsic generator $s_{\sss AB}$ on a field with two superindices is given by
\begin{equation}
\eqnlab{s_Ups}
\begin{split}
s_{\sss CD} \Ups_{\sss AB} &= \frac{1}{2} \Big( I_{\sss DA} \Ups_{\sss CB} - (-1)^{\sss CD} I_{\sss CA} \Ups_{\sss DB} + {} \\ & \quad + (-1)^{\sss AB} I_{\sss DB} \Ups_{\sss CA} - (-1)^{\sss AB+CD} I_{\sss CB} \Ups_{\sss DA} \Big).
\end{split}
\end{equation}
The latter equation is consistent with the commutation relations \eqnref{superalgebra} for the superconformal algebra, as required.

The inverse relation to \Eqnref{ups_H} is
\begin{equation}
\mathcal{H}_{\sss AB} = \ga^2 \exp \left( 2 x^{\ga\de} s_{\ga\de} - 2 \sqrt{2} i \theta^\ga_d \dia{s}{\ga}{d} \right) \Ups_{\sss AB},
\eqnlab{H_Ups}
\end{equation}
which agrees exactly with the general expression in \Eqnref{manifest_Phi}, relating a superconformally covariant field to an ordinary superspace field. This is quite remarkable --- by solving some algebraic constraints on a covariant field, we have recovered the relation between the fields living in the ordinary superspace and the manifestly conformally covariant field $\Ups_{\sss AB}$.

Following the discussion in Section~\ref{sec:manifest_superconf}, this implies that a linear transformation of the field $\Ups_{\sss AB}$ induces a superconformal transformation of the fields contained in $\mathcal{H}_{\sss AB}$ as given by \Eqnref{manifest_piJ}. Explicitly, this yields the same transformation laws for the superfields as in the preceding section. This shows that our anticipation was correct --- the fields used to parametrize the solution to the algebraic constraint \eqnref{subcon} may consistently be interpreted as the superfields of the $(2,0)$ tensor multiplet. This derivation also yields a nice insight into the origins of the different pieces of the transformation laws.

\section{The differential constraint}
\label{sec:diffcon}

What is the purpose of finding a manifestly covariant formalism? An obvious advantage of such a formulation is the possibility to find covariant quantities and transformation laws in a simple way. For example, if we write down a scalar in the superconformal space composed of covariant quantities, we know that its corresponding field in the ordinary superspace will be invariant (in a certain sense, see below) under superconformal transformations.

The simplest non-zero $OSp(8^*|4)$ scalar that we may form from our ingredients is quadratic in the field $\Ups_{\sss AB}(y)$ and written as
\begin{equation}
\eqnlab{tension_scalar}
I^{\sss AD} I^{\sss BC} \Ups_{\sss AB} \Ups_{\sss CD} = -36 \frac{1}{\ga^4} \Om_{ac} \Om_{bd} \Phi^{ab} \Phi^{cd},
\end{equation}
where we used \Eqsref{Ups_H1} and \eqnref{Ups_H2} to translate the fields $\Ups_{\sss AB}(y)$ to $(2,0)$ superfields. We will make use of this quantity when we introduce the interaction between the tensor multiplet and a self-dual string in Chapter~\ref{ch:coupling}.

Usually, a scalar field transforms only differentially, but in the case of superconformal scalars we have to include the homogeneity degree as well. This means that
\begin{equation}
\eqnlab{transf_tension}
\de \left( \Om_{ac} \Om_{bd} \Phi^{ab} \Phi^{cd} \right) = \left[\de x \cdot \pa + \de \theta \cdot \pa + 4 \La(x,\theta) \right] \left( \Om_{ac} \Om_{bd} \Phi^{ab} \Phi^{cd} \right),
\end{equation}
in agreement with the transformation implied by \Eqnref{supertransf_Phi}.

Let us move on to the main purpose of this subsection: to investigate whether the differential constraint~\eqnref{diffcon} for the $(2,0)$ superfield $\Phi^{ab}(x,\theta)$ may be formulated in a manifestly covariant way, with respect to superconformal symmetry. We expect this to be possible, since the differential constraint respects superconformal symmetry and is formulated in terms of superfields and superderivatives.

We do not have very many quantities to build such a covariant constraint from. Considering what the differential constraint~\eqnref{diffcon} and the derived constraints in \Eqnref{diffcon2} look like, we expect the covariant expression to have four free superindices.

Firstly, note that the graded symmetry of $I_{\sss AB}$ and \Eqnref{s_Ups} together imply that
\begin{equation}
\eqnlab{s_Ups=0}
s_{\sss [AB} \Ups_{\sss C]D} = 0,
\end{equation}
where, of course, the antisymmetrization is graded.

Having done this observation, it makes sense to consider the related equation
\begin{equation}
L_{\sss [AB} \Ups_{\sss C]D} = 0,
\end{equation}
which is a differential analogue of \Eqnref{s_Ups=0}. Using the expressions \eqnref{Ups_H1} and \eqnref{Ups_H2} for $\Ups_{\sss AB}(y)$ together with \Eqnref{tildegenerators} for the different pieces of $L_{\sss AB}$, we find that the equation is satisfied exactly when the superfield $\Phi^{ab}(x,\theta)$ obeys the differential constraint~\eqnref{diffcon}, but only if the degree of homogeneity of $\Ups_{\sss AB}$ is $n=-2$. The latter observation is interesting and shows that $n$ cannot be chosen freely.

This means that we may indeed formulate a manifestly superconformally covariant differential constraint, namely
\begin{equation}
\tilde{J}_{\sss [AB} \Ups_{\sss C]D} \equiv \left( L_{\sss [AB} + s_{\sss [AB} \right) \Ups_{\sss C]D} = 0.
\eqnlab{diff_con_super}
\end{equation}
This also implies that the tensor multiplet fields must obey the free equations of motion, since they are a consequence of the differential constraint. So, the constraint~\eqnref{diff_con_super} is both a constraint on the superfield $\Ups_{\sss AB}$ and an equation of motion.

This important result is quite remarkable --- \Eqnref{diff_con_super} contains a lot of information about the tensor multiplet fields in a very simple and manifestly superconformally covariant formulation.

As a final application, it is interesting to consider the supplementary superfields $\Psi^a_\al$ and $H_{\al\be}$. They are defined from contractions of the differential constraint~\eqnref{diffcon}. It is therefore worthwhile to mention the covariant condition
\begin{equation}
I^{\sss CD} \tilde{J}_{\sss AC} \Ups_{\sss D B} = 0,
\end{equation}
which is true provided that the superfields $\Phi^{ab}$, $\Psi^a_\al$ and $H_{\al\be}$ satisfy \Eqnref{Psi_H}. This condition may be split according to
\begin{equation}
\begin{aligned}
I^{\sss CD} L_{\sss AC} \Ups_{\sss D B} &= -2 \Ups_{\sss AB} \\
I^{\sss CD} s_{\sss AC} \Ups_{\sss D B} &= 2 \Ups_{\sss AB},
\end{aligned}
\end{equation}
in contrast to the differential constraint above. This means that the definitions of the superfields $\Psi^a_\al$ and $H_{\al\be}$ are consistent with respect to superconformal symmetry.

\chapter{The self-dual string}
\label{ch:freestring}

Considering the supersymmetry algebra given by \Eqnref{QQ}, we may form a representation by allowing for a non-zero central charge $Z^{ab}_{\al\be}$ while keeping $W^{ab}_{\al\be}=0$. Following Ref.~\cite{Gustavsson:2001sr}, we take $Z^{ab}_{\al\be}$ to be the tensor product of a space-like Lorentz vector $V_{\al \be}$ and an $R$-symmetry vector $\hat{\phi}^{ab}$, such that the scalar product of $V_{\al \be}$ and the momentum $P_{\al \be}$ vanishes (meaning that $V_{\al \be}$ is purely spatial in the frame where $P_{\al \be}$ is purely time-like). This then describes an infinitely long, straight string pointing in the direction given by $V_{\al \be}$, in a background where the moduli parameters are given by $\hat{\phi}^{ab}$. This string is the topic of the present chapter.

\section{Field content}

The presence of the string breaks the six-dimensional translational
invariance and therefore introduces four Goldstone bosons. These are
naturally interpreted as the transverse coordinates of the
string. Similarly, the string breaks half of the sixteen
supersymmetries and therefore gives rise to eight Goldstone fermions;
four left-moving and
four right-moving. It is more convenient, however, to extend these
fields to fill out full $SO(5,1)$ representations. We then
take the bosonic degrees of freedom to form a space-time vector
$X^{\al \be}=X^{\al \be}(\tau,\si)$ living on the string
world-sheet, which is parametrized by the bosonic coordinates $\tau$
and $\si$. Occasionally, we will denote these by $\si^{i}$, $i=(0,1)$. Likewise, the
fermionic world-sheet fields are collected into a single space-time
spinor $\Theta^{\al}_{a}=\Theta^{\al}_{a}(\tau,\si)$ transforming in
the anti-chiral representation
${\bf 4}'$ of the Lorentz group and in the ${\bf 4}$ of the
$R$-symmetry group. The extra degrees of freedom can be removed by
choosing a specific world-sheet parametrization in the bosonic case
and by employing a local fermionic symmetry in the fermionic case. The
latter symmetry, called $\kappa$-symmetry, will play an important role
in the following.

The string degrees of freedom $X^{\al \be}$ and $\Theta^{\al}_{a}$ are naturally interpreted as describing the embedding of the string in the $(2,0)$ superspace introduced in Chapter~\ref{ch:superspace}.

\section{Superconformal transformations}

As embedding fields, $X^{\al \be}$ and $\Theta^{\al}_{a}$ have very natural transformation laws with respect to the superconformal group. In the active picture, they transform in the same way as the coordinates do in the passive picture, but with \emph{opposite sign}. Explicitly, this means that
\begin{align}
\eqnlab{transf_superX}
\begin{split}
\de X^{\al \be} &= - a^{\al \be} + \dia{\om}{\ga}{[\al} X^{\be] \ga} - \la X^{\al \be} - 4 c_{\ga \de} X^{\ga\al} X^{\be\de} + i \Om^{ab} \eta^{[\al}_{a\ph{b}} \Theta^{\be]}_b + {} \\
& \quad + c_{\ga \de} \Theta^\ga \cdot \Theta^{[\al}
\Theta^{\be]} \cdot \Theta^{\de} + i \rho^c_{\ga} \Theta^{[\al}_c
    \left( 2 X^{\be]\ga} - i \Theta^{\be]} \cdot \Theta^{\ga} \right)
\end{split} \\
\eqnlab{transf_superTh}
\begin{split}
\de \Theta^{\al}_a &= - (\dia{\om}{\ga}{\al} - 4 c_{\ga \de} X^{\al \de} - 2i c_{\ga \de} \Theta^{\al} \cdot \Theta^{\de} + 2i
\rho^c_{\ga} \Theta^{\al}_c) \Theta^{\ga}_a - {} \\
& \quad - \frac{1}{2} \la \Theta^{\al}_a - \eta^{\al}_a + \Om_{ac} \rho^c_{\ga} \left( 2 X^{\ga \al} - i \Theta^{\ga} \cdot \Theta^{\al} \right) - v_{ac} \Om^{cd} \Theta^{\al}_d.
\end{split}
\end{align}
In the following, we will also need the transformations of the differentials
\begin{equation}
\eqnlab{EE}
\begin{split}
E^{\al\be} &= \tilde{d}X^{\al\be} + i \Om^{ab} \Theta_{a\ph{b}}^{[\al} \tilde{d} \Theta_b^{\be]} \\
E^\al_a &= \tilde{d} \Theta_a^\al,
\end{split}
\end{equation}
where $\tilde{d} \equiv d \si^i \pa_i$ denotes a differential operator with respect to the world-sheet parameters $\si^i$. The variations of these quantities are
\begin{equation}
\eqnlab{transf_E}
\begin{split}
\de E^{\al \be} &= -\dia{\Om}{\ga}{\al}(X,\Theta) E^{\ga \be} -
\dia{\Om}{\ga}{\be}(X,\Theta) E^{\al \ga} - \La(X,\Theta) E^{\al \be} \\
\de E_a^\al &= -\dia{\Om}{\ga}{\al}(X,\Theta) E_a^\ga -
\frac{1}{2} \La(X,\Theta) E_a^\al - V_{ac}(\Theta) \Om^{cd}
E_d^\al - {} \\
& \quad - 2 \Om_{ac} R^c_{\ga}(\Theta) E^{\al \ga},
\end{split}
\end{equation}
where the superspace dependent parameter functions are the same as in \Eqsref{xi_bos}--\eqnref{Raal} but evaluated on the string world-sheet. Obviously, these transformations are the same as those in \Eqnref{transf_e} but with the coordinates replaced by the embedding fields and reversed signs.

\section{Dynamics}

The dynamics of a free superstring in six dimensions is described in
{\sc Paper II} in a Brink-diVecchia-Howe-Deser-Zumino~\cite{Brink:1976,Deser:1976} fashion
with an auxiliary metric. The construction is analogous to the
Green-Schwarz superstring~\cite{Green:1984,Green:1984b} with a local fermionic
$\kappa$-symmetry acting both on the world-sheet fields ($X^{\al\be}$ and $\Theta^\al_a$) and on the auxiliary metric. In fact, six dimensions is one of the few cases where such superstrings may exist.

It is more illustrative for the following, however, to use an action
of Nambu-Goto type for the string, as we have done in {\sc Paper III}. The string tension is in the full theory, as we will see in Chapter~\ref{ch:coupling}, given by the local value of $\sqrt{\Phi \cdot \Phi}$, where the $SO(5)$ scalar product is defined by
\begin{equation}
\eqnlab{scalarproduct}
\Phi \cdot \Phi = \frac{1}{4} \Om_{ac} \Om_{bd} \Phi^{ab} \Phi^{cd},
\end{equation}
and $\Phi(x,\theta)$ is the superfield discussed above. The string tension is a superconformal scalar, \cf \Eqnref{tension_scalar}. The closest thing to a free string is when the string tension is taken to be a constant \phivev. This amounts to choosing a vacuum
expectation value for the $\phi$ field, thereby choosing moduli parameters, and not allowing the field to deviate from this value. The direction of the constant $\phi$ field is denoted by the $SO(5)$ vector $\hat{\phi}^{ab}$, which evidently obeys $\Om_{ab} \hat{\phi}^{ab}=0$ and has unit length with respect to the scalar product~\eqnref{scalarproduct}. The Nambu-Goto term for the string is then
\begin{equation}
\eqnlab{freeNG}
S_{\sss NG} = - \phivev \int_{\Si} d^2 \si \sqrt{-G},
\end{equation}
where $G$ denotes the determinant of the induced
metric, which is given by
\begin{equation}
\eqnlab{ind_metric}
G_{ij} = \frac{1}{2} \eps_{\al \be \ga \de} E^{\al \be}_i E^{\ga \de}_j.
\end{equation}
The superspace differential is defined in \Eqnref{EE}, such that $E^{\al \be} \equiv d \si^i E^{\al \be}_i$.

In this section, we will only consider how the Nambu-Goto term behaves under supersymmetry and $\ka$-symmetry transformations, postponing the action of the superconformal symmetry group to Chapter~\ref{ch:coupling}, where we discuss the coupled theory.

From \Eqnref{transf_E}, it is clear that the Nambu-Goto term as given in \Eqnref{freeNG} is supersymmetric (neither $E^{\al\be}$ nor \phivev{} are affected by the transformation). However, it cannot also be $\ka$-symmetric, which means that we must add a second supersymmetric term to the action so that the sum of the new term and $S_{\sss NG}$ is invariant under a $\kappa$-transformation. Under such a transformation, the world-sheet fields transform according to
\begin{equation}
\eqnlab{kappa}
\begin{aligned}
\de_{\kappa} X^{\al \be} &= i \Om^{ab} \kappa^{[\al}_a
  \Theta^{\be]}_b \\
\de_{\kappa} \Theta_a^{\al} &= \kappa^{\al}_a,
\end{aligned}
\end{equation}
where $\kappa^{\al}_a=\kappa^{\al}_a(\tau,\si)$ is a local fermionic
parameter obeying the constraint
\begin{equation}
\eqnlab{kappa_cond_bosonic}
\Ga^{\al}_{\ph{\al} \be} \kappa^{\beta}_a = \ga_a^{\ph{a} b}
\kappa^{\al}_b.
\end{equation}
In this expression, the projection operators are given by
\begin{equation}
\begin{aligned}
\Ga^{\al}_{\ph{\al} \be} &=  \frac{1}{2} \frac{1}{\sqrt{- G}}
\eps^{ij} E_i^{\al \ga} E_j^{\de \eps} \eps_{\be \ga \de \eps} \\
\ga_a^{\ph{a} b} &= \Om_{ac} \hat{\phi}^{cb},
\end{aligned}
\end{equation}
where obviously $\Ga^{\al}_{\ph{\al} \al}=0$ and
$\ga_a^{\ph{a} a}=0$. One may also show that $\Ga^{\al}_{\ph{\al} \be}
\Ga^{\be}_{\ph{\be} \ga} = \de^{\al}_{\ph{\al} \ga}$ and
$\ga_a^{\ph{a} b} \ga_b^{\ph{b} c} = \de_a^{\ph{a} c}$. This means
that the condition \eqnref{kappa_cond_bosonic} eliminates half of the
components in $\kappa^\al_a$. This constraint is necessary, since we want $\kappa^\al_a$ to contain exactly eight components, to be used to eliminate the eight surplus components of the embedding field $\Theta^\al_a$.

The new term to be added to $S_{\sss NG}$ is found to be
\begin{equation}
S_{\sss WZ} = - \frac{i}{2} \phivev \int d^2 \sigma \epsilon^{i j}
\epsilon_{\alpha \beta \gamma \delta} \left( E_i^{\alpha \beta} -
\frac{i}{2} \Theta^{\alpha}_c \Om^{cd} \partial_i \Theta^{\beta}_d
\right) \Theta^{\gamma}_a \hat{\phi}^{ab} \partial_j \Theta^{\delta}_b.
\end{equation}
The subscript is for Wess-Zumino, since it has the structure of a Wess-Zumino term~\cite{Henneaux:1985}. It can be shown to be supersymmetric (although not manifestly), meaning that the sum $S_{\sss NG}+S_{\sss WZ}$ is both supersymmetric and $\kappa$-symmetric. These two terms together form the action for a free string, and the relative coefficient is determined by the requirement of $\kappa$-symmetry.

\chapter{The interacting theory}
\label{ch:coupling}

In the preceding two chapters, we introduced and discussed the theories describing a free tensor multiplet and a free string in six dimensions. The obvious next step is to couple them together in a way that respects all the symmetries of the superconformal group. This is the subject of the present chapter.

\section{Background coupling}

Formulating the complete superconformal theory of $(2,0)$ tensor multiplets and self-dual strings is complicated, and perhaps not even the right thing to do. Therefore, it is worthwhile to start with a simpler problem --- the coupling of a self-dual string to a \emph{background} consisting of tensor multiplet fields. This means that the tensor multiplet fields are taken to obey their free equations of motion; our task is to find a theory that shows how strings behave in such a field configuration.


\subsection{The bosonic case}
\label{sec:coupling_bosonic}

It is instructive to start from a purely bosonic theory, involving the scalar fields $\phi^{ab}$ and a three-form $h$, as presented in Chapter~\ref{ch:TM}. There is also a chiral two-form potential $b$ associated to $h$, such that $h=db$. Note that it is only the self-dual part of $h$, written as $f \equiv (h+*h)$, that is allowed to couple to the string; we demand the anti self-dual part, written as $(h-*h)$, to decouple and act as a spectator field. We will employ standard form notation throughout this chapter.

Since $h=db$, the three-form field strength must obey the Bianchi identity
\begin{equation}
 dh=0.
\end{equation}
We also take the tensor multiplet fields to obey the equations of motion for a free tensor multiplet, \ie
\begin{equation}
\begin{aligned}
 \pa \cdot \pa \phi^{ab} &= 0 \\
 d (*h) &= 0,
\end{aligned}
\end{equation}
which are essentially the same as those in \Eqnref{TM_eom}. In a complete theory, the right-hand sides of these equations would be modified.

To incorporate the coupling between the string and the background, we need an interacting version of the Nambu-Goto term in \Eqnref{freeNG}. As indicated in Chapter~\ref{ch:freestring}, this is written as
\begin{equation}
\eqnlab{NG_bos}
S_{\sss NG} = - \int_{\Si} d^2 \si \sqrt{\phi \cdot \phi} \sqrt{-g},
\end{equation}
where the string tension is given by the local value of $\sqrt{\phi \cdot \phi}$. Obviously, $g$ denotes the determinant of the induced bosonic metric, given explicitly by
\begin{equation}
g_{ij} = \frac{1}{2} \eps_{\al\be\ga\de} \pa_i X^{\al\be} \pa_j X^{\ga\de}.
\end{equation}
Note that the Nambu-Goto term is not just a kinetic term for the string; the appearance of $\sqrt{\phi \cdot \phi}$ makes it act as a coupling term.

The electric string coupling is introduced in the standard way through the term
\begin{equation}
S_{\sss E} = - \int_{\Si} b,
\end{equation}
where the integral is over the string world-sheet $\Si$. It is
understood that the integrand is the pull-back of the two-form
$b$ to the string world-sheet. By employing Stokes' theorem, this interaction term can be rewritten as
\begin{equation}
\eqnlab{el_coup}
S_{\sss E} = - \int_{D} h,
\end{equation}
where $D$ is some three-dimensional manifold such that its boundary $\pa D =
\Si$. The choice of $D$ is of course not allowed to affect the
dynamics of the theory; it is similar to the ''Dirac string'' in the
theory for magnetic monopoles in four
dimensions~\cite{Nepomechie:1985,Teitelboim:1986,Dirac:1948,Deser:1998}. We will
call $D$ the world-volume of a ''Dirac membrane'' ending on the
string.

Since the string is self-dual, we expect it to couple both electrically and magnetically to the gauge field. This is accomplished by generalizing the electric coupling term \eqnref{el_coup} according to
\begin{equation}
\eqnlab{WZ_bos}
S_{\sss WZ} = - \int_{D} (h+*h) = - \int_{D} f.
\end{equation}
From this term, it is manifest that only the self-dual part of the field strength couples to the string. This electromagnetic coupling term has the structure of a Wess-Zumino term, hence the labeling. It should be noted that the bosonic theory, because of the electromagnetic coupling term, suffers from a classical anomaly~\cite{Henningson:2004,Henningson:2005}. This anomaly is cancelled in the superconformal theory through the addition of fermionic degrees of freedom.

Note that the three-form $f$ is closed ($df=0$) exactly when the Bianchi identity $dh=0$ and the free equations of motion $d(*h)=0$ are fulfilled. This means that if the tensor multiplet fields obey the free equations of motion, the Wess-Zumino term is independent of the choice of Dirac membrane $D$ according to Stokes' theorem (if we disregard the anomaly discussed above).

Summing up, we have found that
\begin{equation}
S_{\sss INT} = S_{\sss NG} + S_{\sss WZ} = - \int_{\Si} d^2 \si \sqrt{\phi \cdot \phi} \sqrt{-g} - \int_{D} f
\end{equation}
describes a self-dual string in a background consisting of on-shell tensor multiplet fields.

\subsection{The superconformal case}

The obvious next step is to generalize the newly found interaction to the superconformal case, a procedure which is analogous to many similar cases where branes are coupled to background fields~\cite{Bergshoeff:1987,Cederwall:1997_3,Cederwall:1997_p}.
The recipe is to replace the fields in the interaction terms by the corresponding superfields, but in order to find the term corresponding to the Wess-Zumino term~\eqnref{WZ_bos} we also need to use \emph{superforms}. These are described in {\sc Paper III}; we will only state the main results here.

The basis for superforms is the differentials $e^{\al\be}$ and $e^\al_a$ given in \Eqnref{super_diffs}. These are tangent space differentials, and may be collected into a single superdifferential $e^{\sss A}$, where the collective superindex ${}^{\sss A}$ takes values ${}^{[\al\be]}$ (bosonic) and ${}^\al_a$ (fermionic). This superindex should not be confused with the index used in Chapter~\ref{ch:manifest} when discussing manifest superconformal symmetry.

In this notation, a super $p$-form is written as
\begin{equation}
\om = \frac{1}{p!} e^{\sss A_p} \we \ldots \we e^{\sss A_1} \om_{\sss A_1 \ldots A_p},
\end{equation}
where it should be noted that the wedge product is symmetric with respect to two fermionic differentials, otherwise antisymmetric. The exterior superspace derivative acts on a general superform according to
\begin{equation}
\eqnlab{superextder}
d \om = \frac{1}{p!} e^{\sss A_p} \we \ldots \we e^{\sss A_1} \we e^{\sss B} \left( D_{\sss B} \om_{\sss A_1 \ldots A_p} + \frac{p}{2} T_{\sss BA_1}^{\sss \ph{BA_1}C} \om_{\sss C A_2 \ldots A_p} \right),
\end{equation}
where the superderivative $D_{\sss A}$ has the components
\begin{equation}
\begin{aligned}
  D_{[\al\be]} &= \pa_{\al\be} \\
  D^a_\al &= \pa^a_\al + i \Om^{ab} \theta^\be_b \pa_{\al\be}.
\end{aligned}
\end{equation}
Geometrically, these are tangent vectors, dual to the superspace differentials.
We see that the fermionic piece agrees with the superderivative in \Eqnref{superd}. Furthermore, the torsion two-form $T$ has only
one non-zero component, namely
\begin{equation}
 T_{\be \ga}^{bc [\al_1 \al_2]} = -2i \de_{\be}^{\ph{\be} [\al_1}
   \de_{\ga\ph{\be}}^{\ph{\ga} \al_2]} \Om^{bc},
\end{equation}
which will play an important role in the following. Its origin is the relation
\begin{equation}
d e^{\al \be} = i \Om^{ab} d \theta^\al_a \we d\theta^\be_b,
\end{equation}
which is a new ingredient compared to the bosonic theory.

After these preliminaries, we turn to the interaction terms. The Nambu-Goto term in \Eqnref{NG_bos} is easily generalized by replacing
$\sqrt{\phi \cdot \phi}$ by the corresponding superfield $\sqrt{\Phi \cdot \Phi}$, while we replace the bosonic $\sqrt{-g}$ by $\sqrt{-G}$ where $G_{ij}$ is the induced superspace metric given in \Eqnref{ind_metric}.

We know that $\sqrt{\Phi \cdot \Phi}$ is a superconformal scalar and transforms according to~\Eqnref{transf_tension}. Explicitly, this means that
\begin{equation}
\de \left( \sqrt{\Phi \cdot \Phi} \right) = \left[ \xi^{\al\be}(x,\theta) \pa_{\al\be} + \xi^\al_a(x,\theta) \pa^a_\al + 2 \La(x,\theta) \right] \sqrt{\Phi \cdot \Phi},
\end{equation}
but this is the expression in superspace. In the Nambu-Goto term, this field is evaluated on the string world-sheet $\Si$, which means that we have to take into account that the embedding fields  $X^{\al\be}$ and $\Theta^\al_a$ change as well. Putting this together, we find that
\begin{equation}
\de \left( \sqrt{\Phi \cdot \Phi} \right) = 2 \La(X,\Theta) \sqrt{\Phi \cdot \Phi}
\end{equation}
on the world-sheet, since the embedding fields transform as $\de X^{\al\be} = - \xi^{\al\be}(X,\Theta)$ and $\de \Theta^{\al}_a = - \xi^{\al}_a(X,\Theta)$. This observation is very useful; we note that the differential pieces of a superconformal field transformation will always vanish when we transform the pull-back of a field to the string world-sheet.

It should also be emphasized that this calculation is only valid if the tensor multiplet fields obey their free equations of motion, since the superfield formulation is only valid on-shell.

Noting that
\begin{equation}
\de \left( \sqrt{-G} \right) = - 2 \La(X,\Theta) \sqrt{-G},
\end{equation}
which follows from \Eqnref{transf_E}, it is clear that the Nambu-Goto term
\begin{equation}
S_{\sss NG} = \int_{\Si} d^2 \si \sqrt{\Phi \cdot \Phi} \sqrt{-G}
\end{equation}
is invariant under superconformal transformations, as required.

The generalization of the Wess-Zumino term is analogous. We want to replace the three-form $f$ by some super three-form $F$. Since $f$ is closed on-shell, we expect the corresponding super three-form to be closed with respect to the exterior superspace derivative \eqnref{superextder}, such that $dF=0$, when the differential constraint \eqnref{diffcon} is valid. Furthermore, we expect the superconformal Wess-Zumino term to reduce to the bosonic version if all fermionic degrees of freedom are removed.

It turns out that the only possible super three-form $F$ that is consistent with the requirements above is given by
\begin{multline}
\eqnlab{super_F}
F = - \frac{1}{6} e^{\ga_1 \ga_2} \we e^{\be_1 \be_2} \we e^{\al_1 \al_2} \eps_{\al_1 \be_1 \ga_1 \ga_2} H_{\al_2 \be_2} - {} \\ {} - \frac{i}{2} e^{\ga_1 \ga_2} \we e^{\be_1 \be_2} \we e^\al_a \eps_{\al \be_1 \ga_1 \ga_2} \Psi^a_{\be_2} - \frac{i}{2} e^{\ga_1 \ga_2} \we e^{\be}_b \we e^\al_a \eps_{\al \be \ga_1 \ga_2} \Phi^{ab}
\end{multline}
in terms of the superfields of the $(2,0)$ tensor multiplet. A similar superform was constructed in~\cite{Grojean:1998,Howe:2000}. This expression may also be found by solving the Bianchi identity $dF=0$, if certain constraints are imposed on the components of $F$.

Let us see how this superform, or rather, its pull-back to $D$, transforms under superconformal transformations. The superspace differentials and the embedding fields for $D$ transform naturally in the same way as the differentials and the embedding fields for the string world-sheet $\Si$. By considering the relevant transformation laws, it turns out that the only non-trivial thing that we need to check is the terms containing the superspace-dependent parameter function $R^a_{\al}(\theta)$, defined in \Eqnref{Raal}. These connect the superfields to each other and also the variation of $E^\al_a$ to $E^{\al\be}$.

It turns out that the pull-back of the super three-form $F$ to the Dirac membrane world-volume $D$ is superconformally invariant if and only if the coefficients are chosen as in \Eqnref{super_F}, i.e., with the same choice of coefficients that makes it a closed form in superspace! Thus, the Wess-Zumino term
\begin{equation}
S_{\sss WZ} = - \int_{D} F
\end{equation}
is superconformally invariant, but only with this specific three-form. This was the main result of {\sc Paper V}.

So, our candidate for the interaction is
\begin{equation}
\eqnlab{superinteraction}
S_{\sss INT} = - \int_{\Si} d^2 \si \sqrt{\Phi \cdot \Phi} \sqrt{-G} -
\int_{D} F.
\end{equation}
All that remains is to investigate its properties under the local
fermionic $\kappa$-symmetry. The embedding fields transform according
to
\begin{equation}
\begin{aligned}
\de_{\kappa} X^{\al \be} &= i \Om^{ab} \kappa^{[\al}_a
  \Theta^{\be]}_b \\
\de_{\kappa} \Theta_a^{\al} &=  \kappa^{\al}_a,
\end{aligned}
\end{equation}
where the parameter $\kappa^\al_a$ obeys
\begin{equation}
\eqnlab{kappa_cond}
\Ga^{\al}_{\ph{\al} \be} \kappa^{\beta}_a = \ga_a^{\ph{a} b}
\kappa^{\al}_b.
\end{equation}
Here, the projection operators are
\begin{equation}
\begin{aligned}
\Ga^{\al}_{\ph{\al} \be} &= \frac{1}{2} \frac{1}{\sqrt{- G}}
\eps^{ij} E_i^{\al \ga} E_j^{\de \eps} \eps_{\be \ga \de \eps} \\
\ga_a^{\ph{a} b} &= \frac{1}{\sqrt{\Phi \cdot \Phi}} \Om_{ac}
\Phi^{cb},
\end{aligned}
\end{equation}
which should be compared with the corresponding expressions in Chapter~\ref{ch:freestring}. The relation \eqnref{kappa_cond} eliminates half of the components in $\kappa^\al_a$, as before.

By using the closedness of $F$, it is straight-forward to show that the
interaction \eqnref{superinteraction} indeed is $\kappa$-symmetric. The details may be found in Chapter 4 of {\sc Paper III}. It should be stressed that this calculation also determines the relative coefficient in the interaction.

Thus, the superconformally invariant interaction between a self-dual string and a background consisting of $(2,0)$ tensor multiplet fields is described by \Eqnref{superinteraction}. It should be emphasized that this theory does not suffer from the anomaly mentioned in the bosonic case~\cite{Henningson:2004}. The classical anomaly associated with the electromagnetic coupling is cancelled by a quantum anomaly from the chiral fermions on the string world-sheet. The cancellation demands that the model obeys an $ADE$-classification.

The background coupling described here may be used~\cite{Flink:2005} to calculate supersymmetric amplitudes for tensor multiplet particles scattering off an infinitely long straight string. The corresponding calculation in the bosonic case was done in {\sc Paper I}.

Finally, let us consider the limit when $\Phi^{ab}$ is
constant and purely bosonic, \ie $\Phi^{ab}=\phivev
\hat{\phi}^{ab}$. This means that the fermionic component field
$\psi^a_{\al}$ is zero and we may take also $h_{\al \be}=0$.
In this limit, the interaction
\eqnref{superinteraction} reduces exactly to the free string action
obtained in Chapter~\ref{ch:freestring}. This also motivates the labeling of the
Wess-Zumino term appearing there.

\section{Off-shell coupling}

In the previous section, we formulated a model for a self-dual string interacting with a background consisting of free tensor multiplet fields. The purpose of the present section is to discuss how this model can be generalized to incorporate coupling terms in the equations of motion for the tensor multiplet fields as well.

{\sc Paper IV} outlines a way to construct a unique $(2,0)$ supersymmetric action in six dimensions, describing a tensor multiplet interacting with a self-dual string. This involves a so called Dirac-Dirac term, describing a direct string-string interaction. The purpose of this term is to make the action both supersymmetric and invariant under a local symmetry. This symmetry is a generalization of the fermionic $\kappa$-symmetry and allows us to choose the Dirac membrane world volume freely and eliminate half the fermionic degrees of freedom on the string world-sheet.

In retrospect, the path taken in {\sc Paper IV} may not be the best to formulate the full interacting theory. In fact, our model may not even be a consistent theory for high energies. When the excitation energies approach the same order of magnitude as the square root of the string tension, additional effects may come into play. This is also indicated from the higher-dimensional origin described in Chapter~\ref{ch:origin}. Therefore, we will not discuss the complete interacting theory any further in this thesis. A better way to obtain clues to what the final theory should look like is perhaps through a superspace generalization of the results obtained in Ref.~\cite{Henningson:2005}.


%
%

\clearpage

\pagestyle{plain}
\def\href#1#2{#2}
\bibliographystyle{utphysmod3b}
\addcontentsline{toc}{part}{\sffamily\bfseries Bibliography}
\bibliography{biblio}

\end{document}